\documentclass[a4paper,fleqn,usenatbib,useAMS]{mnras}

\usepackage{newtxtext,newtxmath}
\usepackage{graphicx}  
\usepackage{multicol}        
\usepackage{bm}    
\usepackage{pdflscape}  
\usepackage{multirow}
\usepackage{xcolor}
\usepackage{cuted}
\usepackage{hyperref}
\usepackage{booktabs}
\usepackage{subfigure}
\usepackage{academicons}
\usepackage{natbib}
\usepackage{orcidlink}
\usepackage{xspace}
\usepackage{amsmath}
\usepackage{tikz}
\usetikzlibrary{positioning,calc}

\numberwithin{equation}{section}

\newcommand{\orcid}[1]{\href{https://orcid.org/#1}{\textcolor[HTML]{A6CE39}{\aiOrcid}}}

\newcommand\dd{\mathrm d}
\newcommand\mg{\mathrm g}
\newcommand\mh{\mathrm h}
\newcommand\me{\mathrm e}

\title[Inflow-driven galaxy evolution -- I]{Inflow-driven galaxy evolution -- I. Revealing the physics of the fundamental metallicity relation}

\author[Wang et al.]{Kai Wang,$^{1, 2}$\thanks{Contact e-mail: wkcosmology@gmail.com}\orcidlink{0000-0002-3775-0484}
  Carlton Baugh,$^{1,3}$\orcidlink{0000-0002-9935-9755}
  N. F. Boardman,$^{4}$\orcidlink{0000-0001-8197-5798}
  Sownak Bose,$^{1}$\orcidlink{0000-0002-0974-5266}
  Zheng Cai,$^{5}$\orcidlink{0000-0002-5954-7903}
  Shaun Cole,$^{1}$\orcidlink{0000-0002-5954-7903}
  \newauthor
  Carlos S. Frenk,$^{1}$\orcidlink{0000-0002-2338-716X}
  Cedric Lacey,$^{1}$\orcidlink{0000-0001-9016-5332}
  Peder Norberg,$^{1,2}$\orcidlink{0000-0002-5875-0440}
  Yingjie Peng,$^{6,7}$\orcidlink{0000-0003-0939-9671}
  Isabel Santos-Santos,$^{8}$
  \newauthor
  Amélie Saintonge,$^{9, 10}$\orcidlink{0000-0003-4357-3450}
  Tom Theuns,$^{1}$\orcidlink{0000-0002-3790-9520}
  and Enci Wang$^{11, 12}$\orcidlink{0000-0003-1588-9394}
  \\
  $^{1}$Institute for Computational Cosmology, Department of Physics, Durham University, South Road, Durham, DH1 3LE, UK\\
  $^{2}$Centre for Extragalactic Astronomy, Department of Physics, Durham University, South Road, Durham DH1 3LE, UK\\
  $^{3}$Institute for Data Science, Durham University, South Road, Durham DH1 3LE, UK\\
  $^{4}$School of Physics and Astronomy, University of St Andrews, North Haugh, St Andrews KY16 9SS, UK\\
  $^{5}$Department of Astronomy, Tsinghua University, Beijing 100084, People's Republic of China\\
  $^{6}$Department of Astronomy, School of Physics, Peking University, Beijing 100871, China\\
  $^{7}$Kavli Institute for Astronomy and Astrophysics, Peking University, Beijing 100871, China\\
  $^{8}$Leibniz-Institut für Astrophysik Potsdam (AIP), An der Sternwarte 16, 14482 Potsdam, Germany\\
  $^{9}$Department of Physics \& Astronomy, University College London, London WC1E 6BT, UK\\
  $^{10}$Max Planck Institute for Radio Astronomy, Auf dem Hügel 69, D-53121 Bonn, Germany\\
  $^{11}$Department of Astronomy, University of Science and Technology of China, Hefei 230026, China\\
  $^{12}$School of Astronomy and Space Science, University of Science and Technology of China, Hefei 230026, China\\
}

\date{Accepted XXX. Received YYY; in original form ZZZ}

\pubyear{\the\year{}}

\begin{document}
\raggedbottom
\label{firstpage}
\pagerange{\pageref{firstpage}--\pageref{lastpage}}
\maketitle

\begin{abstract}
  We present a unified physical framework for the fundamental metallicity relation (FMR), based on the mass-continuity equations governing the baryon cycle in galaxies.
  The FMR is not merely the anti-correlation between star formation rate (SFR) and gas metallicity ($Z_\mg$) at fixed stellar mass ($M_\star$); it is a redshift-invariant surface in the $(M_\star, \mathrm{SFR}, Z_\mg)$ space.
  We construct a minimal cosmological gas flow model, calibrated to reproduce the mass--metallicity relation, the star-forming main sequence, and the stellar-to-halo mass relation from $z \sim 0$ to $\sim 3$, and show that the FMR emerges as a prediction of the calibrated physics.
  Through controlled experiments that progressively simplify the model, we reveal that in a universe where both the star formation efficiency ($\epsilon$) and mass-loading factor ($\eta$) are constants, the FMR reduces to a single, universal scaling between $Z_{\rm g}$ and $M_\star/\mathrm{SFR}$, whose shape traces the transition from the inflow-driven regime to equilibrium.
  The specific parameterisation of the observed FMR is not a fundamental symmetry but a contingent consequence of how $\epsilon$ and $\eta$ depend on stellar mass and redshift.
  We show that the gaseous FMR (gFMR), defined in the $(M_\star, M_\mg, Z_\mg)$ space, is more fundamental than the standard FMR: in the inflow-driven limit, $Z_\mg$ is proportional to $M_\star/M_\mg$ independently of $\epsilon$, and the approach to equilibrium is governed by $M_\star/M_\mg$ and $\eta$ alone.
  We derive an analytic solution for an idealised version of the model that provides closed-form expressions relating $Z_{\rm g}$, $M_{\rm g}/M_\star$, and $\eta$, enabling any one of these three quantities to be inferred when the other two are known, and show that this framework accurately reproduces the minimal cosmological gas flow model.
  By establishing the physical origin of the FMR and its connection to the more fundamental gFMR, we provide the theoretical foundation to turn metallicity scaling relations into precision probes of the baryon cycle over cosmic history.
\end{abstract}

\begin{keywords}
  galaxies: fundamental parameters - galaxies: evolution - galaxies: ISM - ISM: evolution
\end{keywords}



\section{Introduction}

The metal content of galaxies encodes the integrated history of gas inflows, star formation, and feedback-driven outflows over cosmic time \citep{coleHierarchicalGalaxyFormation2000, deluciaChemicalEnrichmentIntracluster2004, baughCanFaintSubmillimetre2005, nagashimaMetalEnrichmentElliptical2005, moGalaxyFormationEvolution2010, hirschmannGalaxyAssemblyStellar2016}.
Metals are synthesised in stellar interiors and returned to the interstellar medium (ISM) through supernova explosions and stellar winds, from where they may be ejected into the circumgalactic medium or intergalactic medium by feedback-driven outflows, or locked into subsequent generations of stars.
Metallicity does not vary in isolation: it participates simultaneously in multiple tight scaling relations with stellar mass \citep[e.g.][]{tremontiOriginMassMetallicityRelation2004}, star formation rate (SFR) \citep[e.g.][]{ellisonCluesOriginMassMetallicity2008, mannucciFundamentalRelationMass2010, curtiMassmetallicityFundamentalMetallicity2020, looserStellarFundamentalMetallicity2024}, gas mass \citep[e.g.][]{bothwellFundamentalRelationMetallicity2013, maOriginEvolutionGalaxy2016, lagosFundamentalPlaneStar2016, deluciaGasAccretionRegulates2020, zuHIGasContent2020}, galaxy size \citep[e.g.][]{ellisonCluesOriginMassMetallicity2008, deugenioGasphaseMetallicitiesStarforming2018, maRevisitingFundamentalMetallicity2024, sanchez-menguianoMoreFundamentalFundamental2024, boardmanCompetingEffectsRecent2025, jiaPotentialdrivenMetalCycling2025, liCentralVelocityDispersion2025, wangOriginGalaxySizestellar2026}, and environment \citep[e.g.][]{pasqualiGasphaseMetallicityCentral2012, pengStrangulationPrimaryMechanism2015, baheOriginEnhancedMetallicity2017, gallazziGalaxyEvolutionEnvironments2021, wangMassMetallicityRelationCosmic2022, wangEnvironmentalDependenceMassMetallicity2023}, in both observations and models.
This web of correlations makes chemical abundance one of the most powerful diagnostics of galaxy formation and evolution, encoding information about the baryon cycle that is difficult to access by other means.

The most studied of these scaling relations is the stellar mass--metallicity relation \citep[MZR;][]{tremontiOriginMassMetallicityRelation2004, gallazziAgesMetallicitiesGalaxies2005}, which indicates  that at fixed cosmic epoch, more massive galaxies are systematically more metal-rich.
The MZR rises steeply as a power law toward higher stellar mass for $M_\star \lesssim 10^{10.5}\,\mathrm{M}_\odot$, before flattening at higher masses still.
Crucially, the MZR evolves with redshift: at fixed stellar mass, galaxies at higher redshift are systematically more metal-poor in both observations \citep{maiolinoAMAZEEvolutionMassmetallicity2008, zahidUniversalRelationGalactic2014, sandersMOSDEFSurveyEvolution2021, liMassMetallicityRelationDwarf2023, jainUniformAnalysisGasphase2025} and hydrodynamical simulations \citep{derossiGalaxyMetallicityScaling2017, torreyEvolutionMassmetallicityRelation2019, garciaDoesFundamentalMetallicity2025}.

The scatter about the MZR is itself structured.
Galaxies with higher SFR at fixed stellar mass tend to be more metal-poor \citep{ellisonCluesOriginMassMetallicity2008}, and \citet{mannucciFundamentalRelationMass2010} demonstrated that stellar mass, SFR, and gas metallicity together define a tight three-dimensional surface, the fundamental metallicity relation (FMR), with a residual scatter of only $\sim 0.05$~dex \citep[see also][]{yatesRelationMetallicityStellar2012, andrewsMassMetallicityRelationDirect2013, salimCriticalLookMassMetallicityStar2014, curtiMassmetallicityFundamentalMetallicity2020, sandersMOSDEFSurveyEvolution2021, jainUniformAnalysisGasphase2025}.
Strikingly, this surface appears redshift-invariant out to at least $z \sim 3$--$3.5$ \citep{sandersMOSDEFSurveyEvolution2021, curtiJADESInsightsLowmass2024, jainUniformAnalysisGasphase2025}, suggesting a universal mechanism linking these three quantities.
Whether the FMR extends to higher redshifts remains an open question: some JWST studies report systematic offsets from the locally defined surface at $z \gtrsim 5$ \citep{curtiJADESInsightsLowmass2024}.
Meanwhile, because the star-forming main sequence (SFMS) evolves with redshift such that high-$z$ galaxies have systematically higher SFR at fixed stellar mass \citep[see also][]{whitakerStarFormationMass2012, speagleHighlyConsistentFramework2014}, the anti-correlation between SFR and metallicity encoded in the FMR naturally connects the SFMS evolution to the observed evolution of the MZR.
This suggests that the MZR, the FMR, and the SFMS are all governed by the same underlying physical process.

\citet{bothwellFundamentalRelationMetallicity2013} showed that replacing SFR with the atomic hydrogen mass yields an equally tight or tighter correlation, suggesting that the gas content may be the more fundamental second parameter \citep[see also][]{santiniEvolutionDustGas2014, zahidUniversalRelationGalactic2014, bothwellMolecularGasDriver2016, bothwellGalaxyMetallicitiesDepend2016, brownRoleAtomicHydrogen2018, scholteAtomicGasSequence2024, boardmanFundamentalMetallicityRelation2026}.
In this picture, the SFR enters the standard FMR only as a proxy for the gas mass through the star formation efficiency $\epsilon \equiv \mathrm{SFR}/M_{\rm g}$.
If confirmed, this gaseous FMR (gFMR) would point to a direct physical link between metallicity and the gas reservoir rather than an indirect one mediated by star formation.
However, the physical origin of the gFMR and its relationship to the standard FMR have not been established within a self-consistent theoretical framework.

Several analytic frameworks have sought to explain these scaling relations.
The gas regulator class of models, in which a galaxy is treated as a reservoir regulated by the balance of inflows, star formation, and outflows \citep{boucheImpactColdGas2010, daveAnalyticModelEvolution2012, lillyGASREGULATIONGALAXIES2013}, provides the most developed theoretical picture.
These models reproduce the average trends in galaxy evolution, but are inherently incapable of explaining the FMR.
In equilibrium, the metallicity approaches a fixed value determined solely by the mass-loading factor and yield, independently of other galaxy properties, and therefore no anti-correlation between SFR and metallicity can arise.
\citet{lillyGASREGULATIONGALAXIES2013} attempted to circumvent this by assuming $\mathrm{d}Z_\mg/\mathrm{d}t = 0$ while allowing the gas reservoir to vary.
As \citet{forbesOriginFundamentalMetallicity2014} pointed out, this is inconsistent: the metallicity equilibrates on a timescale longer than that of the gas mass, so assuming it is in steady state while the gas mass is not is physically unjustified (see \S\,\ref{sub:revisiting_the_equilibrium_model} for more discussion of the gas regulator model).
\citet{forbesOriginFundamentalMetallicity2014} showed instead that stochastic fluctuations in the gas accretion rate, which drive galaxies away from equilibrium, naturally produce the anti-correlation between SFR and metallicity at fixed stellar mass \citep[see also][]{torreySimilarStarFormation2018, deluciaGasAccretionRegulates2020, wangGasphaseMetallicityDiagnostic2021, maRevisitingFundamentalMetallicity2024}.
However, as we will argue, the FMR is not merely the anti-correlation between SFR and metallicity: it requires that galaxies populate a redshift-invariant surface in the $(M_\star, \mathrm{SFR}, Z_{\rm g})$ space, and stochastic fluctuations alone do not explain why this surface exists or why it does not evolve (see \S\,\ref{sub:other_processes_in_regulating_metallicity_evolution} for more discussion).
\citet{dayalPhysicsFundamentalMetallicity2013} reproduced the shape of the local FMR by assuming that the gas inflow rate is proportional to the SFR, a prescription that lacks physical justification since the inflow rate is set by cosmological accretion and halo-scale cooling rather than by the instantaneous star formation activity.
This assumption allows the FMR to be fitted but does not explain why it exists or why it remains redshift-invariant while the MZR evolves.
\citet{linConstraintsGalacticOutflows2023} reproduced the redshift-invariant FMR without imposing equilibrium, attributing its emergence to the self-similarity of star formation histories and the coherent enrichment of galaxies across cosmic time, but did not identify the physical origin of the relation or the conditions under which it holds (see \S\,\ref{sub:degeneracy} for more discussion).
Beyond this framework, \citet{daveAnalyticModelEvolution2012} anticipated that metal-enriched recycling of previously ejected gas could drive the redshift evolution of the MZR.
However, semi-analytic models that implement such recycling still predict a nearly redshift-independent MZR \citep[e.g.][]{luSemianalyticModelsCANDELS2014, somervilleStarFormationSemianalytic2015, guoGalaxiesEAGLEHydrodynamical2016}, and reproducing the observed evolution required \citet{yatesLGALAXIES2020Evolution2021} to invoke direct ejection of metals into the CGM alongside significant parameter re-tuning.
None of these approaches provides a self-consistent, unified account of the redshift-invariant FMR together with the redshift-evolving MZR and SFMS.

A common thread in all of the above models is the assumption that galaxies reside close to a quasi-equilibrium state, in which metallicity is set by the instantaneous balance between inflows, star formation, and outflows rather than by the history of star formation and gas processing.
In this work, we depart from this assumption by considering the opposite limit: the inflow-driven regime \citep[as first introduced by][]{wangOriginGalaxySizestellar2026}, in which gas accretion dominates over depletion and galaxies have not yet had time to reach equilibrium \footnote{
  Some authors describe any regime that is not equilibrium as ``non-equilibrium'', but this term suggests a disordered state.
  We prefer ``inflow-driven'' to emphasise that it is a well-defined regime with clean analytic properties, not merely the absence of equilibrium.}.
We show that the transition from this inflow-driven regime to equilibrium provides a unified framework for understanding the MZR, the FMR, and the gFMR simultaneously.

The central question we address is why the FMR exists as a redshift-invariant surface, and whether its parameterisation encodes fundamental physics or merely reflects the contingent dependence of star formation efficiency ($\epsilon$) and loading factor ($\eta$) on stellar mass and redshift.
Our approach proceeds in two stages.
First, we construct a cosmological gas flow model that tracks gas accretion from dark matter halo growth, cooling, star formation, and feedback-driven outflows, and calibrate it to reproduce three observational constraints simultaneously: the redshift-evolving MZR, the SFMS, and the stellar mass--halo mass relation.
Without further tuning, this model predicts both the standard FMR and the gFMR as emergent consequences of the calibrated physics.
We then perform controlled experiments, systematically varying the assumptions about the star formation efficiency and mass-loading factor, to isolate the physical origin of both relations.
Second, we derive an analytic solution for an idealised version of the model with constant inflow rate, star formation efficiency, and mass-loading factor.
This analytic solution provides a closed-form expression for the gas metallicity as a function of the gas fraction and the evolutionary stage $t/\tau_{\rm eq}$, and interpolates continuously between the inflow-driven limit and equilibrium.

The remainder of this paper is organised as follows.
\S~\ref{sec:gas_flow_model} introduces the cosmological gas flow model and its calibration to observed scaling relations.
\S~\ref{sec:fundamental_metallicity_relation} demonstrates the emergence of the FMR and gFMR from the calibrated model, and uses controlled experiments to reveal the physical origin of both relations.
\S~\ref{sec:analytic_approximate} derives the analytic solution under the ideal model approximation and validates it against the full cosmological calculation.
\S~\ref{sec:discussion} revisits previous analytic frameworks in light of our results and discusses the broader implications of our findings.
\S~\ref{sec:summary} summarises our conclusions. Throughout this work we adopt a flat $\Lambda$CDM cosmology with $H_0 = 70\,\mathrm{km}\,\mathrm{s}^{-1}\,\mathrm{Mpc}^{-1}$, $\Omega_{\rm m} = 0.3$, and $\Omega_\Lambda = 0.7$, and that stars are formed with a mass distribution given by a \citet{chabrierGalacticStellarSubstellar2003} stellar initial mass function.

\section{The gas flow model}
\label{sec:gas_flow_model}

\begin{figure}
  \begin{center}
    \includegraphics[width=0.9\linewidth]{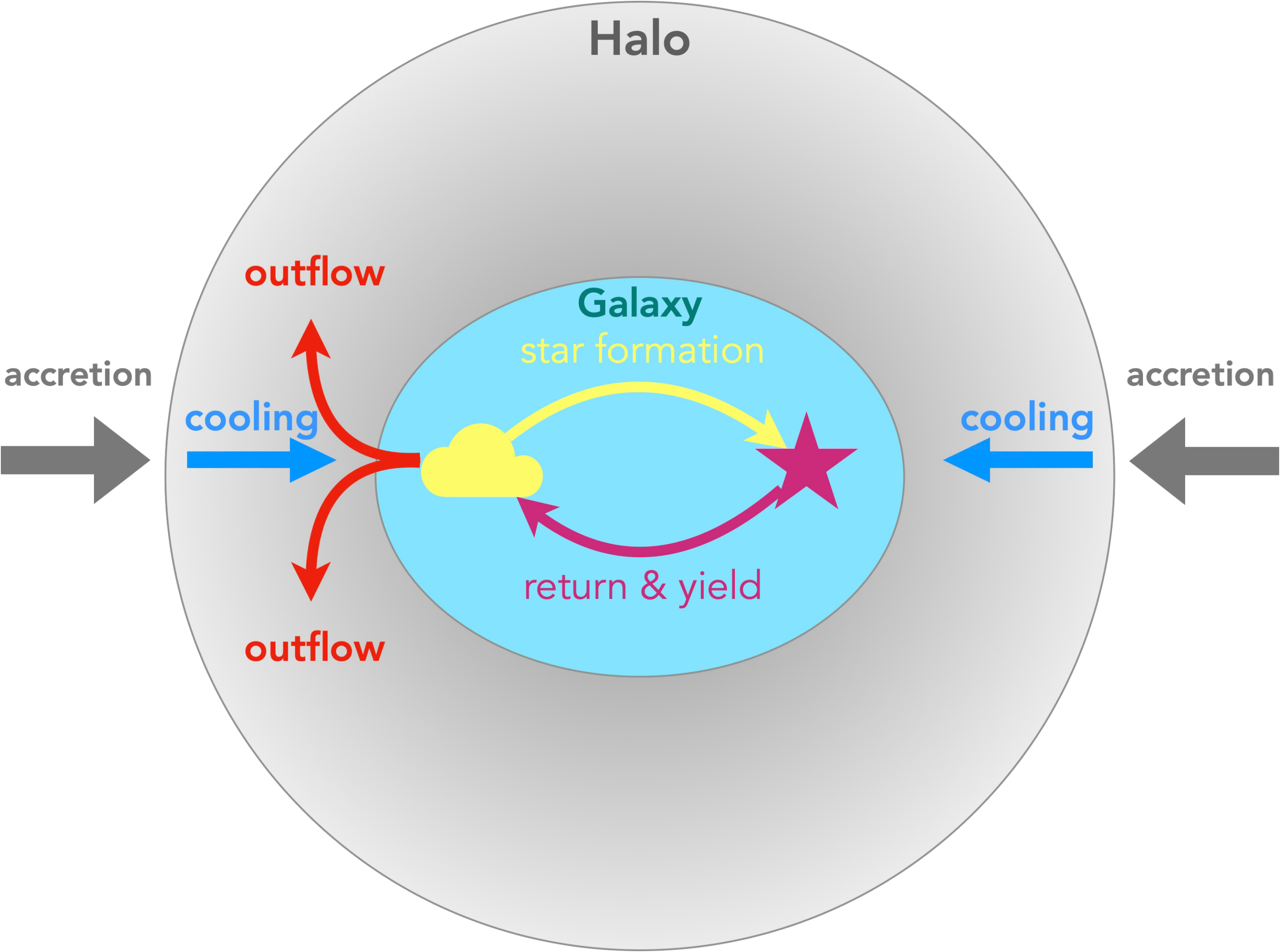}
  \end{center}
  \caption{
    Schematic illustration of the gas flow model adopted in this work.
    Gas is accreted from the intergalactic medium onto the dark matter halo, where it is shock-heated and added to the hot halo atmosphere.
    A fraction of this gas cools and settles onto the central galaxy, providing the fuel for star formation.
    Massive stars return a fraction $R$ of the newly formed stellar mass to the ISM along with freshly synthesised metals (yield $y$).
    Stellar feedback drives outflows that eject gas from the galaxy at a rate $\eta$ times SFR.
    The model is governed by four key processes: cosmological accretion onto the halo, radiative cooling onto the galaxy, star formation and stellar recycling within the ISM, and feedback-driven outflows.
  }
  \label{fig:demo_gas_flow}
\end{figure}

Galaxy evolution begins with the accretion of baryons during the growth of the host dark matter halo.
The accreted gas cools and condenses at the halo centre to assemble the interstellar medium (ISM), with a corresponding mass inflow rate $\Phi$.
Star formation proceeds at a rate $\mathrm{SFR} = \epsilon\,M_{\rm g}$, where $\epsilon$ is the star formation efficiency and $M_{\rm g}$ is the ISM gas mass\footnote{The ISM gas mass is the sum of the atomic and molecular gas in the galaxy. Observationally, it is traced by $M_\mg = 1.36\,M_{\rm H_{\rm I}} + M_{\rm H_2}$, where the factor of 1.36 accounts for helium and heavier elements. The molecular gas community conventionally applies this correction to the molecular hydrogen mass \citep{saintongeColdInterstellarMedium2022a}, so we only apply it to the atomic hydrogen mass here.}.
Stellar evolution returns a fraction $R$ of newly formed stellar mass to the ISM, so that the net mass growth rate of long-lived stars is $(1-R)\,\mathrm{SFR}$, accompanied by metal production at a rate $y\times\mathrm{SFR}$, where $y$ is the metal yield \textit{per unit star formation}.
Both $R$ and $y$ depend only on the IMF and are treated as constants throughout.
This description assumes that newly produced metals are \textit{instantaneously and uniformly} mixed within the ISM.
Stellar feedback drives galactic outflows that eject gas at a rate $\eta\times\mathrm{SFR}$, where $\eta$ is the mass-loading factor.
We assume that the outflowing gas carries the ISM metallicity.
Higher-order processes such as gas recycling from the CGM are neglected in order to keep the model intentionally minimal; their potential effects are discussed in \S\,\ref{sub:other_processes_in_regulating_metallicity_evolution}.
Fig.~\ref{fig:demo_gas_flow} illustrates the four key processes governing the model: cosmological accretion onto the halo, radiative cooling onto the galaxy, star formation and stellar recycling within the ISM, and feedback-driven outflows.

These processes are captured by two mass-continuity equations \citep{coleHierarchicalGalaxyFormation2000},
\begin{align}
  \frac{\mathrm{d} M_{\rm g}}{\mathrm{d} t}             & = \Phi - (1-R+\eta)\,\epsilon\,M_{\rm g} , \label{eq:mgas}                             \\[4pt]
  \frac{\mathrm{d}(M_{\rm g}\,Z_{\rm g})}{\mathrm{d} t} & = y\,\epsilon\,M_{\rm g} - Z_{\rm g}\,(1-R+\eta)\,\epsilon\,M_{\rm g} ,\label{eq:zgas}
\end{align}
where $Z_{\rm g} \equiv M_Z/M_{\rm g}$ is the mass fraction of metals in the ISM, also known as the gas metallicity, and we have substituted $\mathrm{SFR} = \epsilon\,M_{\rm g}$.
Equation~\eqref{eq:mgas} states that the gas reservoir grows by inflow and is depleted by star formation (net of recycling) and outflows.
Equation~\eqref{eq:zgas} states that the metal content of the ISM increases through stellar nucleosynthesis and decreases through the same two channels, star formation and outflows, with each channel removing gas at the current ISM metallicity.
These equations have appeared in the literature under various names: \textit{bathtub model} \citep{dekelToyModelsGalaxy2013} or \textit{gas regulator model} \citep{lillyGASREGULATIONGALAXIES2013}, reflecting the historical focus on their equilibrium solutions.
Here we adopt the more general term \textit{gas flow model} to emphasise that they are fundamentally mass-continuity equations whose solutions need not be restricted to equilibrium.

Given the three input functions $\Phi(t)$, $\epsilon(t)$, and $\eta(t)$, equations~\eqref{eq:mgas} and \eqref{eq:zgas} determine the full time evolution of $M_{\rm g}(t)$ and $Z_{\rm g}(t)$.
The stellar mass and mass-weighted stellar metallicity then follow from integrating the implied star formation and enrichment histories,
\begin{align}
  M_\star(t) & = \int_0^t (1-R)\,\epsilon\,M_{\rm g}(t')\,\mathrm{d}t' \label{eq:mstar}                  \\[4pt]
  Z_\star(t) & = \frac{(1-R)}{M_\star(t)} \int_0^t \epsilon\,M_{\rm g}(t')\,Z_{\rm g}(t')\,\mathrm{d}t'.
  \label{eq:zstar}
\end{align}

\subsection{The ideal gas flow model}
\label{sub:the_ideal_gas_flow_model}

\begin{figure}
  \begin{center}
    \includegraphics[width=0.9\linewidth]{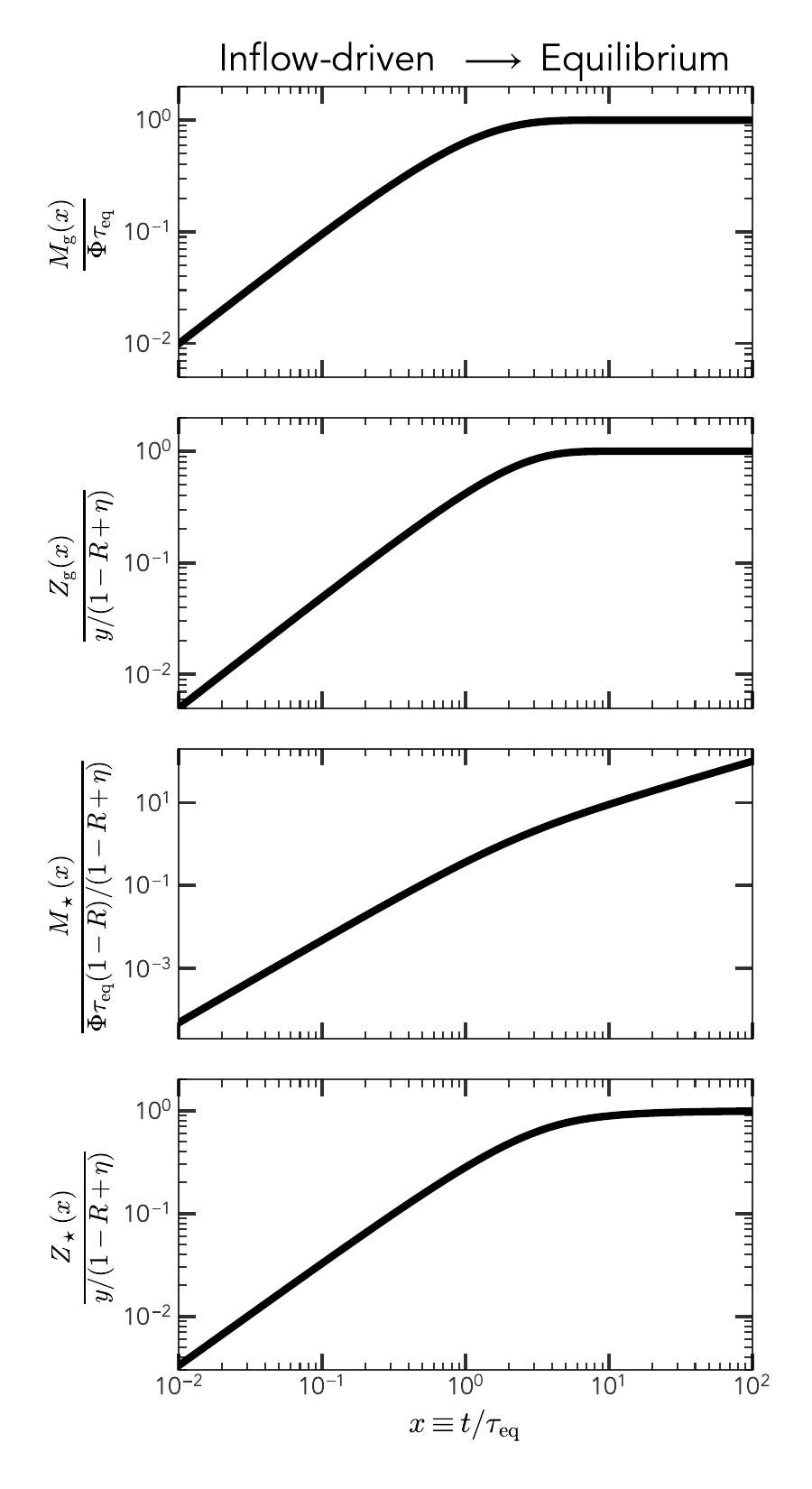}
  \end{center}
  \caption{
    Time evolution of the gas mass $M_{\rm g}(t)$, gas phase metallicity $Z_{\rm g}(t)$, stellar mass $M_\star(t)$, and stellar metallicity $Z_\star(t)$ for the idealised system with constant inflow rate $\Phi$, obtained from the analytic solution in equations.~\eqref{eq:mgas_ana}--\eqref{eq:zstar_ana}.
    Each quantity is normalised properly ($\Phi\tau_{\rm eq}$ for $M_{\rm g}$, $y/(1-R+\eta)$ for $Z_{\rm g}$ and $Z_\star$, and $\Phi\tau_{\rm eq}(1-R)/(1-R+\eta)$ for $M_\star$) and shown as a function of $t/\tau_{\rm eq}$ on logarithmic axes.
    The system transitions from the inflow-driven regime at $t\lesssim\tau_{\rm eq}$, with analytic behaviour described in \S\,\ref{sec:inflow_driven_regime}, to the equilibrium regime at $t\gtrsim\tau_{\rm eq}$, as studied in \S\,\ref{sec:equilibrium_limit}.
  }
  \label{fig:figure/ideal_model_evolution}
\end{figure}

A fully realistic gas flow model would require modelling cosmological accretion and gas cooling to determine the inflow rate $\Phi$, while allowing the star formation efficiency and mass-loading factor to vary with stellar mass and redshift.
Although such treatments are necessary for quantitative comparison with observations, they obscure the simple underlying behaviour of the equations.

We therefore begin with an \textit{ideal gas flow model}, in which $\Phi$, $\epsilon$, and $\eta$ are all constant and the initial conditions are $M_{\rm g}(t = 0) = Z_{\rm g}(t = 0) = 0$.
Under these assumptions, the model admits a closed-form analytic solution \citep[see also][]{coleHierarchicalGalaxyFormation2000},
\begin{align}
  M_{\rm g}(t) & = \Phi\,\tau_{\rm eq} \left(1 - \me^{-t/\tau_{\rm eq}}\right),
  \label{eq:mgas_ana}                                                                                                                                                                  \\[4pt]
  Z_{\rm g}(t) & = y\,\epsilon\left(\tau_{\rm eq}
  - \frac{t}{\me^{t/\tau_{\rm eq}} - 1}\right),
  \label{eq:zgas_ana}                                                                                                                                                                  \\[4pt]
  M_\star(t)   & = (1-R)\,\epsilon\,\tau_{\rm eq}\,\Phi \left[t - \tau_{\rm eq} \left(1 - \me^{-t/\tau_{\rm eq}}\right)\right],
  \label{eq:mstar_ana}                                                                                                                                                                 \\[4pt]
  Z_\star(t)   & = y\,\epsilon\,\tau_{\rm eq}\, \frac{t - 2\tau_{\rm eq} + (t + 2\tau_{\rm eq})\, \me^{-t/\tau_{\rm eq}}} {t - \tau_{\rm eq} + \tau_{\rm eq}\, \me^{-t/\tau_{\rm eq}}},
  \label{eq:zstar_ana}
\end{align}
where $\tau_{\rm eq} \equiv 1/[(1-R+\eta)\,\epsilon]$ is the equilibrium timescale \citep[see also][]{coleHierarchicalGalaxyFormation2000, pengHaloesGalaxiesDynamics2014}, the characteristic time required for the gas reservoir to grow large enough that consumption through star formation and outflows balances the gas inflow rate.
The full derivation is presented in Appendix~\ref{sec:ideal_model_derivation}.

Fig.~\ref{fig:figure/ideal_model_evolution} shows the evolution of four galaxy properties as a function of evolution time normalised by the equilibrium timescale, illustrating the transition between the two regimes of the model.
The gas mass interpolates smoothly between linear growth, $M_{\rm g} \approx \Phi\,t$, at early times ($t\ll \tau_{\rm eq}$) and a constant equilibrium value, $\Phi\,\tau_{\rm eq}$, at late times ($t\gg \tau_{\rm eq}$).
The equilibrium timescale $\tau_{\rm eq}$ therefore has a clear physical meaning: it is the time required for star formation and outflows, the two gas consumption channels, to balance the inflow rate.
Galaxies with high star formation efficiency or strong outflows reach equilibrium quickly, while those with low star formation efficiency or weak outflows remain in the inflow-driven regime for longer.
Similarly, the gas metallicity rises monotonically from zero and asymptotes to $y/(1-R+\eta)$ at late times, with the rate of approach again set by $\tau_{\rm eq}$.
Importantly, the metallicity at early time ($t \lesssim \tau_{\rm eq}$)\footnote{Here $t$ is the galaxy evolution time, rather than cosmic time. Taking the approximation $t\approx M_\star/[(1 - R)\,{\rm SFR}]$ up to a factor of order unity, the inflow-driven regime criterion becomes $M_\mg /M_\star \gg (1 - R + \eta)/(1 - R)$: whenever a galaxy's gas fraction exceeds the mass-loading factor, the galaxy is in the inflow-driven regime. We explore this further in \S\,\ref{sec:analytic_approximate}.} depends on the full history of star formation and gas processing, not merely on the instantaneous balance of inflows and outflows.
This is precisely the information that is lost when the equilibrium limit is imposed from the outset, as in \citet{lillyGASREGULATIONGALAXIES2013} and \citet{feldmannLessonsCosmicHistory2013}.

The ratio $t/\tau_{\rm eq}$ serves as the single dimensionless parameter governing the evolutionary state of the system: $t/\tau_{\rm eq} \ll 1$ corresponds to the inflow-driven regime \citep[see also][]{wangOriginGalaxySizestellar2026}, while $t/\tau_{\rm eq} \gg 1$ corresponds to equilibrium.
Since $\tau_{\rm eq}$ depends on $\epsilon$ and $\eta$, both of which vary with stellar mass and redshift for realistic galaxies, different galaxies occupy different positions along this continuum at any given epoch.
This motivates the two limiting cases we now examine in turn.

\subsubsection{Equilibrium limit}
\label{sec:equilibrium_limit}

We first examine the equilibrium limit, $t \gg \tau_{\rm eq}$, which corresponds to the quasi-equilibrium assumption adopted by the gas regulator class of models \citep{boucheImpactColdGas2010, daveAnalyticModelEvolution2012, lillyGASREGULATIONGALAXIES2013, feldmannLessonsCosmicHistory2013}.
In this limit the gas mass and gas metallicity converge to constant values,
\begin{align}
  M_{\rm g}  & = \Phi \tau_{\rm eq} ,     \\
  Z_{\rm g}  & = y / (1-R+\eta) \label{eq:zgas_eq} , \\
  M_\star(t) & = \Phi\,t\,(1-R) / (1-R+\eta)  ,      \\
  Z_\star    & = y / (1-R+\eta) .\label{eq:zstar_eq}
\end{align}
The gas mass is set by the balance between inflow and consumption, while the metallicity is determined entirely by the ratio of the yield to the effective mass-loss rate per unit star formation rate, $(1-R+\eta)$, independent of the inflow rate, stellar mass, or cosmic time.
The convergence of gas and stellar metallicity to the same equilibrium value reflects the fact that once the gas metallicity has saturated, every generation of stars forms from gas of identical composition.

The equilibrium limit reveals two fundamental shortcomings of previous models.
First, the equilibrium metallicity depends only on $\eta$, and $\eta$ is commonly parameterised in galaxy formation models as a function of stellar mass or halo circular velocity alone, with no redshift dependence.
Consequently, the equilibrium metallicity carries no redshift dependence, predicting an MZR that does not evolve with cosmic time; this has been a persistent problem in many semi-analytic models \citep{luSemianalyticModelsCANDELS2014, somervilleStarFormationSemianalytic2015, guoGalaxiesEAGLEHydrodynamical2016, hirschmannGalaxyAssemblyStellar2016}, in direct tension with observations \citep{maiolinoAMAZEEvolutionMassmetallicity2008, andrewsMassMetallicityRelationDirect2013, curtiJADESInsightsLowmass2024, jainUniformAnalysisGasphase2025}.
Second, no anti-correlation between SFR and metallicity can arise at fixed stellar mass, meaning the FMR cannot be reproduced within a strict equilibrium framework \citep[see also][]{forbesOriginFundamentalMetallicity2014}.
\citet{lillyGASREGULATIONGALAXIES2013} attempted to circumvent this limitation, but their approach relies on assumptions that are internally inconsistent, as we discuss in \S\,\ref{sub:revisiting_the_equilibrium_model}.
These two issues share a common origin: the equilibrium assumption erases all memory of the accretion history, so the metallicity retains no information about galaxy formation history.
The inflow-driven regime, to which we now turn, resolves both problems simultaneously.

\subsubsection{Inflow-driven regime}
\label{sec:inflow_driven_regime}

In the inflow-driven limit, $t \ll \tau_{\rm eq}$, the system has not yet reached equilibrium and the analytic solutions simplify considerably again.
Expanding equations~\eqref{eq:mgas_ana}--\eqref{eq:zstar_ana} to leading order in $t/\tau_{\rm eq}$ yields
\begin{align}
  M_{\rm g}(t) & = \Phi\,t ,\label{eq:mgas_inflow}                         \\
  Z_{\rm g}(t) & = y\,\epsilon\,t/2 ,\label{eq:zgas_inflow}                \\
  M_\star(t)   & = (1-R)\,\Phi\,\epsilon\,t^2/2  \label{eq:mstar_inflow} , \\
  Z_\star(t)   & = y\,\epsilon\,t/3 . \label{eq:zstar_inflow}
\end{align}
In this regime, the gas mass grows linearly with time because star formation and outflows both scale with $M_{\rm g}$ and therefore remain negligible before a substantial gas reservoir has been established.
The stellar mass grows as $t^2$ because the SFR itself increases linearly with the growing gas supply.
The gas and stellar metallicities grow linearly with time, with a fixed ratio $Z_\star/Z_{\rm g} = 2/3$: the stellar metallicity is the star-formation-rate-weighted average of a linearly rising enrichment history, and therefore lags the instantaneous gas value.

The simple power-law behaviour of the inflow-driven solutions gives rise directly to three metallicity scaling relations, each with a transparent physical interpretation.
Taken together, they form the central analytic results of this paper.

\paragraph*{Redshift evolution of the MZR.}
The gas metallicity grows as
\begin{equation}
  Z_{\rm g} = \frac{y\,\epsilon\,t}{2}.
\end{equation}
Writing this as $y\epsilon t/2 = [yM_\star/(1-R)]/M_\mg$ shows that the metallicity reflects the steady accumulation of newly synthesised metals, $yM_\star/(1-R)$, diluted by the growing reservoir of pristine inflowing gas, $M_\mg$, with negligible loss to outflows or stellar locking.
This behaviour, in which the metallicity retains a direct imprint of the accretion history, is precisely what the equilibrium regime cannot produce, since there the metallicity depends only on $\eta$ and carries no memory of the accretion history.

\paragraph*{Redshift evolution of SFMS.}
The specific SFR evolves as $\mathrm{sSFR} = \epsilon M_\mathrm{g}/M_\star = 2/[(1-R)\,t]$, with no dependence on stellar mass.
Even in the equilibrium regime the sSFR retains this independence: $\mathrm{sSFR} = 1/[(1-R)\,t]$, differing only in the numerical prefactor.
The evolution of the sSFR is therefore highly robust, insensitive to the details of gas accretion, cooling, star formation efficiency, and outflow strength \citep[see also][]{pengHaloesGalaxiesDynamics2014}.

\paragraph*{The fundamental metallicity relation.}
Combining equations~\eqref{eq:mgas_inflow}, \eqref{eq:mstar_inflow}, and \eqref{eq:zgas_inflow} to eliminate $t$ and $\Phi$ yields
\begin{equation}
  Z_{\rm g} = \frac{y\,\epsilon}{1-R}\, \frac{M_\star}{\mathrm{SFR}}.
  \label{eq:fmr_ideal}
\end{equation}
At fixed stellar mass, equation~\eqref{eq:fmr_ideal} predicts that galaxies with higher SFR have lower gas metallicity, reproducing the anti-correlation that underlies the FMR.
This anti-correlation is not driven by stochastic fluctuations in accretion, as proposed by \citet{forbesOriginFundamentalMetallicity2014}, but is a deterministic consequence of the inflow-driven scaling: at fixed $M_\star$, galaxies with higher SFR have assembled their stellar mass more rapidly, leaving less time for chemical enrichment.

\paragraph*{The gaseous fundamental metallicity relation.}
With the definition that $\epsilon\equiv {\rm SFR}/M_\mg$, equation~\eqref{eq:fmr_ideal} becomes
\begin{equation}
  Z_{\rm g} = \frac{y}{1-R}\,\frac{M_\star}{M_{\rm g}},
  \label{eq:gas_fmr_ideal}
\end{equation}
which is the gFMR.
Unlike the FMR (equation~\ref{eq:fmr_ideal}), which retains a dependence on the star formation efficiency $\epsilon$, the gFMR depends only on $y$ and $R$, two quantities determined solely by the IMF.
The physical origin is transparent: in the inflow-driven regime both the gas mass and the metal mass grow in proportion to the cumulative inflow, so their ratio is set entirely by stellar nucleosynthesis.
This makes the gFMR more fundamental than the FMR.

\paragraph*{}
All three relations emerge from the same underlying physics: the linear growth of gas mass and metallicity with time in the inflow-driven regime.
Their persistence in the full cosmological model, which we demonstrate in the following section, confirms that the inflow-driven limit captures the dominant physics of galaxy chemical evolution for the majority of the star-forming population.

\subsection{The cosmological gas flow model}
\label{sub:the_cosmological_gas_flow_model}

The ideal gas flow model demonstrates that the inflow-driven limit naturally produces the redshift evolution of the MZR, the FMR, and the gFMR from first principles.
However, it assumes constant inflow rate, star formation efficiency, and mass-loading factor, which are clearly  oversimplifications.
In reality, the inflow rate is set by cosmological accretion and gas cooling, the star formation efficiency varies with stellar mass and redshift, and the mass-loading factor depends on the depth of the gravitational potential.
To test whether the scaling relations derived in \S\,\ref{sec:inflow_driven_regime} persist under more realistic conditions, we now construct a cosmological gas flow model in which these quantities are allowed to vary.
The model retains the analytic simplicity of the one-zone framework (equations~\ref{eq:mgas} and \ref{eq:zgas}) while incorporating the key physical dependencies needed to reproduce observed galaxy scaling relations across cosmic time.

\subsubsection{Inflow rate}
\label{sec:inflow_rate}

The gas inflow rate onto the central galaxy can be factorised into two components: the baryonic accretion rate onto the halo, and the cooling efficiency that determines what fraction of accreted baryons reaches the ISM.
We write
\begin{equation}
  \Phi(M_\mh, z) = \lambda(M_\mh, z)\, f_{\rm b}\,\dot{M}_\mh ,
  \label{eq:cooling}
\end{equation}
where $f_{\rm b} \equiv \Omega_{\rm b}/\Omega_{\rm m}$ is the cosmic baryon fraction, $\dot{M}_\mh$ is the dark matter accretion rate, and $\lambda(M_\mh, z)$ is the cooling efficiency that depends on halo mass and redshift.

We compute the halo accretion rate from the halo mass growth history, which we model following \citet{wechslerConcentrationsDarkHalos2002} as
\begin{equation}
  M_\mh(z) = M_{\rm h,0}\,\exp\!\left[ -\frac{8\,a_0}{c}\left(\frac{a_0}{a} - 1\right)\right] ,
\end{equation}
where $a \equiv 1/(1+z)$, $a_0 = 1$ is the scale factor at $z = 0$, and $c$ is the concentration parameter.
The concentration controls the formation epoch of the halo: haloes with higher concentration formed earlier and have slower late-time accretion rates.
The dark matter accretion rate follows as $\dot{M}_\mh = (\mathrm{d}M_\mh/\mathrm{d}z)\,(\mathrm{d}z/\mathrm{d}t)$.
We adopt $c = 12$ \citep[see also][]{wechslerConcentrationsDarkHalos2002, netoStatisticsCDMHalo2007, wangEfficientRobustMethod2024}, which closely reproduces the mean halo growth histories in \citet{fakhouriMergerRatesMass2010} (see Appendix~\ref{sec:halo_accretion_rate}), and add a $0.15$~dex log-normal scatter to $c$ to capture the diversity of halo formation histories \citep[see also][]{netoStatisticsCDMHalo2007, wangEfficientRobustMethod2024}.

The cooling efficiency encapsulates the baryonic physics that regulates how much of the accreted gas reaches the ISM.
At low halo masses, stellar feedback suppresses gas accretion by heating the halo atmosphere \citep{daveAnalyticModelEvolution2012, mitchellGalacticOutflowRates2020, wrightImpactStellarAGN2020}.
At high halo masses, virial shock heating prevents efficient cold-mode accretion \citep{dekelColdStreamsEarly2009}, and radio-mode AGN feedback can further suppress gas cooling by injecting energy into the hot halo atmosphere \citep{bowerBreakingHierarchyGalaxy2006, bowerDarkNemesisGalaxy2017}.
Therefore, we model the net cooling efficiency as
\begin{equation}
  \lambda(M_\mh, z) = \lambda_0\, \frac{(1+z)^{\lambda_{z}}} {(M_\mh/M_{\lambda})^{\lambda_{m_1}} + (M_\mh/M_{\lambda})^{\lambda_{m_2}}} ,
  \label{eq:cooling_pre}
\end{equation}
where $M_{\lambda} = 10^{12}\,\mathrm{M_\odot}$ is the anchoring halo mass, and the redshift factor $(1+z)^{\lambda_{z}}$ with $\lambda_{z} > 0$ captures the increase in cooling efficiency at high redshift, driven by higher gas densities \citep[see also][]{moGalaxyFormationEvolution2010, moTwophaseModelGalaxy2024} and the prevalence of cold-mode accretion that feeds gas directly to the ISM \citep[see also][]{keresHowGalaxiesGet2005, dekelColdStreamsEarly2009}.
The denominator implements a double power law: the term $(M_\mh/M_{\lambda})^{\lambda_{m_1}}$ with $\lambda_{m_1} < 0$ suppresses cooling at low halo masses, representing inefficient gas cooling and preventative feedback \citep{daveAnalyticModelEvolution2012, luImportancePreventiveFeedback2017, wrightImpactStellarAGN2020}, while the term $(M_\mh/M_{\lambda})^{\lambda_{m_2}}$ with $\lambda_{m_2} > 0$ suppresses cooling at high halo masses, representing virial shock heating and AGN feedback \citep{whiteCoreCondensationHeavy1978, bowerBreakingHierarchyGalaxy2006, dekelColdStreamsEarly2009, bowerDarkNemesisGalaxy2017}.
The cooling efficiency peaks near $M_{\lambda}$ and declines on both sides, producing the characteristic shape of the stellar mass--halo mass relation \citep{yangConstrainingGalaxyFormation2003, wechslerConnectionGalaxiesTheir2018, wangTestingGalaxyFormation2025}.
This parameterisation is essential for decoupling the mass--metallicity relation from the stellar mass--halo mass relation, which would otherwise be tightly linked to each other \citep[see also][and \S\,\ref{sub:mzr_shmr}]{lillyGASREGULATIONGALAXIES2013}.

\subsubsection{Star formation efficiency}
\label{ssub:star_formation_efficiency}

The star formation efficiency, defined as $\epsilon \equiv \mathrm{SFR}/M_\mathrm{g}$ with $M_\mathrm{g}$ includes both atomic and molecular hydrogen, governs how rapidly the total cold gas reservoir is converted into stars.
We parameterise it as a separable power law in stellar mass and redshift,
\begin{equation}
  \epsilon(M_\star, z) = \epsilon_0 \left(\frac{M_\star}{M_{\epsilon}}\right)^{\epsilon_m}(1+z)^{\epsilon_z},
  \label{eq:sfe}
\end{equation}
where $M_{\epsilon}=10^{10}\,\rm M_\odot$ is the anchoring stellar mass, $\epsilon_0$ sets the overall normalisation, $\epsilon_m$ encodes the stellar-mass dependence at fixed redshift, and $\epsilon_z$ controls the redshift evolution at fixed stellar mass.

Before calibrating these parameters against observational data, we first derive their plausible ranges from available empirical constraints.
In the local Universe, the global gas depletion timescale for normal star-forming disc galaxies is $M_\mg / {\rm SFR} \approx 2.1\,\mathrm{Gyr}$ \citep{kennicuttStarFormationGalaxies1998}, corresponding to $\epsilon_0 \sim 0.5\,\mathrm{Gyr}^{-1}$.

The stellar-mass dependence of $\epsilon$ can be understood by decomposing it as the product of the specific SFR and the inverse gas fraction,
\begin{equation}
  \epsilon = \frac{\mathrm{SFR}}{M_\star} \times \frac{M_\star}{M_{\rm g}} = \mathrm{sSFR} \times \mu^{-1}.
  \label{eq:sfe_mass}
\end{equation}
where ${\rm sSFR} \equiv {\rm SFR} / M_\star$ and $\mu \equiv M_\mg / M_\star$.
On the star-forming main sequence at $z \approx 0$, the sSFR scales weakly with stellar mass, $\mathrm{sSFR} \propto M_\star^{-0.2}$ \citep{speagleHighlyConsistentFramework2014}.
The total gas fraction is dominated by the atomic component across the stellar mass range $M_\star = 10^{9}$--$10^{11.5}\,\mathrm{M}_\odot$ probed by the xGASS survey \citep{catinellaXGASSTotalCold2018}, with the gas-to-stellar mass ratio scaling as $\mu \propto M_\star^{-0.6}$.
Combining these two scaling relations yields $\epsilon \propto M_\star^{+0.4}$.

The redshift dependence is most transparently seen through an alternative decomposition into the molecular star formation efficiency and the molecular gas fraction,
\begin{equation}
  \epsilon = \frac{\mathrm{SFR}}{M_{\rm H_2}} \times \frac{M_{\rm H_2}}{M_{\rm HI} + M_{\rm H_2}} = \epsilon_{\rm mol} \times f_{\rm mol}.
  \label{eq:sfe_redshift}
\end{equation}
The molecular star formation efficiency evolves mildly, $\epsilon_\mathrm{mol} \propto (1+z)^{0.6}$, with only a weak stellar-mass dependence \citep{tacconiPHIBSSUnifiedScaling2018}.
The molecular fraction $f_\mathrm{mol}$ provides the dominant lever: at high redshift, galaxies are more gas-rich and compact, driving up the midplane pressure and hence the molecular-to-atomic ratio \citep{blitzRolePressureGMC2006}.
The total-gas star formation efficiency therefore rises more steeply with redshift than $(1+z)^{0.6}$ alone, motivating a positive $\epsilon_z$ driven by the combination of a mildly increasing $\epsilon_\mathrm{mol}$ and a more rapidly increasing $f_\mathrm{mol}$ with redshift.

\subsubsection{Mass-loading factor}
\label{ssub:loading_factor}

Galactic outflows driven by stellar feedback are a key regulator of both the gas content and the metallicity of galaxies.
The mass-loading factor $\eta$ quantifies the efficiency of these outflows relative to SFR.
Both theoretical expectations from momentum-driven and energy-driven wind models \citep[see also][]{finlatorOriginGalaxyMassmetallicity2008, daveAnalyticModelEvolution2012} and observational constraints from UV absorption-line studies and background quasar spectroscopy \citep{heckmanSystematicPropertiesWarm2015, schroetterMusEGAsFLOw2024} indicate that $\eta$ decreases with increasing stellar mass, reflecting the deeper gravitational potentials of more massive galaxies.
We parameterise this dependence as
\begin{equation}
  \eta(M_\star) = \eta_0\left(\frac{M_\star}{M_\eta}\right)^{\eta_m},
  \label{eq:eta}
\end{equation}
where $M_\eta=10^{10}\,\rm M_\odot$ is the anchoring stellar mass, and $\eta_m < 0$.
We do not include an explicit redshift dependence, as observational constraints from $z \approx 0$ to $z \approx 1.5$ show no significant evolution of $\eta$ at fixed stellar mass \citep[see also][]{schroetterMusEGAsFLOw2019, schroetterMusEGAsFLOw2024}.

\subsubsection{The initial mass function}
\label{sec:the_initial_mass_function}

The mass return fraction $R$ and the metal yield $y$ depend on the stellar mass distribution and are therefore set by the initial mass function (IMF).
For a \citet{chabrierGalacticStellarSubstellar2003} IMF, standard stellar evolution models give $R \approx 0.40$--$0.46$ and $y \approx 0.02$--$0.04$ for the total metal yield \citep{vincenzoModernYieldsStellar2016}.
We fix $R = 0.44$ and treat $y$ as the sole free parameter related to the IMF.
The primary reason is that observationally inferred oxygen abundances suffer from systematic uncertainties of up to $\sim 0.2$~dex depending on the strong-line calibration adopted \citep{curtiMassmetallicityFundamentalMetallicity2020}, comparable to the range of $y/(1-R)$ values predicted across plausible IMF choices.
Moreover, \citet{griffithImpactBlackHole2021} showed that the yield of $\alpha$-elements can vary by up to a factor of three at fixed IMF, depending on the assumed treatment of black hole formation during stellar collapse.
Fixing $y$ to a value derived from a specific IMF and stellar evolution model would impose a false sense of precision on the metallicity normalisation.
By treating $y$ as a free parameter, we allow the model to absorb these calibration systematics and focus on the relative behaviour of the scaling relations rather than their absolute normalisation.

\subsubsection{Calibration}
\label{ssub:calibration}

\begin{figure*}
  \begin{center}
    \includegraphics[width=0.95\linewidth]{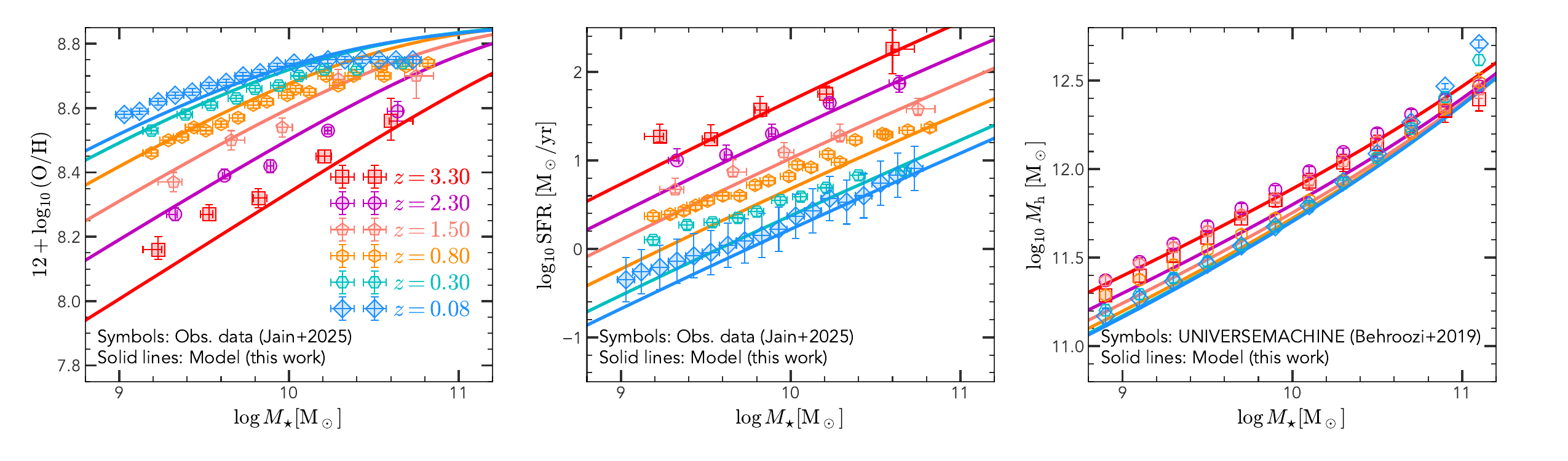}
  \end{center}
  \caption{
    Calibration of the cosmological gas flow model against three observational scaling relations, each shown at six redshifts from $z \approx 0.08$ to $z = 3.30$.
    \textbf{Left panel:} the gas-phase mass--metallicity relation from \citet{jainUniformAnalysisGasphase2025}.
    \textbf{Middle panel:} the star-forming main sequence from \citet{jainUniformAnalysisGasphase2025}.
    \textbf{Right panel:} the stellar mass--halo mass relation from the \textsc{UniverseMachine} empirical model \citep{behrooziUNIVERSEMACHINECorrelationGalaxy2019}.
    Solid lines show the output of the cosmological gas flow model.
  }
  \label{fig:mzr_redshift_evolution}
\end{figure*}

\begin{table}
  \centering
  \caption{
    Parameters of the cosmological gas flow model.
    Anchoring masses are $M_\lambda = 10^{12}\,\rm M_\odot$ and $M_\epsilon = M_\eta = 10^{10}\,\mathrm{M}_\odot$.}
  \label{tab:parameters}
  \begin{tabular}{llll}
    \hline
    Symbol          & Parameter                       & Best-fit value                     & Equation                                                                                    \\
    \hline
    \multicolumn{4}{l}{\textbf{Cooling efficiency}: $\lambda=\frac{\lambda_0 (1 + z)^{\lambda_z}}{(M_\mh / M_\lambda)^{\lambda_{m_1}}  + (M_\mh / M_\lambda)^{\lambda_{m_2}}}$} \\[10pt]
    $\lambda_{0}$   & Cooling normalisation           & $0.35$                    & (\ref{eq:cooling_pre})                                                                          \\
    $\lambda_{m_1}$ & Low-mass suppression slope      & $-0.60$                   & (\ref{eq:cooling_pre})                                                                          \\
    $\lambda_{m_2}$ & High-mass suppression slope     & $0.20$                    & (\ref{eq:cooling_pre})                                                                          \\
    $\lambda_{z}$   & Cooling redshift exponent       & $0.20$                    & (\ref{eq:cooling_pre})                                                                          \\[8pt]
    \multicolumn{4}{l}{\textbf{Star formation efficiency (SFE)}: $\epsilon=\epsilon_0\left(\frac{M_\star}{M_\epsilon}\right)^{\epsilon_m} (1 + z)^{\epsilon_z}$}                  \\[10pt]
    $\epsilon_0$    & SFE normalisation               & $0.38\;\mathrm{Gyr}^{-1}$ & (\ref{eq:sfe})                                                                              \\
    $\epsilon_m$    & SFE stellar mass slope          & $0.33$                    & (\ref{eq:sfe})                                                                              \\
    $\epsilon_z$    & SFE redshift exponent           & $1.1$                     & (\ref{eq:sfe})                                                                              \\[8pt]
    \multicolumn{4}{l}{\textbf{Mass-loading}: $\eta = \eta_0\left(\frac{M_\star}{M_\eta}\right)^{\eta_m}$}                                                                        \\[10pt]
    $\eta_0$        & Mass-loading normalisation      & $0.22$                    & (\ref{eq:eta})                                                                              \\
    $\eta_m$        & Mass-loading stellar mass slope & $-0.30$                   & (\ref{eq:eta})                                                                              \\[4pt]
    \multicolumn{4}{l}{\textbf{IMF and yield}}                                                                                                                                  \\[2pt]
    $R$             & Mass return fraction            & $0.44$                    &                                                                                             \\
    $y$             & Metal yield                     & $0.012$                   &                                                                                             \\
    \hline
  \end{tabular}
\end{table}

The cosmological gas flow model contains 10 free parameters: $\{\lambda_0,\, \lambda_{m_1},\, \lambda_{m_2},\, \lambda_{z},\, \epsilon_0,\, \epsilon_m,\, \epsilon_z,\, \eta_0,\, \eta_m,\, y\}$.
Of these, four $(\lambda_0,\, \lambda_{m_1},\, \lambda_{m_2},\, \lambda_{z})$ control gas cooling and affect only the stellar-to-halo mass relation; even substantial changes to these parameters do not alter any results related to galaxy properties other than halo mass (see \S\,\ref{sub:mzr_shmr} for detailed discussion).
The yield $y$ serves primarily to absorb the systematic uncertainty in the observational metallicity calibration and the yield uncertainty.
This leaves \textit{five parameters} that directly affect the main results of this work: \textit{three} for the normalisation, mass dependence, and redshift dependence of the star formation efficiency $(\epsilon_0,\, \epsilon_m,\, \epsilon_z)$, and \textit{two} for the normalisation and mass dependence of the mass-loading factor $(\eta_0,\, \eta_m)$.

We calibrate these parameters by simultaneously fitting three observational constraints within $M_\star \approx 10^9-10^{10.5}\,\rm M_\odot$: the redshift-evolving gas-phase mass--metallicity relation from \citet{jainUniformAnalysisGasphase2025}, the star-forming main sequence from \citet{jainUniformAnalysisGasphase2025}, and the stellar mass--halo mass relation from the \textsc{UniverseMachine} empirical model \citep{behrooziUNIVERSEMACHINECorrelationGalaxy2019}.
All three constraints are evaluated at six redshifts from $z \approx 0.08$ to $z = 3.30$. Fig.~\ref{fig:mzr_redshift_evolution} compares the calibrated model with the data.

We calibrate the model by manually tuning the parameters rather than employing an automated fitting procedure, for two reasons.
First, given the deliberately minimal nature of our model, which omits several physical processes discussed in \S\,\ref{sub:other_processes_in_regulating_metallicity_evolution}, we do not expect a precise fit to the data, nor would a formally optimal fit imply that the inferred parameters carry physical meaning to the corresponding precision.
Our goal is to capture the main characteristics of the observed scaling relations rather than to minimise residuals.
Second, although \citet{jainUniformAnalysisGasphase2025} made considerable effort to measure gas-phase abundances uniformly across the redshift range $z \sim 0$ to $\sim 3$, the underlying data are drawn from different surveys with different selection functions, and the resulting systematic uncertainties are difficult to propagate into a formal likelihood.
This is reflected in the non-uniform error estimates across surveys.

We integrate the system of equations~(\ref{eq:mgas})--(\ref{eq:zstar}) numerically from $z = 20$ to $z = 0$ for a grid of 200,000 present-day halo masses spaced logarithmically from $M_{\rm h,0} = 10^{9}$ to $10^{15}\,\mathrm{M}_\odot$.
Each halo grows according to the \citet{wechslerConcentrationsDarkHalos2002} mass accretion history, with the concentration parameter drawn from a log-normal distribution centred on $c = 12$ with a dispersion of $0.15$~dex.
Given $\Phi(z)$, $\epsilon(M_\star, z)$, and $\eta(M_\star)$, the state vector $(M_{\rm h}, M_\star, M_\mathrm{g}, M_{Z,\star}, M_{Z,\mathrm{g}})$ is evolved using an adaptive implicit solver.
The adopted parameter values are listed in Table~\ref{tab:parameters}.

\section{The physics of the fundamental metallicity relation}
\label{sec:fundamental_metallicity_relation}

The fundamental metallicity relation states that star-forming galaxies populate a well-defined surface in the three-dimensional $(M_\star, \mathrm{SFR}, Z_{\rm g})$ space, and that this surface is approximately redshift-invariant out to at least $z \sim 3$ \citep{mannucciFundamentalRelationMass2010, curtiMassmetallicityFundamentalMetallicity2020, sandersMOSDEFSurveyEvolution2021}.
One manifestation of this relation is the anti-correlation between SFR and gas metallicity at fixed stellar mass, which is sometimes taken to be the entire content of the FMR \citep[e.g.][]{forbesOriginFundamentalMetallicity2014, torreySimilarStarFormation2018}.
However, the anti-correlation alone does not capture the fact that galaxies lie on a single, non-evolving surface rather than simply exhibiting correlated scatter at each epoch.
In what follows, we first demonstrate that our cosmological gas flow model reproduces both the FMR and the gFMR, in which the gas mass replaces the SFR as the second parameter.
We then manipulate the model assumptions to reveal the physical origin of both relations.

\subsection{FMR in the cosmological gas flow model}
\label{sub:fmr_in_the_cosmological_gas_flow_model} 

\begin{figure}
  \begin{center}
    \includegraphics[width=0.95\linewidth]{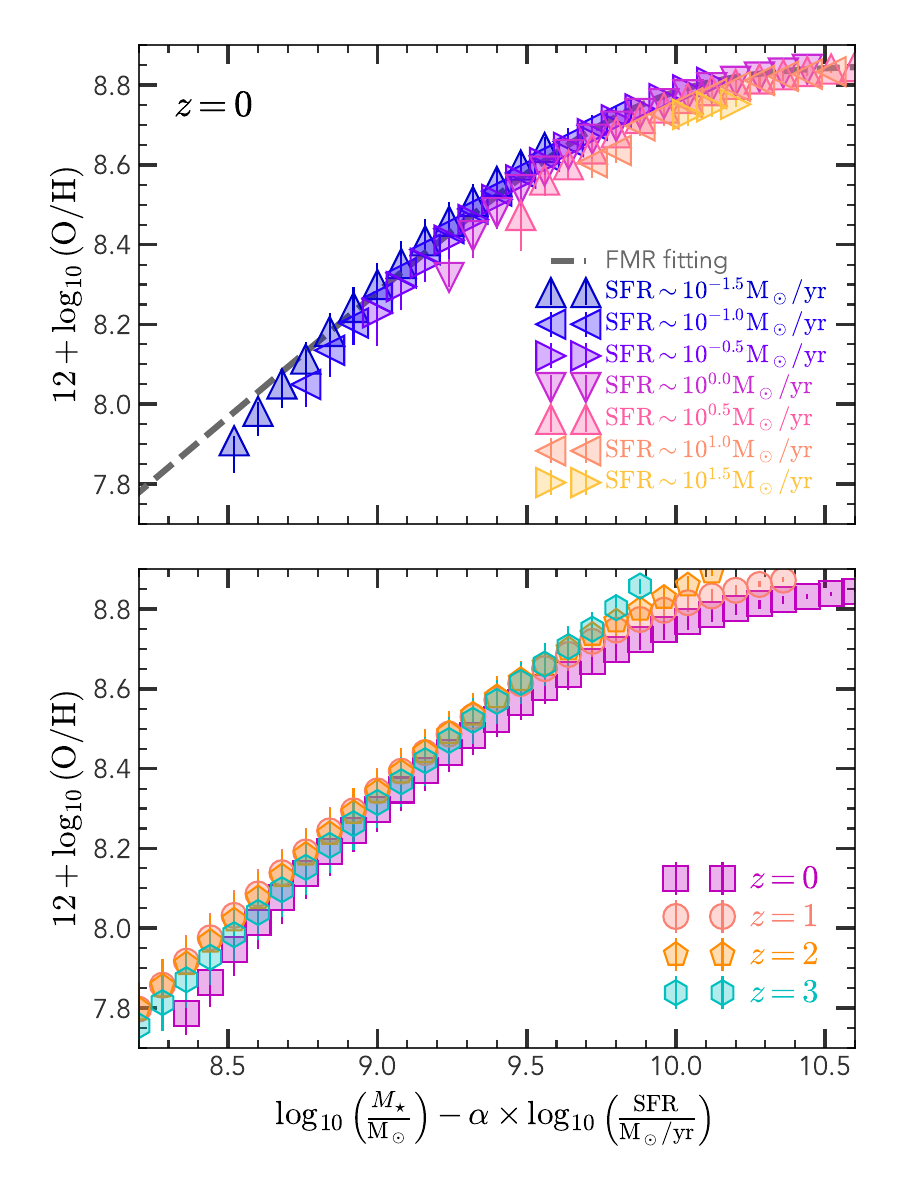}
  \end{center}
  \caption{
    Gas metallicity as a function of the FMR parameter, $\log_{10}(M_\star/{\rm M_\odot}) - \alpha\log_{10}({\rm SFR}/{\rm M_\odot\,yr^{-1}})$, with $\alpha = 0.55$, predicted by the cosmological gas flow model.
    \textbf{Upper panel:} results at $z = 0$, colour-coded by SFR.
    Different SFR bins collapse onto a single sequence, consistent with the existence of a fundamental metallicity relation.
    The dashed line shows the best-fitting relation using the parametrization in \citet{curtiMassmetallicityFundamentalMetallicity2020}.
    \textbf{Lower panel:} the same projection applied across redshifts $z = 0$--$3$.
    Data from all epochs fall on a common locus, confirming that the FMR predicted by the model is approximately redshift-invariant, consistent with the observational result.
  }
  \label{fig:fmr_parameter_fitting}
\end{figure}

Having calibrated the cosmological gas flow model to reproduce three observational constraints (the mass--metallicity relation, the star-forming main sequence, and the stellar-to-halo mass relation; Fig.~\ref{fig:mzr_redshift_evolution}), we now examine whether the model also reproduces the fundamental metallicity relation.
The FMR was not used as a calibration target, so its emergence from the model constitutes a non-trivial prediction.

Fig.~\ref{fig:fmr_parameter_fitting} shows the gas metallicity predicted by the model as a function of the FMR parameter
\begin{equation}
  \xi \equiv \log_{10}\left(\frac{M_\star}{\mathrm{M}_\odot}\right) - \alpha\,\log_{10}\left(\frac{\mathrm{SFR}}{\mathrm{M}_\odot\,\mathrm{yr}^{-1}}\right)
\end{equation}
with $\alpha = 0.55$ following \citet{curtiMassmetallicityFundamentalMetallicity2020}.
In the upper panel, galaxies at $z = 0$ with SFR spanning over three orders of magnitude collapse onto a single sequence when projected along this combination of stellar mass and SFR.
In the lower panel, the same projection is applied across redshifts from $z = 0$ to $z = 3$, and galaxies at all epochs continue to follow a common locus.
The model therefore predicts a redshift-invariant FMR as a natural outcome of the calibrated physics, without any additional parameter tuning, consistent with observational findings \citep{mannucciFundamentalRelationMass2010, curtiMassmetallicityFundamentalMetallicity2020}.

We fit the predicted relation using the functional form of \citet{curtiMassmetallicityFundamentalMetallicity2020},
\begin{equation}
  \log_{10}(\mathrm{O/H})  = \log_{10}(\mathrm{O/H})_0 - \frac{\gamma}{\beta}\log_{10}\left[1 + 10^{-\beta\,(\xi - \xi_0)}\right]                                                
  \label{eq:fmr_fitting}
\end{equation}
where $12 + \log_{10}(\mathrm{O/H})_0 = 8.85$, $\beta = 2.0$, $\gamma = 0.64$, and $\xi_0 = 9.9$.
The best-fitting slope $\gamma = 0.64$ is steeper than the value of $0.30$ reported by \citet{curtiMassmetallicityFundamentalMetallicity2020}.
This difference arises from Eddington bias \citep{eddingtonFormulaCorrectingStatistics1913}: uncertainty in the SFR measurement preferentially scatters low-$\xi$ galaxies, predominantly low-mass galaxies, into the high-$\xi$ population rather than the reverse, which flattens the observed slope, as we demonstrate in Appendix~\ref{sec:the_impact_of_eddington_bias_on_the_slope_of_fmr}.

\subsection{gFMR in the cosmological gas flow model}
\label{sub:gaseous_fmr_in_the_cosmological_gas_flow_model} 

\begin{figure}
  \begin{center}
    \includegraphics[width=0.95\linewidth]{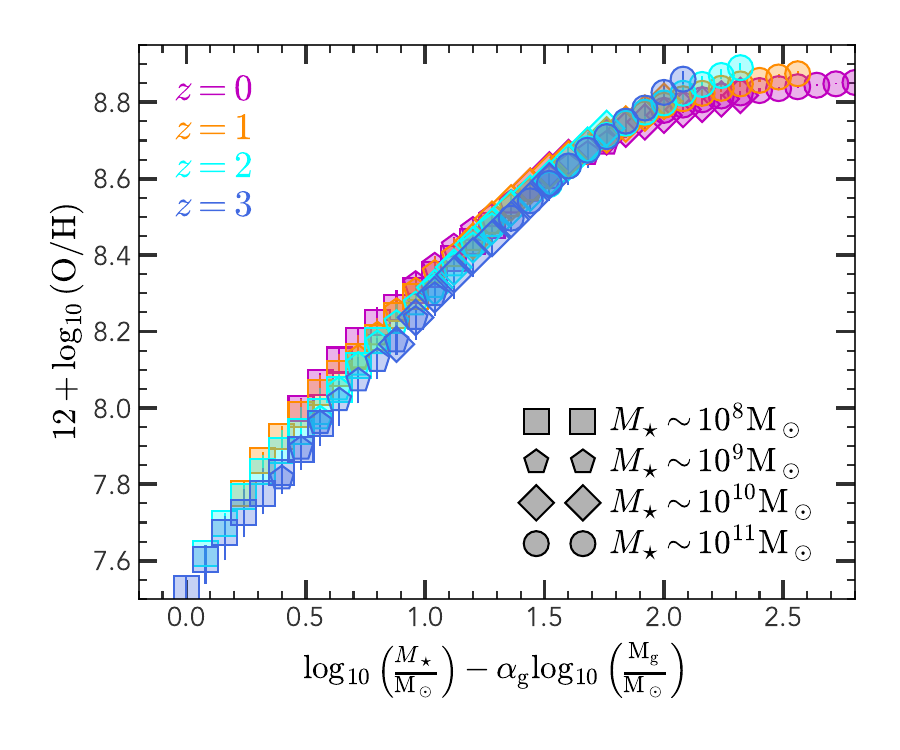}
  \end{center}
  \caption{
    Gas metallicity as a function of the gFMR parameter $\log_{10}(M_\star/\mathrm{M}_\odot) - \alpha_{\rm g}\,\log_{10}(M_\mg / {\rm M_\odot})$, predicted by the cosmological gas flow model.
    Marker shapes distinguish four stellar mass bins from $M_\star \sim 10^{8}$ to $10^{11}\,\mathrm{M}_\odot$, and different shading denotes redshifts $z = 0$--$3$.
    All stellar mass bins and redshifts collapse onto a single sequence, in which $Z_{\rm g}$ depends on the combination of $M_\star$ and $M_{\rm g}$ rather than on each independently. This relation is approximately invariant in redshift.
  }
  \label{fig:gaseous_fmr}
\end{figure}

While the standard FMR uses SFR as the second parameter alongside stellar mass, \citet{bothwellFundamentalRelationMetallicity2013} showed observationally that replacing SFR with the atomic hydrogen mass yields an equally tight or tighter relation \citep[see also][]{bothwellMolecularGasDriver2016, bothwellGalaxyMetallicitiesDepend2016, brownRoleAtomicHydrogen2018, scholteAtomicGasSequence2024, boardmanFundamentalMetallicityRelation2026}.
Here we use the total gas mass $M_\mathrm{g}$ and examine this gFMR in our model.

Fig.~\ref{fig:gaseous_fmr} shows the gas metallicity predicted by the model as a function of the gFMR parameter $\log_{10}(M_\star/\mathrm{M}_\odot) - \alpha_{\rm g}\,\log_{10}(M_{\rm g}/\mathrm{M}_\odot)$ with $\alpha_\mg = 0.85$.
All stellar mass bins from $M_\star \sim 10^{8}$ to $10^{11}\,\mathrm{M}_\odot$ and all redshifts from $z = 0$ to $z = 3$ collapse onto a single sequence, confirming the existence of a gFMR in the model.
In the following subsection, we investigate the connection between FMR and gFMR, as well as the underlying physics of both relations in detail.

\subsection{The physics of the FMR and gFMR}
\label{sub:the_physics_of_fmr} 

\begin{figure*}
  \begin{center}
    \includegraphics[width=0.98\linewidth]{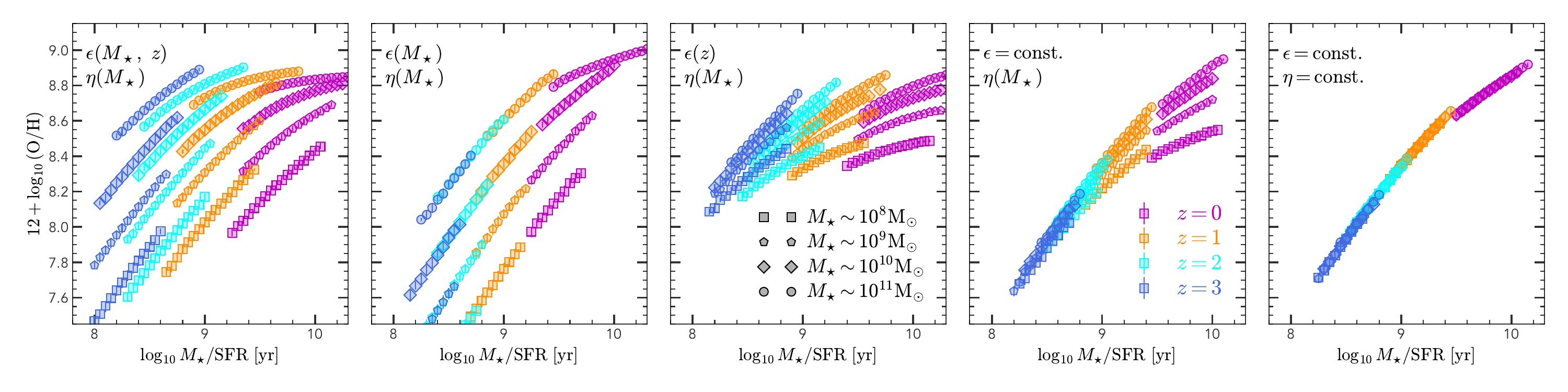}
  \end{center}
  \caption{
  Gas metallicity, $12 + \log_{10}(\rm O/H)$, versus the inverse of sSFR, $\log_{10}M_\star/{\rm SFR}$, predicted by the cosmological gas flow model under five assumptions about the star formation efficiency $\epsilon \equiv \mathrm{SFR}/M_{\rm g}$ and the mass-loading factor $\eta$.
  Marker shapes denote stellar mass bins ($M_\star \sim 10^{8}$--$10^{11}\,\mathrm{M}_\odot$) and shading denotes redshift ($z = 0$--$3$).
  From left to right: both $\epsilon$ and $\eta$ depend on stellar mass, with $\epsilon$ also depending on redshift; $\epsilon$ and $\eta$ depend on stellar mass only; $\epsilon$ depends on redshift only while $\eta$ depends on stellar mass; $\epsilon$ is constant while $\eta$ depends on stellar mass; and both $\epsilon$ and $\eta$ are constant.
  In the simplest case (rightmost panel), all galaxies follow a single universal sequence whose shape traces the transition from the inflow-driven regime (steep, low $M_\star/\mathrm{SFR}$) to equilibrium (flat, high $M_\star/\mathrm{SFR}$).
  Progressively reintroducing the mass and redshift dependence of $\epsilon$ and $\eta$ produces the offsets among stellar mass bins and redshifts seen in the leftmost panel.
  See \S\,\ref{sub:the_physics_of_fmr} for a detailed discussion.
  }
  \label{fig:fmr_physics}
\end{figure*}

\begin{figure*}
  \begin{center}
    \includegraphics[width=0.98\linewidth]{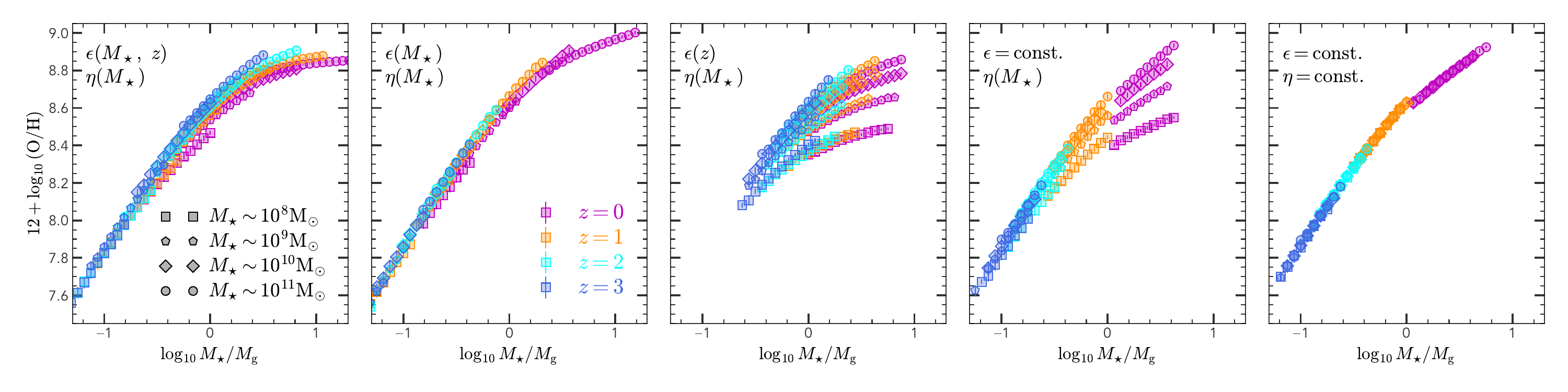}
  \end{center}
  \caption{
  Same as Fig.~\ref{fig:fmr_physics}, but showing gas metallicity versus the inverse of gas fraction, $\log_{10}M_\star/M_{\rm g}$.
  The last two panels are qualitatively similar to their FMR counterparts, but the first two panels reveal a substantially tighter relation than in Fig.~\ref{fig:fmr_physics}: varying $\epsilon$ with stellar mass or redshift has little effect on the $Z_{\rm g}$--$M_\star/M_{\rm g}$ relation.
  This tightness arises because the inflow-driven limit predicts $Z_{\rm g} \propto M_\star/M_{\rm g}$ with no dependence on $\epsilon$, and because the opposing mass dependences of $\eta$ and $\epsilon$ partially cancel in the equilibrium timescale $\tau_{\rm eq} \propto 1/[(1-R+\eta)\epsilon]$, so that galaxies of different masses approach equilibrium at a similar pace.
  See \S\,\ref{sub:the_physics_of_fmr} for a detailed discussion.
  }
  \label{fig:gfmr_physics}
\end{figure*}

There are three features of the FMR that call for a physical explanation.
First, the best-fitting FMR parameter combines stellar mass and SFR as $\xi\equiv\log_{10}(M_\star/\mathrm{M}_\odot) - \alpha\,\log_{10}({\rm SFR}/\mathrm{M}_\odot\,{\rm yr}^{-1})$ with $\alpha < 1$, whereas $\alpha = 1$ would correspond to metallicity depending solely on the sSFR.
What sets $\alpha$ to a value below unity?
Second, the $\log_{10}Z_{\rm g}$--$\xi$ relation is steep at low $\xi$ and flattens at high $\xi$.
What determines this shape?
Finally, why is the scaling relation between $12 + \log_{10}({\rm O/H})$ and $\xi$ redshift invariant?
To answer these questions, we perform a series of controlled experiments in which we vary the assumptions about star formation efficiency ($\epsilon$) and mass-loading factor ($\eta$) in the cosmological gas flow model and examine the resulting $Z_{\rm g}$--$M_\star/\mathrm{SFR}$ relation.

\subsubsection{Reducing the FMR to its simplest form}

Fig.~\ref{fig:fmr_physics} presents the gas metallicity as a function of $\log_{10}\,M_\star/\mathrm{SFR}$ in bins of stellar mass and redshift, under five progressively simpler prescriptions for star formation efficiency ($\epsilon$) and mass-loading factor ($\eta$).

In the fiducial model (leftmost panel), where $\epsilon$ depends on both stellar mass and redshift and $\eta$ depends on stellar mass, the metallicity anti-correlates with SFR at fixed stellar mass and redshift, but the sequences are offset among different mass bins and redshifts. The standard FMR asserts that a single parameter $\alpha$ can absorb both types of offset (see Fig.~\ref{fig:fmr_parameter_fitting}).

Removing the redshift dependence of $\epsilon$ by setting $\epsilon_z = 0$ in equation~\eqref{eq:sfe} while retaining its mass dependence (second panel) eliminates the offsets among redshifts at fixed stellar mass, indicating that the redshift-dependent star formation efficiency is responsible for the offset between epochs.

Retaining only the redshift dependence of $\epsilon$ (third panel) by only setting $\epsilon_m = 0$ in equation~\eqref{eq:sfe} restores the redshift offsets.
The offsets among stellar mass bins shrink considerably but do not vanish entirely, consistent with the expectation that both $\epsilon$ and $\eta$ contribute to the mass-dependent structure, as predicted by the equilibrium limit where metallicity depends on $\eta$ (see equation~\ref{eq:zgas_eq}).

Setting $\epsilon$ to a constant ($\epsilon_m = \epsilon_z = 0$) while keeping a mass-dependent $\eta$ (fourth panel) removes the redshift offsets entirely.
The mass-dependent offsets largely disappear at high redshift but persist at low redshift, consistent with the picture that low-redshift, massive galaxies have had time to approach equilibrium where the mass-dependent $\eta$ imprints itself on the metallicity.

Finally, setting both $\epsilon$ and $\eta$ to constants by setting $\epsilon_m = \epsilon_z = \eta_m = 0$ (rightmost panel), all offsets vanish.
Every galaxy, regardless of stellar mass or redshift, falls on a single, universal sequence in the $Z_{\rm g}$--$M_\star/\mathrm{SFR}$ plane.
This is a non-trivial result: it demonstrates that in a universe where star formation efficiency and mass-loading factor are universal constants, the FMR reduces to a one-dimensional relation between $Z_\mathrm{g}$ and $M_\star/\mathrm{SFR}$ with no need for the $\alpha$ parameter.

\subsubsection{Building up the physical picture}

We now reconstruct the full model by progressively reintroducing the mass and redshift dependence of $\epsilon$ and $\eta$, using the two limiting regimes of chemical evolution to interpret each step.

The universal sequence in the rightmost panel of Fig.~\ref{fig:fmr_physics} has a characteristic shape: steep at low $M_\star/\mathrm{SFR}$ and flat at high $M_\star/\mathrm{SFR}$.
This shape directly solves the second puzzle.
Since $M_\star/\mathrm{SFR} \propto t$ across the whole evolution stage, and $\tau_{\rm eq}$ is a universal constant in this experiment, galaxies with low $M_\star/\mathrm{SFR}$ necessarily have low $t/\tau_{\rm eq}$ and therefore reside in the inflow-driven regime, where $Z_{\rm g} \propto \epsilon\,M_\star/\mathrm{SFR}$ (equation~\ref{eq:fmr_ideal}), giving a steep, linear scaling.
Galaxies with high $M_\star/\mathrm{SFR}$ have had time to approach equilibrium, where $Z_{\rm g} \to y/(1-R+\eta)$ (equation~\ref{eq:zgas_eq}) and the metallicity becomes insensitive to further increases in $M_\star/\mathrm{SFR}$.
\textit{The shape of the universal sequence is therefore a direct manifestation of the transition from inflow-driven evolution to equilibrium, governed by the ratio $t/\tau_{\rm eq}$.}

Allowing $\eta$ to depend on stellar mass (fourth panel) introduces offsets among mass bins, but only at the low-redshift, high-$M_\star/\mathrm{SFR}$ end.
This is precisely what the two-limits picture predicts.
In the inflow-driven regime, the metallicity is $Z_{\rm g} \propto \epsilon\,M_\star/\mathrm{SFR}$ and is independent of $\eta$; galaxies of different masses still follow the same sequence.
In the equilibrium regime, the metallicity is $Z_{\rm g} = y/(1-R+\eta)$, which depends on $\eta$; since $\eta$ now varies with mass, different mass bins saturate at different metallicities.
Moreover, a mass-dependent $\eta$ also modifies the equilibrium timescale $\tau_{\rm eq} = 1/[(1-R+\eta)\epsilon]$, so that galaxies of different masses approach equilibrium at different rates.
The net effect is that the mass-dependent offsets grow from high redshift to low redshift as an increasing fraction of the galaxy population enters the equilibrium regime.

Introducing a mass-dependent $\epsilon$ (second panel) adds offsets among mass bins that are now visible across the full redshift range, including at high redshift where galaxies are in the inflow-driven regime.
Again, this follows directly from the inflow-driven scaling: $Z_{\rm g} \propto \epsilon\,M_\star/\mathrm{SFR}$, so a mass-dependent $\epsilon$ shifts different mass bins vertically at fixed $M_\star/\mathrm{SFR}$.
Meanwhile, a mass-dependent $\epsilon$ also changes $\tau_{\rm eq}$, further modifying the approach to equilibrium for different masses.

Finally, adding a redshift dependence to $\epsilon$ (leftmost panel) introduces offsets among different epochs.
Since high-redshift galaxies reside predominantly in the inflow-driven regime, where $Z_{\rm g} \propto \epsilon\,M_\star/\mathrm{SFR}$ (equation~\ref{eq:fmr_ideal}), a higher $\epsilon$ at high redshift directly elevates the metallicity at fixed $M_\star/\mathrm{SFR}$, shifting high-redshift sequences above their low-redshift counterparts.
This produces the full spread seen in the fiducial model on the leftmost panel.
It is the combination of all these offsets that the FMR parameter $\xi=\log_{10}(M_\star/\mathrm{M}_\odot) - \alpha\,\log_{10}({\rm SFR}/{\rm M_\odot}\,{\rm yr}^{-1})$ must absorb to produce a redshift-invariant hypersurface in Fig.~\ref{fig:fmr_parameter_fitting}.

\subsubsection{Why $\alpha < 1$ and the nature of the FMR}

We are now in a position to answer the first puzzle.
In the simplest case (rightmost panel of Fig.~\ref{fig:fmr_physics}), $\alpha$ is unnecessary: the metallicity depends only on $M_\star/\mathrm{SFR}$, which corresponds to $\alpha = 1$.
Once $\epsilon$ and $\eta$ acquire their mass and redshift dependence, the sequences for different masses and redshifts are offset from one another in the $Z_{\rm g}$--$M_\star/\mathrm{SFR}$ plane.
These sequences have similar shapes because, at fixed stellar mass and redshift, galaxies with different SFR span a range of evolutionary stages from the inflow-driven limit to equilibrium, tracing the same underlying transition but shifted horizontally by the mass and redshift dependence of $\epsilon$ and $\eta$.
The FMR parameter $\xi=\log_{10}(M_\star/\mathrm{M}_\odot) - \alpha\,\log_{10}({\rm SFR}/{\rm M_\odot}\,{\rm yr}^{-1})$ acts as a horizontal shift that realigns these sequences.
Since massive galaxies have higher SFR than low-mass galaxies, and high-redshift galaxies have higher SFR than low-redshift galaxies at fixed mass, a value of $\alpha < 1$ shifts massive and high-redshift galaxies further to the right by an amount of $(1 - \alpha)\log_{10}{\rm SFR}$, compensating for their offsets.
\textit{The specific value of $\alpha$ that achieves the best collapse encodes the combined effect of the mass and redshift dependence of $\epsilon$ and $\eta$.}

We emphasise that there is no fundamental reason for galaxies to populate a redshift-invariant surface in the $(M_\star, \mathrm{SFR}, Z_{\rm g})$ space.
If $\epsilon$ and $\eta$ had arbitrary dependences on stellar mass and redshift, no single $\alpha$ would absorb all the offsets.
The observed FMR works because the dependences of $\epsilon$ and $\eta$ on stellar mass are well approximated by power laws, producing sequences of similar shape that can be aligned by a single horizontal shift, and because the redshift evolution of $\epsilon$ is sufficiently regular that the same $\alpha$ absorbs both the mass and redshift offsets simultaneously.
The tightness of the observed FMR is therefore not a fundamental symmetry of galaxy evolution but a contingent consequence of how star formation efficiency and mass-loading happen to depend on stellar mass and redshift in the real universe.
It is precisely this contingent nature that makes the FMR a potentially powerful constraint on the mass and redshift dependence of $\epsilon$ and $\eta$.

\subsubsection{The physics of gFMR}

We repeat the same set of experiments for the gFMR, plotting $Z_\mg$ against $\log_{10}\,M_\star/M_\mg$ in Fig.~\ref{fig:gfmr_physics}.
The last two panels, where $\epsilon$ is constant, are qualitatively similar to their counterparts in Fig.~\ref{fig:fmr_physics}: a universal sequence emerges when both $\epsilon$ and $\eta$ are constant, and a mass-dependent $\eta$ introduces offsets at the equilibrium end.
This is expected, since when $\epsilon$ is constant the mapping between $M_{\rm g}$ and SFR is exact and the two projections carry the same information.

The first two panels, however, reveal a striking difference.
When $\epsilon$ depends on stellar mass (with or without a redshift dependence), the $Z_{\rm g}$--$M_\star/M_{\rm g}$ relation remains far tighter than the corresponding $Z_{\rm g}$--$M_\star/\mathrm{SFR}$ relation in Fig.~\ref{fig:fmr_physics}.
Two effects contribute to this tightness.

First, in the inflow-driven regime the gas metallicity scales as $Z_\mg \propto M_\star/M_\mg$ (equation~\ref{eq:gas_fmr_ideal}), with no modulation from the star formation efficiency.
The standard FMR, by contrast, involves $Z_\mg \propto \epsilon\,M_\star/{\rm SFR}$, so any mass or redshift dependence of $\epsilon$ directly imprints itself as offsets among different populations.
The gFMR bypasses this entirely: variations in $\epsilon$ affect the SFR but do not enter the relationship between metallicity and gas fraction.

Second, even the transition from the inflow-driven regime to equilibrium, quantified by $t/\tau_{\rm eq}$, proceeds at a similar pace across different stellar masses.
The galaxy evolution time $t \propto M_\star/\mathrm{SFR}$, so that $t/\tau_{\rm eq} \propto (1 - R + \eta)\,M_\star/M_\mathrm{g}$, since $\mathrm{SFR} = \epsilon\,M_\mathrm{g}$ and the star formation efficiency cancels between the numerator and denominator.
As a result, the pace at which galaxies evolve from the inflow-driven regime to equilibrium is determined by the mass-loading factor ($\eta$) and the gas fraction ($M_\mathrm{g}/M_\star$) alone, with no residual dependence on $\epsilon$.

Together, these two effects explain why the gFMR is substantially more robust than the standard FMR: the inflow-driven limit is inherently independent of $\epsilon$, and the approach to equilibrium depends only on the gas fraction and mass-loading factor.

A cautious reader may wonder, if the star formation efficiency does not affect the relation between gas fraction and gas metallicity, why is the scatter smaller in the second panel of Fig.~\ref{fig:gfmr_physics}, with $\epsilon(M_\star)$ and $\eta(M_\star)$, than in the fourth panel, which assumes $\epsilon={\rm const.}$.
The fourth panel is straightforward to understand: the metallicity in the equilibrium regime is regulated by the mass-loading factor, so an offset is expected there, since the mass-loading factor is itself a function of stellar mass.
The tightness of the second panel instead arises because low-mass galaxies ($M_\star \sim 10^8\,\rm M_\odot$) now have higher gas fractions and are shifted leftward in the panel.
The reason these low-mass galaxies have higher gas fractions, once the star formation efficiency is allowed to depend on stellar mass, is that the sSFR of galaxies evolves robustly as a function of cosmic time alone ($\propto 1/t$).
If star formation efficiency instead depends on stellar mass and is lower for low-mass galaxies, those galaxies must carry higher gas content to keep the sSFR only weakly dependent on stellar mass.

\section{The analytic structure of galaxy chemical evolution}
\label{sec:analytic_approximate}

\begin{figure}
  \begin{center}
    \includegraphics[width=0.95\linewidth]{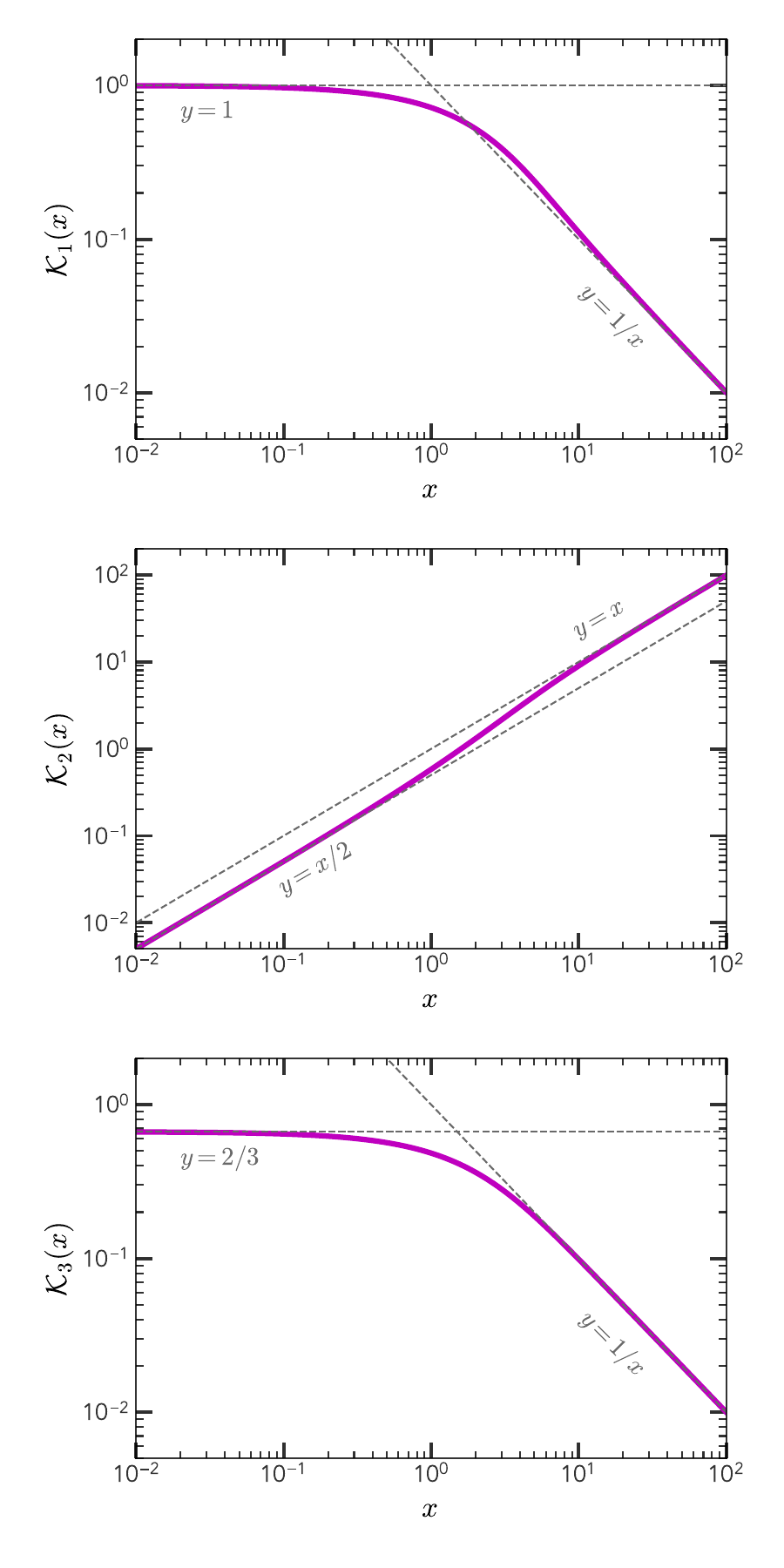}
  \end{center}
  \caption{
    The three functions $\mathcal{K}_1$, $\mathcal{K}_2$, and $\mathcal{K}_3$ that govern the analytic solution of the ideal gas flow model (equations~\ref{eq:master}--\ref{eq:zstar_master}), plotted as functions of the evolutionary stage $x \equiv t/\tau_{\rm eq}$.
    Dashed lines show the asymptotic limits.
    In the inflow-driven regime ($x \ll 1$): $\mathcal{K}_1 \to 1$, $\mathcal{K}_2 \to x/2$, and $\mathcal{K}_3 \to 2/3$.
    In the equilibrium limit ($x \gg 1$): $\mathcal{K}_1 \to 1/x$, $\mathcal{K}_2 \to x$, and $\mathcal{K}_3 \to 1/x$.
    The transition between the two regimes occurs near $x \sim 1$, corresponding to $t \sim \tau_{\rm eq}$.
  }
  \label{fig:K123}
\end{figure}

The controlled experiments of \S\,\ref{sec:fundamental_metallicity_relation} established a hierarchy: the gFMR is primary, the standard FMR is its projection into SFR space, and the FMR parameterisation $\alpha$ encodes the mass and redshift dependence of $\epsilon$ and $\eta$ rather than a fundamental symmetry.
What the experiments cannot provide is a closed-form expression relating $Z_{\rm g}$, $M_{\rm g}/M_\star$, and $\eta$ that makes this hierarchy explicit and quantitative.
We now derive such an expression from the ideal gas flow model, in which $\Phi$, $\epsilon$, and $\eta$ are all constants.

\subsection{Gas metallicity}

Combining equations~\eqref{eq:mgas_ana}--\eqref{eq:mstar_ana}, we can relate the gas metallicity, gas fraction, and mass-loading factor (see Appendix~\ref{sec:k1k2k3} for detailed derivation):
\begin{align}
  &Z_{\rm g} = \frac{y}{1 - R} \left(\frac{M_{\rm g}}{M_\star}\right)^{-1} \mathcal{K}_1\!\left(\frac{t}{\tau_{\rm eq}}\right), \quad
  \mathcal{K}_1(x) = \frac{1 - \me^{-x} - x\,\me^{-x}}
  {x - 1 + \me^{-x}}
  \label{eq:master}\\
  &\frac{1 - R + \eta}{1 - R} \left(\frac{M_{\rm g}}{M_\star}\right)^{-1} = \mathcal{K}_2\!\left(\frac{t}{\tau_{\rm eq}}\right), \quad
  \mathcal{K}_2(x) = \frac{x - 1 + \me^{-x}}{1 - \me^{-x}}.
  \label{eq:t2tau}
\end{align}
\textit{
  Together, equations~\eqref{eq:master} and \eqref{eq:t2tau} form a closed-form system that relates the gas metallicity, the gas fraction, and the mass-loading factor.
  Given any two of these three quantities, the third can be determined.
}
The functions $\mathcal{K}_1$ and $\mathcal{K}_2$ are plotted in Fig.~\ref{fig:K123}, together with their asymptotic limits in the inflow-driven and equilibrium regimes.

These equations encode the full transition from the inflow-driven regime to equilibrium. In the inflow-driven limit ($t/\tau_{\rm eq} \to 0$), $\mathcal{K}_1 \to 1$ and the gas metallicity reduces to equation~\eqref{eq:gas_fmr_ideal}, which depends only on the gas fraction and is independent of both $\epsilon$ and $\eta$.
This is the gFMR in its simplest form: the metallicity is set by the ratio of metals produced (proportional to $M_\star$) to the gas reservoir that dilutes them.
In the equilibrium limit ($t/\tau_{\rm eq} \to \infty$), $\mathcal{K}_1 \to \tau_{\rm eq}/t$ and the metallicity saturates at the value given by equation~\eqref{eq:zgas_eq}, which depends only on the mass-loading factor and is insensitive to the gas fraction.
The function $\mathcal{K}_1$ interpolates smoothly between these two limits, with its argument $t/\tau_{\rm eq}$ governing where along this transition a given galaxy sits.

\subsection{Stellar metallicity}

The ideal model also yields an analytic expression for the stellar metallicity. Following the same procedure, we obtain
\begin{equation}
  Z_\star = \frac{y}{1 - R} \left(\frac{M_{\rm g}}{M_\star}\right)^{-1} \mathcal{K}_3\!\left(\frac{t}{\tau_{\rm eq}}\right), \quad
  \mathcal{K}_3(x) = \frac{1 - \mathcal{K}_1(x)}{\mathcal{K}_2(x)},
  \label{eq:zstar_master}
\end{equation}
which has the same prefactor as the gas relation but with a different function of the evolutionary stage.
The function $\mathcal K_3$ is plotted in the lower panel of Fig.~\ref{fig:K123}.

In the inflow-driven limit, $\mathcal{K}_3 \to 2/3$, giving the result in equation~\eqref{eq:zstar_inflow}.
In the equilibrium limit, both metallicities converge to the same value as $Z_\mg$: once the gas metallicity has saturated, all newly formed stars inherit the same equilibrium abundance, and the mass-weighted stellar metallicity asymptotically catches up.

\subsection{Validation with the cosmological gas flow model}
\label{sub:validation}

\begin{figure*}
  \begin{center}
    \includegraphics[width=0.85\linewidth]{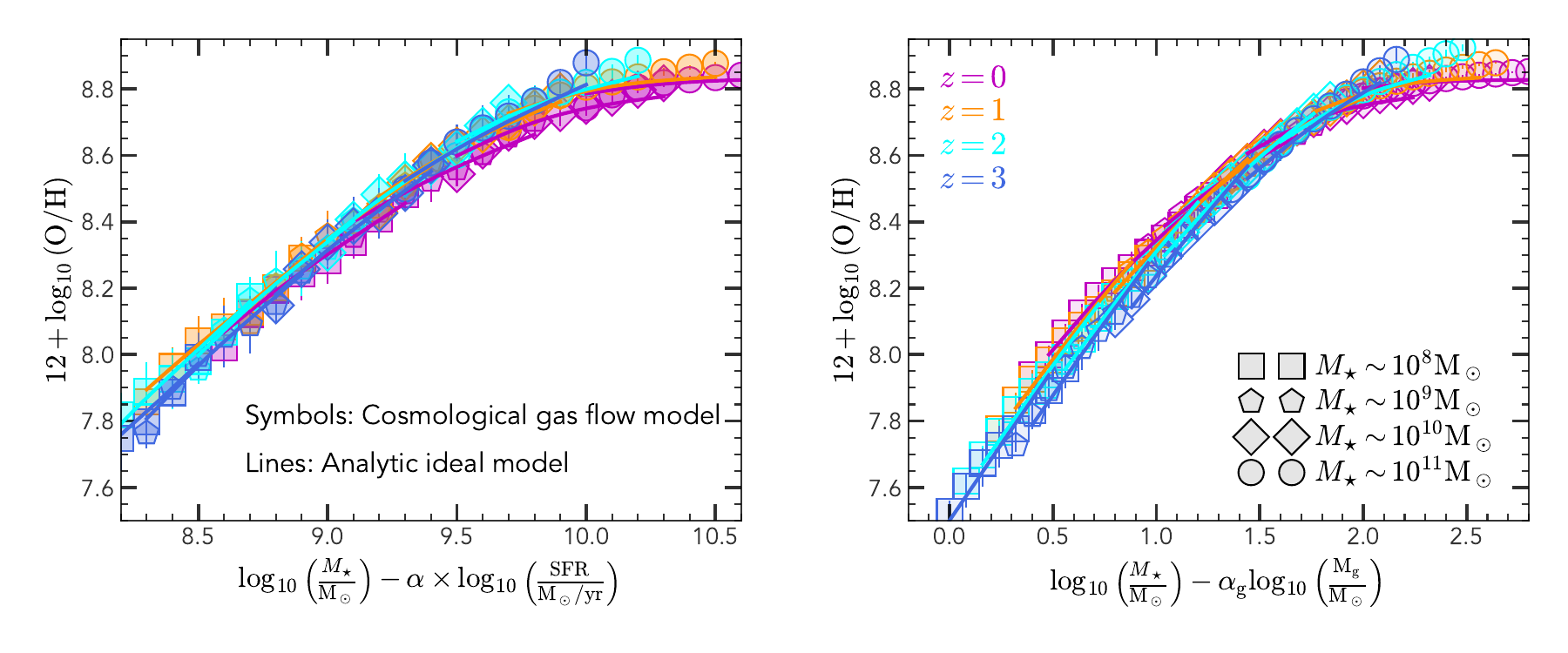}
  \end{center}
  \caption{
    Comparison between the cosmological gas flow model (symbols) and the analytic ideal model (solid lines).
    \textbf{Left panel:} the FMR, showing gas metallicity as a function of the FMR parameter $\log_{10}(M_\star/\mathrm{M}_\odot) - \alpha\,\log_{10}(\mathrm{SFR}/\mathrm{M}_\odot\,\mathrm{yr}^{-1})$ with $\alpha=0.55$.
    \textbf{Right panel:} the gFMR, showing gas metallicity as a function of $\log_{10}(M_\star/\mathrm{M}_\odot) - \alpha_\mg\,\log_{10}(M_{\rm g}/\mathrm{M}_\odot)$ with $\alpha_\mg = 0.85$.
    In both panels, marker shapes distinguish stellar mass bins and shading denotes redshift.
    The solid lines show the median metallicity predicted by equations~\eqref{eq:master} and \eqref{eq:t2tau}, using the gas fraction and mass-loading factor from the cosmological model as inputs.
    The excellent agreement confirms that the analytic ideal model captures the essential physics of the full cosmological calculation.
  }
  \label{fig:ideal_model_explain}
\end{figure*}

To test whether the analytic expressions derived from the ideal model can approximate the more realistic cosmological gas flow model, we use the gas fraction ($M_\mg/M_\star$) and mass-loading factor ($\eta$) from the cosmological model as inputs to equations~\eqref{eq:master} and \eqref{eq:t2tau}, and predict the gas metallicity ($Z_\mg$).
Fig.~\ref{fig:ideal_model_explain} compares the result with the cosmological model output for both the FMR (left panel) and the gFMR (right panel).
The symbols show the cosmological model output and the solid lines show the analytic prediction using equations~\eqref{eq:master} and \eqref{eq:t2tau}.
The agreement is excellent across the full range of stellar masses and redshifts, confirming that the ideal model captures the essential physics governing the metallicity scaling relations despite its simplifying assumptions of constant inflow rate and constant $\epsilon$ and $\eta$.

The agreement between the analytic ideal model and the full cosmological calculation (Fig.~\ref{fig:ideal_model_explain}) can be understood by examining which input quantities affect the metallicity scaling relations and how rapidly they vary.
First, as demonstrated in \S\,\ref{sub:mzr_shmr} (Fig.~\ref{fig:inflow_rate_sensitivity}), the gas inflow rate $\Phi$ does not enter any scaling relation among $M_\star$, SFR, $M_\mathrm{g}$, and $Z_\mathrm{g}$, so its time variation does not affect the metallicity scaling relations, at least for the smooth accretion histories considered here.
For the remaining quantities, consider a galaxy growing from $M_\star/2$ to $M_\star$: this final doubling accounts for half of the total metal production and therefore dominates the metal budget.
Over this factor-of-two growth in stellar mass, the star formation efficiency $\epsilon \propto M_\star^{0.33}$ changes by only a factor of $2^{0.33} \approx 1.26$, and the mass-loading factor $\eta \propto M_\star^{-0.30}$ changes by a factor of $2^{-0.30} \approx 0.81$.
Both variations are modest, justifying the approximation of constant values over the interval that matters most for the chemical enrichment history.
Furthermore, the redshift evolution of $\epsilon$ over this same interval is also mild, since the time required for a galaxy to double in stellar mass is short compared to the Hubble time, so that $(1+z)$ changes little.
The ideal model therefore succeeds not by coincidence but because the quantities that govern the metallicity, $\epsilon$ and $\eta$, vary slowly over the mass-doubling interval that dominates the metal budget.

\subsection{Towards an analytic understanding of the FMR and gFMR}
\label{sub:analytic_understanding_fmr}

\begin{figure}
  \begin{center}
    \includegraphics[width=0.95\linewidth]{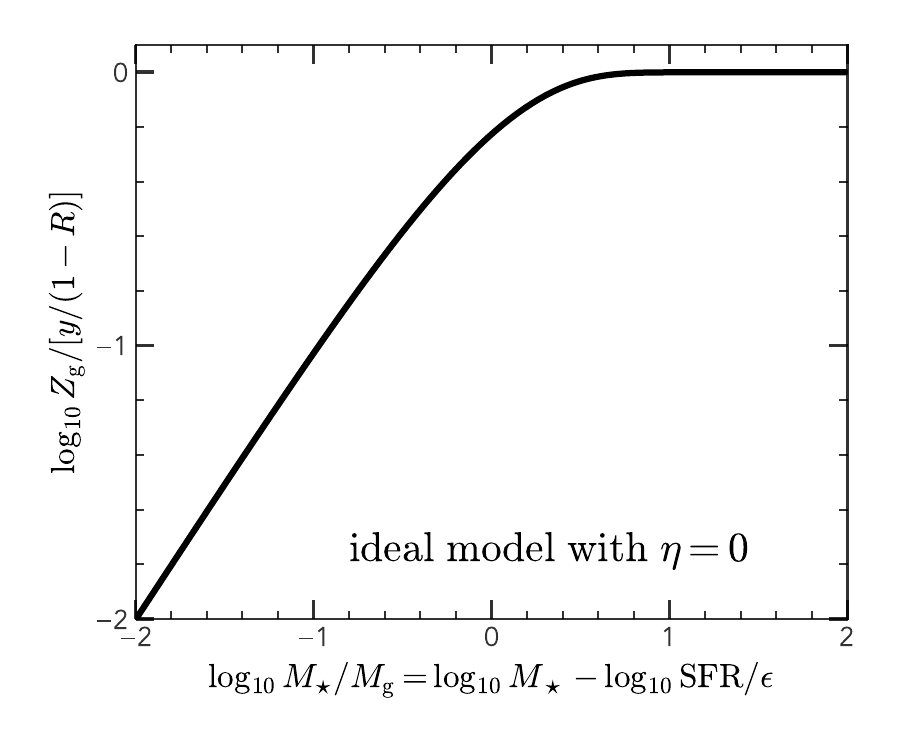}
  \end{center}
  \caption{
    Gas metallicity as a function of the stellar-to-gas mass ratio $M_\star/M_\mg$, predicted by the ideal gas flow model with $\eta = 0$.
    The metallicity is normalised by the maximum (outflow-free) yield $y/(1-R)$, and the horizontal axis is equivalently $\log_{10} M_\star - \log_{10} \mathrm{SFR}/\epsilon$, showing that when $\epsilon$ is constant the gFMR and FMR projections carry identical information.
    The curve traces the transition from the inflow-driven regime, where $Z_\mg \propto M_\star/M_\mg$, to equilibrium, where $Z_\mg \to y/(1-R)$.
  }
  \label{fig:no_eta}
\end{figure}

The analytic solution derived above provides a framework for understanding why the gFMR is more fundamental than the standard FMR, and under what conditions the standard FMR is expected to be redshift-invariant.

Equations~\eqref{eq:master} and \eqref{eq:t2tau} together establish that, for a given mass-loading factor $\eta$, there is a unique mapping between the gas fraction $M_{\rm g}/M_\star$ and the gas metallicity $Z_{\rm g}$.
This uniqueness arises because the gas fraction and $\eta$ jointly determine the evolutionary stage $t/\tau_{\rm eq}$ through equation~\eqref{eq:t2tau}, which in turn fixes $\mathcal{K}_1$ and hence $Z_{\rm g}$ through equation~\eqref{eq:master}.
The mapping is well-defined across the full evolutionary continuum: in the inflow-driven limit $Z_{\rm g}$ depends only on the gas fraction (equation~\ref{eq:gas_fmr_ideal}), while in the equilibrium limit $Z_{\rm g}$ depends only on $\eta$ (equation~\ref{eq:zgas_eq}).
\textit{If $\eta$ depends only on stellar mass, then equations~\eqref{eq:master} and \eqref{eq:t2tau} define a surface in the $(M_\star, M_{\rm g}, Z_{\rm g})$ space: at any given $M_\star$ and $M_{\rm g}$, the metallicity is uniquely determined.}
This is the gFMR, and its existence follows directly from the structure of the analytic solution.

The standard FMR emerges from the gFMR by replacing $M_{\rm g}$ with $\mathrm{SFR}/\epsilon$.
If both $\eta$ and $\epsilon$ depend only on stellar mass, then at any given cosmic epoch the substitution is exact and galaxies occupy a well-defined surface in the $(M_\star, \mathrm{SFR}, Z_{\rm g})$ space.
If $\epsilon$ also depends on redshift, the mapping between $M_{\rm g}$ and $\mathrm{SFR}$ changes from epoch to epoch, and it is not immediately obvious that a redshift-invariant surface persists.

To see whether redshift invariance survives, consider the median evolution of the star-forming population.
In both the inflow-driven and equilibrium limits the specific SFR evolves as $\mathrm{sSFR} \propto 1/t$ (\S\,\ref{sec:inflow_driven_regime}), providing a monotonic mapping between cosmic time and an observable galaxy property at the population level.
Since any redshift dependence of $\epsilon$ can therefore be re-expressed as a dependence on the median $\mathrm{sSFR}$, the median gas metallicity at fixed $M_\star$ depends on $\mathrm{SFR}$ alone with no residual redshift dependence, and a redshift-invariant median FMR surface is guaranteed regardless of the specific functional form of $\epsilon(z)$.

It is important to note, however, that this argument is strictly a population-level statement.
The relation $\mathrm{sSFR} \propto 1/t$ describes the median of the star-forming main sequence; at fixed $M_\star$ and fixed cosmic time, individual galaxies scatter around this median due to variations in their accretion histories.
For an individual galaxy, $\mathrm{sSFR}$ is not uniquely determined by $t$, so the mapping between the redshift dependence of $\epsilon$ and $\mathrm{sSFR}$ breaks down object by object.
The redshift-invariance argument therefore guarantees only that the \textit{median} FMR surface does not evolve with redshift; it does not require the scatter around that surface to be redshift-invariant, nor does it correctly describe the position of individual objects on the surface.

Whether the median surface can be captured by the standard FMR parameterisation $\xi = \log_{10} M_\star - \alpha \log_{10} \mathrm{SFR}$ with a single value of $\alpha$ is a separate question that does not admit a simple analytic answer.
We therefore approach the problem in two limiting cases: the case where the mass-loading factor dominates the metallicity evolution, and the case where the gas fraction dominates and the mass-loading factor can be neglected.

In the limiting case where galaxies are all in equilibrium due to a high mass-loading factor, the gas metallicity reduces to $Z_\mg = y/(1-R+\eta) \approx (y/\eta_0)(M_\star/M_\eta)^{-\eta_m}$, which depends only on stellar mass and is independent of SFR.
Galaxies then form a sequence in the $Z_\mg$--$M_\star$ plane with no SFR dependence at all, corresponding to a flat sequence in Fig.~\ref{fig:fmr_physics} at each stellar mass.
In this limit, setting $\alpha = 0$ trivially collapses all galaxies onto a single locus, but the resulting relation is simply the MZR with no scatter: the FMR reduces to a two-dimensional relation and carries no additional information beyond the MZR itself.

In the second limiting case, the gas fraction dominates the metallicity evolution and the mass-loading factor can be neglected.
To see this, we turn to the ideal model, which has been shown to reproduce the cosmological model with excellent accuracy (\S\,\ref{sub:validation}).
If $\eta$ can be neglected\footnote{Neglecting $\eta$ does not require $\eta \ll 1-R$; it is sufficient that $M_\mg/M_\star \gg (1-R+\eta)/(1-R)$, so that galaxies remain in the inflow-driven limit and $\eta$ is dynamically unimportant even if its value is not small.}, equations~\eqref{eq:master} and \eqref{eq:t2tau} reduce to a monotonic, universal relation between $Z_\mg$ and $M_\star/M_\mg$, shown in Fig.~\ref{fig:no_eta}.
This condition is satisfied in the real Universe, since the galaxies closest to equilibrium are massive and low-redshift, and these are precisely the galaxies with the smallest mass-loading factors due to their deep gravitational potential wells.

The universal curve in Fig.~\ref{fig:no_eta} is the backbone of the FMR: any parameterisation that maps galaxies of different masses and redshifts onto a common value of $M_\star/M_\mg$ will collapse them onto this curve.
If the star formation efficiency can be approximated as a power law in stellar mass and SFR, $\epsilon \propto M_\star^a\,\mathrm{SFR}^b$, then
\begin{equation}
  \log_{10}\!\left(\frac{M_\star}{M_\mg}\right) =
  (1+a)\left(\log_{10} M_\star -
  \frac{1-b}{1+a}\log_{10}\mathrm{SFR}\right) + \mathrm{const.},
  \label{eq:fmr_projection}
\end{equation}
and a single FMR parameter $\alpha = (1-b)/(1+a)$ collapses all galaxies onto the universal curve regardless of mass or redshift.

In practice, $\epsilon$ depends on both stellar mass and redshift rather than on SFR directly, so converting between the two introduces an additional dependence on cosmic time.
This means there is not one but two distinct optimisation problems that each determine a preferred value of $\alpha$: one that minimises the scatter within a single epoch, and one that minimises the offset between epochs at different redshifts.

The first is the best collapse \textit{at a single epoch}.
At fixed redshift, the factor $(1+z)^{\epsilon_z}$ is a constant that shifts all galaxies equally and drops out of the FMR projection, leaving only the stellar-mass dependence $\epsilon \propto M_\star^{\epsilon_m}$.
So the value of $\alpha$ that perfectly collapses a single epoch is
\begin{equation}
  \alpha_{m} = \frac{1}{1 + \epsilon_m}.
  \label{eq:alpha_snap}
\end{equation}
For the fiducial value $\epsilon_m = 0.33$ this gives $\alpha_m \approx 0.75$.

The second is the best \textit{redshift invariance} across all epochs.
Now the redshift dependence of $\epsilon$ must be absorbed into $\xi$.
Re-expressing the redshift evolution as a time dependence via $t \propto (1+z)^{-3/2}$ in the matter-dominated approximation gives $\epsilon \propto M_\star^{\epsilon_m}\, t^{-\epsilon_z'}$ where $\epsilon_z' \equiv 2\epsilon_z/3$.
Substituting into the inflow-driven scaling and using $\mathrm{sSFR} \propto 1/t$ yields $\epsilon\propto M_\star^{\epsilon_{\rm m} -\epsilon_z'}\,{\rm SFR}^{\epsilon_z'}$, from which the value of $\alpha$ that achieves redshift invariance is
\begin{equation}
  \alpha_z =
  \frac{1 - \epsilon_z'}{1 + \epsilon_m - \epsilon_z'}.
  \label{eq:alpha_inv}
\end{equation}
For the fiducial values $\epsilon_m = 0.33$ and $\epsilon_z = 1.1$ ($\epsilon_z' \approx 0.73$), equation~\eqref{eq:alpha_inv} yields $\alpha_z \approx 0.45$.
One can verify directly from equations~\eqref{eq:alpha_snap} and \eqref{eq:alpha_inv} that $\alpha_z < \alpha_m$ whenever $\epsilon_z' > 0$, i.e.\ whenever $\epsilon$ increases with redshift at fixed $M_\star$.
The reason is that a positive $\epsilon_z'$ means high-redshift galaxies are systematically more efficient than the single-epoch expectation, so a smaller $\alpha$ is needed to bring them into alignment with the low-redshift population.

The observed best-fit value $\alpha \approx 0.55$ lies strictly between these two limits,
\begin{equation}
  \alpha_z \lesssim \alpha
  \lesssim \alpha_m,
  \label{eq:alpha_bracket}
\end{equation}
as it must: it is the compromise that simultaneously minimises scatter within each epoch and scatter across redshifts.
This bracketing is not a coincidence but a direct consequence of $\epsilon$ having both a positive mass dependence ($\epsilon_m > 0$) and a positive redshift dependence ($\epsilon_z > 0$).
The residual offset of $\alpha$ above $\alpha_z$ further reflects the contribution of massive, low-redshift galaxies that have entered the equilibrium regime, where the mass-dependent $\eta$ pulls $\alpha$ upward from the pure inflow-driven prediction.

\textit{In summary, the existence of a redshift-invariant median FMR surface is ensured to the degree that $\mathrm{sSFR} \propto 1/t$ faithfully describes the median evolution of the star-forming population.
  The further requirement that this surface can be parameterised by a single $\alpha$ is more restrictive.
  When the mass-loading factor dominates the metallicity evolution, $\alpha = 0$ trivially works but the FMR reduces to the MZR with no SFR dependence.
  When the gas fraction dominates instead, $\alpha$ takes a value between 0 and 1, set by the power-law dependences of the star formation efficiency on stellar mass and SFR through equation~\eqref{eq:fmr_projection}, so that the universal $Z_\mg$--$M_\star/M_\mg$ relation of Fig.~\ref{fig:no_eta} can be projected onto the observable $\log_{10}M_\star - \alpha\log_{10}\mathrm{SFR}$ plane.
}
The tightness of the observed FMR therefore reflects not only the existence of a universal enrichment pathway from the inflow-driven regime to equilibrium, but also the regularity of the star formation efficiency as a function of stellar mass and redshift.
Precise measurements of $\alpha$, combined with independent constraints on $\epsilon_m$, could in turn be used to infer $\epsilon_z$ through equation~\eqref{eq:alpha_inv}, turning the FMR parameterisation into a quantitative diagnostic of gas consumption physics across cosmic time.

\section{Discussion}
\label{sec:discussion}

\subsection{Revisiting the equilibrium model}
\label{sub:revisiting_the_equilibrium_model} 

\begin{figure}
  \begin{center}
    \includegraphics[width=0.95\linewidth]{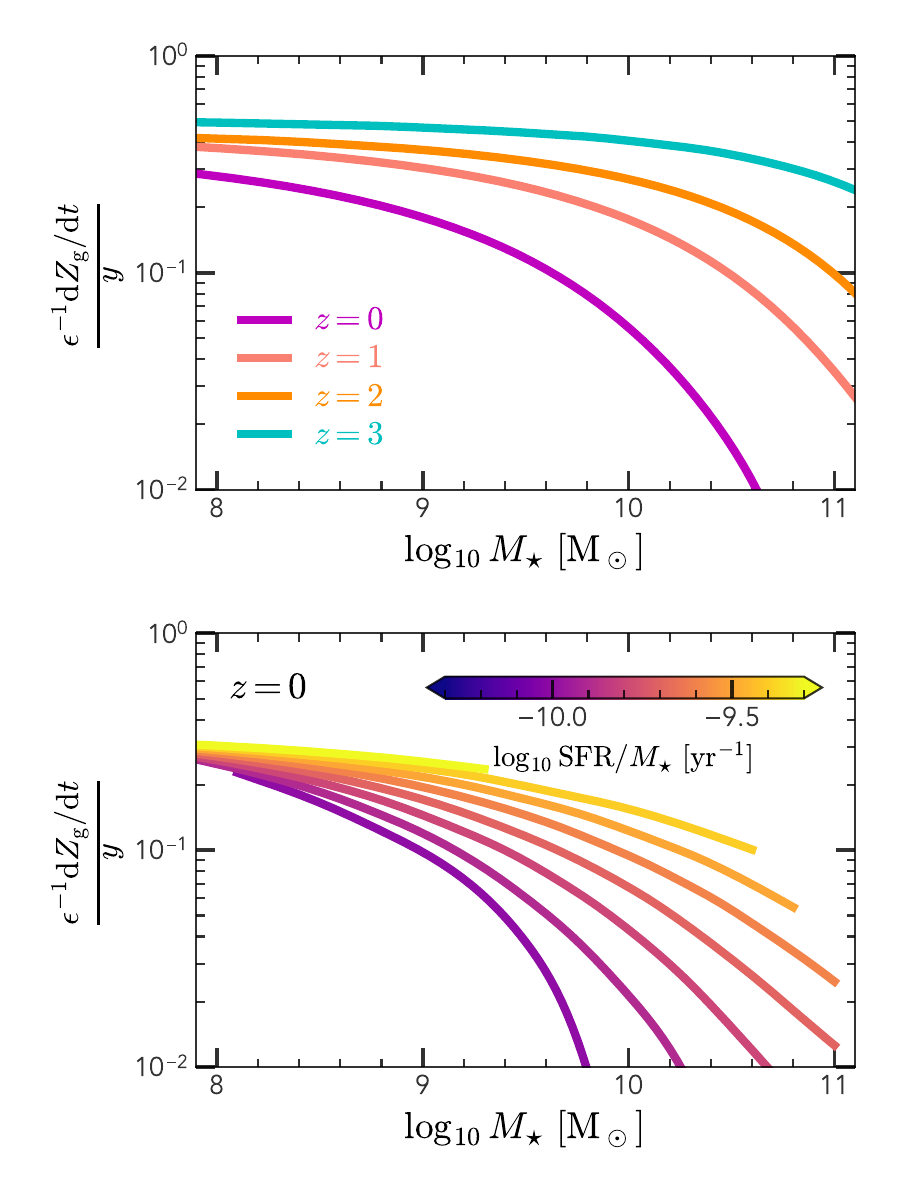}
  \end{center}
  \caption{
    Diagnostic of the equilibrium approximation in the gas flow model.
    Both panels show the ratio $(\epsilon^{-1}\,\dd{Z}_\mg/\dd t)/y$, which measures the relative importance of the time-derivative term in equation~\eqref{eq:zgas_lilly} to the yield.
    The equilibrium approximation requires this ratio to be small, so that the time-derivative of the gas metallicity can be neglected.
    \textbf{Upper panel:} the ratio evaluated at $z=0$, 1, 2, and 3.
    At all redshifts, the ratio decreases with stellar mass as more massive galaxies have higher star formation efficiencies, placing them closer to equilibrium.
    The ratio increases with increasing redshift, reflecting the shorter time to form the galaxy.
    All curves asymptote to an upper limit of $0.5$ at low masses, corresponding to the inflow-driven prediction $(\epsilon^{-1}\,\dd Z_\mg/\dd t)/y = 0.5$.
    \textbf{Lower panel:} the ratio at $z=0$, with individual model tracks colour-coded by their present-day specific SFR.
    At fixed stellar mass, galaxies with higher sSFR are closer to the inflow-driven regime, because they have larger gas fractions and correspondingly smaller $t/\tau_{\rm eq}$ (equation~\ref{eq:t2tau}).
  }
  \label{fig:dZdt}
\end{figure}

Previous models based on the equilibrium assumption \citep[e.g.][]{lillyGASREGULATIONGALAXIES2013, feldmannLessonsCosmicHistory2013, feldmannEquilibriumViewDust2015, bassiniInflowOutflowProperties2024} take as their starting point the expression
\begin{align}
  Z_{\mg} = \frac{y - \epsilon^{-1}\,{\dd Z_{\mg}}/{\dd t}}{(1 - R + \eta) + (1 - R)\mu + \epsilon^{-1}\,{\dd\ln\mu}/{\dd t}},
  \label{eq:zgas_lilly}
\end{align}
where $\mu\equiv M_\mg/M_\star$ is the gas fraction.
This expression is exact and follows directly from the two mass-continuity equations (see Appendix~\ref{sec:the_derivation_of_equation_eqref_eq_zgas_lilly_}).
The equilibrium approximation consists of dropping the $\epsilon^{-1}\,\dd Z_\mg/\dd t$ term in the numerator, so that the metallicity can be inferred from the star-forming main sequence and gas fraction, together with their temporal evolution.
However, the validity of this approximation has never been quantitatively demonstrated.
The only justification offered in the literature is a linear stability analysis showing that perturbations away from equilibrium decay on a timescale of  $\tau_\text{eq}$ \citep{lillyGASREGULATIONGALAXIES2013, feldmannLessonsCosmicHistory2013}; this establishes that an equilibrium \emph{exists}, but does not show that galaxies have had sufficient time to reach it \citep[see also][]{pengHaloesGalaxiesDynamics2014}.

We test this assumption directly by evaluating the ratio $(\epsilon^{-1}\,\mathrm{d}Z_{\rm g}/\mathrm{d}t)/y$ as a function of stellar mass and redshift from our cosmological gas-flow model.
The results are shown in the upper panel of Fig.~\ref{fig:dZdt}.
The equilibrium approximation requires this ratio to be small ($\ll 1$).
The ratio decreases monotonically with stellar mass and increases with redshift: more massive galaxies have higher star formation efficiencies, placing them closer to equilibrium.
At the opposite extreme, the ratio asymptotes to an upper limit of 0.5 at low masses and high redshifts.
This is precisely the prediction of the inflow-driven regime: from equation~\eqref{eq:zgas_inflow}, $Z_{\rm g} = y\epsilon t/2$, giving $\epsilon^{-1}\,\mathrm{d}Z_{\rm g}/\mathrm{d}t = y/2$, so that $(\epsilon^{-1}\,\mathrm{d}Z_{\rm g}/\mathrm{d}t)/y = 0.5$.
The fact that this upper bound is reached over a wide range of stellar masses at $z \geq 2$ confirms that the majority of star-forming galaxies at high redshift have not yet approached equilibrium.

The lower panel of Fig.~\ref{fig:dZdt} shows the same diagnostic for galaxies at $z = 0$, colour-coded by their present-day sSFR.
At fixed stellar mass, galaxies with higher sSFR lie closer to the inflow-driven regime.
This follows directly from the analytic framework developed in \S\,\ref{sec:analytic_approximate}: at fixed stellar mass and mass-loading factor, a higher gas fraction $M_{\rm g}/M_\star$ corresponds to a smaller evolutionary stage $t/\tau_{\rm eq}$ through equation~\eqref{eq:t2tau}, placing the galaxy closer to the inflow-driven limit.
Since $\mathrm{sSFR} = \epsilon(M_\star)M_\mg /M_\star$, galaxies with higher sSFR at fixed stellar mass are more gas-rich and therefore less evolved.

We note that dropping the time-derivative term in the inflow-driven limit introduces a systematic offset in the gas-phase metallicity of a factor of 2 ($\approx 0.3$~dex).
For context, the observed gas metallicity difference between the highest and lowest SFR galaxies at fixed stellar mass is $\lesssim 0.3$~dex \citep{curtiMassmetallicityFundamentalMetallicity2020}, and the redshift evolution at fixed stellar mass from $z \sim 0$ to $z \sim 3.3$ amounts to only $\lesssim 0.5$~dex \citep[see also][]{jainUniformAnalysisGasphase2025}.
The equilibrium approximation $y \gg \epsilon^{-1}\,\mathrm{d}Z_{\rm g}/\mathrm{d}t$ is therefore quantitatively justified only for massive galaxies at low redshift.
For the bulk of the star-forming population, the time-derivative term remains significant, and the full framework developed in this paper, which encompasses both the inflow-driven and equilibrium regimes, provides a more accurate description of the metallicity evolution.

This inconsistency is also visible within the results of \citet{lillyGASREGULATIONGALAXIES2013}.
Despite assuming $\mathrm{d}Z_\mg/\mathrm{d}t = 0$ for all galaxies, they derive a mass--metallicity relation with a positive slope and a normalisation that increases with cosmic time at fixed stellar mass (their fig.~7).
This implies that the star-forming population, growing in stellar mass and evolving forward in time, must also increase in metallicity, i.e. $\mathrm{d}Z_\mg/\mathrm{d}t > 0$, directly contradicting the equilibrium assumption used to derive the relation in the first place.

A further limitation of the scheme introduced by \citet{lillyGASREGULATIONGALAXIES2013} concerns their treatment of the ideal regulator, which they define as the case of having constant star formation efficiency and mass-loading factor.
Under this definition, they set $\dd\ln\mu/\dd t = 0$, concluding that the gas-phase metallicity is determined solely by the instantaneous state of the system with no memory of its past evolution.
However, this does not follow from constant efficiency and loading alone.
$\dd\ln\mu/\dd t = 0$ requires the additional condition that the specific inflow rate is constant, a condition that is invoked implicitly in their derivations but which is absent from their definition of the ideal model, and is not justified in general.

Their response to this is an appeal to timescales: they argue from observations (their fig.~4) that the specific inflow rate varies on a timescale roughly three times longer than the gas consumption timescale, so that realistic galaxies remain close to the no-memory limit.
However, this timescale comparison is compromised by two factors that systematically elevate the true gas consumption timescale.
First, their estimate of the gas consumption timescale is based on molecular gas alone, whereas atomic gas dominates the total gas reservoir at least to $z\sim1$ \citep{saintongeColdInterstellarMedium2022a,chowdhuryAtomicGasDominates2022}, increasing the effective gas consumption timescale by a factor of approximately two to six.
Second, mass return from stellar evolution replenishes the gas reservoir, further elevating the effective gas consumption timescale by a factor of approximately two.
Correcting for both effects brings the gas consumption timescale into the same range as the timescale on which the specific inflow rate evolves, undermining the timescale separation on which their argument rests.

Even setting aside these observational corrections, justifying a slowly varying specific inflow rate is itself a non-trivial task.
\citet{lillyGASREGULATIONGALAXIES2013} implicitly assume a proportionality between the dark matter accretion rate and the cold gas inflow rate, with the ratio between the two encoding the cooling efficiency.
This efficiency is a complex function of halo mass and redshift, reflecting the temperature and metallicity dependence of radiative cooling, the varying importance of preventative feedback, and the transition between cold-stream and hot-mode accretion at different epochs.
Consequently, the assumption $\dd\ln\mu/\dd t \approx 0$ is not justified for realistic galaxies, and the gas regulator model of \citet{lillyGASREGULATIONGALAXIES2013} does not provide a valid foundation for modelling the chemical evolution of galaxies.

\subsection{Revisiting \citet{peeplesConstraintsStarFormation2011}}
\label{sub:revisiting_citep_peeplesconstraintsstarformation_} 

\begin{figure}
  \begin{center}
    \includegraphics[width=0.95\linewidth]{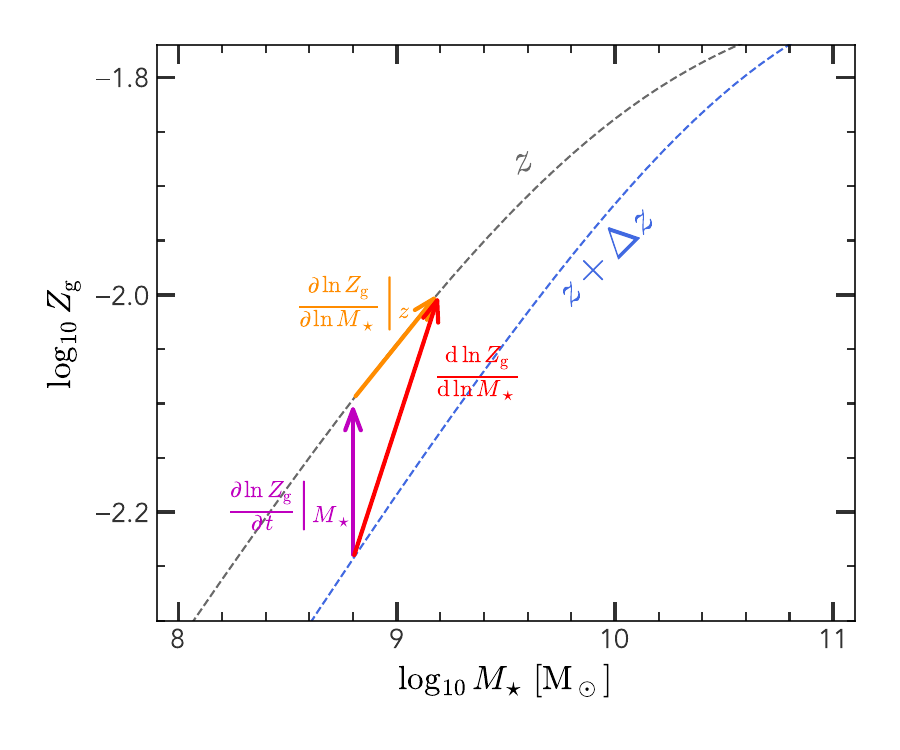}
  \end{center}
  \caption{
    Geometric illustration of the distinction between total (Lagrangian) and partial (Eulerian) derivatives of metallicity with respect to stellar mass.
    The dashed curves show the mass--metallicity relation at two epochs separated by $\Delta z$, with the lower-redshift MZR (grey) lying above the higher-redshift MZR (blue) at fixed stellar mass.
    The orange arrow shows the partial derivative $\partial\ln Z_\mg/\partial\ln M_\star|_t$, i.e.\ the slope along the MZR at fixed cosmic time.
    The magenta arrow shows the MZR normalisation evolution at fixed stellar mass, $\partial\ln Z_\mg/\partial t|_{M_\star}$.
    The red arrow shows the total derivative $\dd\ln Z_\mg/\dd\ln M_\star$, i.e.\ the slope of the actual evolutionary track of an individual galaxy as it grows in both stellar mass and metallicity between the two epochs.
    The total derivative is steeper than the epoch MZR slope because the galaxy's metallicity increases both by moving along the MZR and by the MZR itself shifting upward with time, as given by equation~\eqref{eq:material_derivative}.
  }
  \label{fig:demo_derivative}
\end{figure}

\begin{figure}
  \begin{center}
    \includegraphics[width=0.95\linewidth]{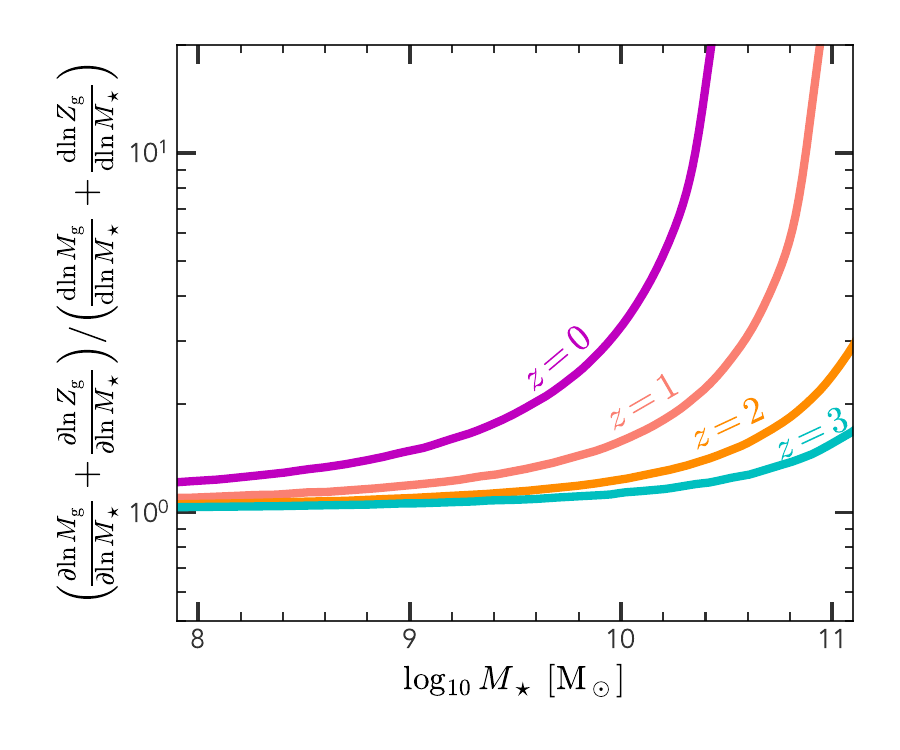}
  \end{center}
  \caption{
    Ratio of the sum of partial (Eulerian) derivatives to the sum of total (Lagrangian) derivatives entering the $\alpha$ parameter (equation~\ref{eq:alpha_peeples}), evaluated from the cosmological gas flow model at $z=0$ , 1 , 2 , and 3 .
    A ratio of unity indicates that the observed slopes of the mass--metallicity and mass--gas mass relations can be used interchangeably with the Lagrangian derivatives.
    This condition is satisfied at low stellar masses and high redshifts, where galaxies reside in the inflow-driven regime and the gas-phase metal mass is a universal function of $M_\star$ alone.
    The ratio rises steeply toward higher masses and lower redshifts as galaxies approach equilibrium: the Lagrangian derivatives tend to zero while the Eulerian slopes remain finite, causing the \citet{peeplesConstraintsStarFormation2011} framework to progressively overestimate $\alpha$.
  }
  \label{fig:derivative_alpha_ratio}
\end{figure}

\citet{peeplesConstraintsStarFormation2011} rewrite equation~\eqref{eq:zgas_lilly} in the form (see Appendix~\ref{sec:the_derivation_of_equation_eqref_eq_zgas_lilly_} for a detailed derivation)
\begin{align}
  Z_{\mg}         & = \frac{y}{(1 - R + \eta) + \alpha_{\rm PS} \,\mu}, \label{eq:zgas_peeples}                                \\[4pt]
  \alpha_{\rm PS} & \equiv (1- R)\!\left[\frac{\dd \ln M_\mg}{\dd \ln M_\star} + \frac{\dd \ln Z_\mg}{\dd \ln M_\star}\right],
  \label{eq:alpha_peeples}
\end{align}
where $\mu\equiv M_\mg /M_\star$.
Because equation~\eqref{eq:alpha_peeples} is derived from the time-evolution equations~\eqref{eq:mgas} and \eqref{eq:zgas}, the derivatives in $\alpha$ are \emph{total} (Lagrangian) derivatives: they describe how gas mass and metallicity change as an individual galaxy grows in stellar mass.
\citet{peeplesConstraintsStarFormation2011} evaluate these derivatives using the observed slopes of the stellar mass--gas mass relation and the mass--metallicity relation at $z=0$, an approach that has been widely adopted to interpret observational results \citep[see also][]{zahidUniversalRelationGalactic2014, sandersMOSDEFSurveyEvolution2021, bassiniInflowOutflowProperties2024}.
However, these observed slopes are \emph{partial} (Eulerian) derivatives: they describe how galaxy properties vary across the population at fixed cosmic time, not how an individual galaxy evolves.
The two are related by the material derivative,
\begin{equation}
  \frac{\dd \ln Z_\mg}{\dd \ln M_\star} = \frac{\partial \ln Z_\mg}{\partial \ln M_\star}\bigg|_{t} + \frac{\dd t}{\dd \ln M_\star}\frac{\partial \ln Z_\mg}{\partial t}\bigg|_{M_\star},
  \label{eq:material_derivative}
\end{equation}
and analogously for $M_\mg$.
The first term on the right-hand side is the slope of the mass--metallicity relation at fixed redshift; the second accounts for the evolution of the MZR normalisation at fixed stellar mass, projected along the galaxy's mass-growth trajectory, as demonstrated in Fig.~\ref{fig:demo_derivative}.
The two coincide only when the scaling relations do not evolve.

To quantify the magnitude of this distinction, we evaluate the ratio between the $\alpha_{\rm PS}$ factor evaluated using partial derivative and total derivative from our cosmological gas flow model; the results are shown in Fig.~\ref{fig:derivative_alpha_ratio}.
The partial and total derivatives agree for low-mass galaxies and at high redshift---precisely where our model predicts galaxies to reside in the inflow-driven regime.
This agreement can be understood analytically in the ideal gas flow model.
In the inflow-driven limit, the population spread in $M_\star$ at fixed time arises from differences in the inflow rate $\Phi$, and both $M_\mg = \Phi\, t$ and $M_\star = (1-R)\,\Phi\,\epsilon\, t^2\!/2$ scale linearly with $\Phi$, giving $\partial\ln M_\mg/\partial\ln M_\star|_t = 1$.
Meanwhile, $Z_\mg = y\epsilon t/2$ is independent of $\Phi$, so $\partial\ln Z_\mg/\partial\ln M_\star|_t = 0$. Both routes give the same sum,
\begin{equation}
  \frac{\dd \ln M_\mg}{\dd \ln M_\star} + \frac{\dd \ln Z_\mg}{\dd \ln M_\star}  =1 ~~{\rm and} ~~ \frac{\partial \ln M_\mg}{\partial \ln M_\star}\bigg|_{t} + \frac{\partial \ln Z_\mg}{\partial \ln M_\star}\bigg|_{t} = 1.
\end{equation}
Physically, the cancellation occurs because the gas-phase metal mass in the inflow-driven regime is a function of stellar mass alone, $M_\mg Z_\mg = y\,M_\star/(1-R)$, so that $\partial\!\ln(M_\mg Z_\mg)/\partial t\,|_{M_\star} = 0$; the two correction terms in equation~\eqref{eq:material_derivative} and the analogue equation for $M_\mg$ are equal and opposite.

The agreement breaks down for massive, low-redshift galaxies that have entered the equilibrium regime.
In the ideal equilibrium limit, both $M_\mg = \Phi\tau_{\rm eq}$ and $Z_\mg = y/(1-R+\eta)$ become constant along an individual galaxy's track, independent of $t$, so the total derivatives vanish identically,
\begin{equation}
  \frac{\dd \ln M_\mg}{\dd \ln M_\star} =
  \frac{\dd \ln Z_\mg}{\dd \ln M_\star} = 0.
\end{equation}
The partial derivatives, by contrast, remain nonzero across the population at fixed $t$: since $M_\star \propto \Phi$ at fixed $\eta$ and $t$ while $Z_\mg$ is independent of $\Phi$,
\begin{equation}
  \frac{\partial \ln M_\mg}{\partial \ln M_\star}\bigg|_{t} = 1,
  \qquad
  \frac{\partial \ln Z_\mg}{\partial \ln M_\star}\bigg|_{t} = 0,
\end{equation}
The ratio of the two derivatives therefore diverges in sharp contrast to the inflow-driven regime where both derivatives agree as shown above.

In summary, the identification of observed scaling-relation slopes with the Lagrangian derivatives in $\alpha_\mathrm{PS}$ is justified in the inflow-driven regime but not in equilibrium.
In the former case, however, the analytic solution gives $\alpha_\mathrm{PS} = 1 - R$, and equation~(\ref{eq:t2tau}) shows directly that $(1-R+\eta)\ll (1-R) \mu$ in this limit, so the outflow term $(1-R+\eta)$ is negligible compared to the gas fraction term $(1-R)\,\mu$ in the denominator of equation~(\ref{eq:zgas_peeples}).
The metallicity is therefore insensitive to $\eta$, and no useful constraint on the mass-loading factor can be extracted.
In the equilibrium regime, where outflow constraints are most sought, the substitution is invalid.
Outflow properties ($\eta$) inferred using this framework should therefore be interpreted with caution.

\subsection{Revisiting \citet{zahidUniversalRelationGalactic2014}}
\label{sub:revisiting_citet_zahiduniversalrelationgalactic_} 

\begin{figure}
  \begin{center}
    \includegraphics[width=0.95\linewidth]{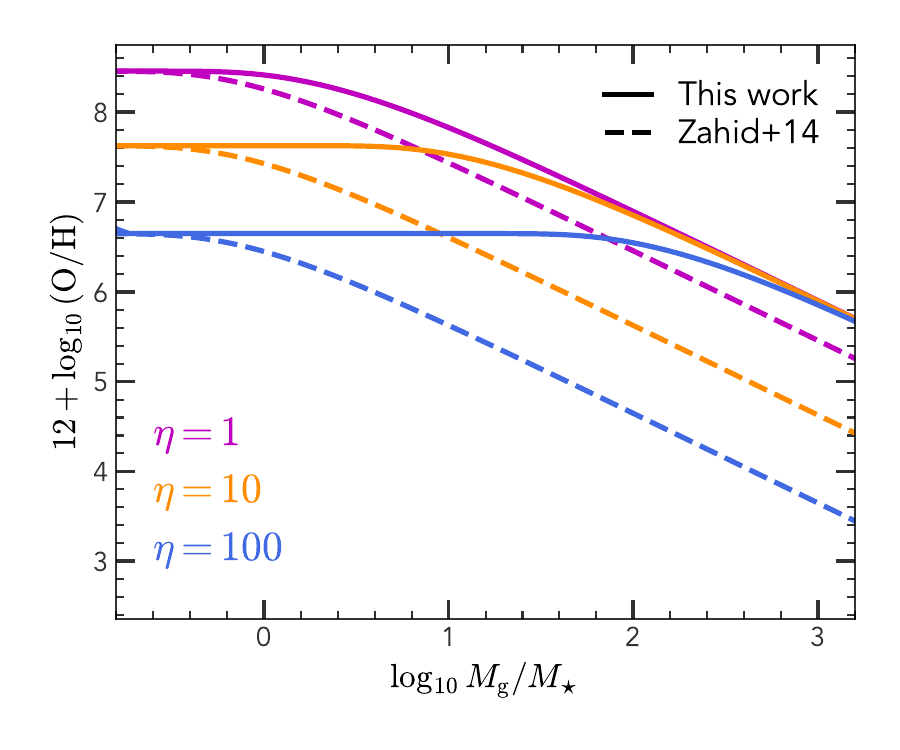}
  \end{center}
  \caption{
    Comparison between the gas metallicity predicted by the $\mathcal{K}_1$--$\mathcal{K}_2$ framework derived in this work (solid lines) and the universal metallicity relation of \citet{zahidUniversalRelationGalactic2014} (dashed lines; equation~\ref{eq:zahid_universal}), plotted as a function of the gas fraction $M_\mathrm{g}/M_\star$ for three values of the mass-loading factor: $\eta = 1$, 10, and 100.
    Both frameworks converge to the same equilibrium metallicity $y/(1-R+\eta)$ at low gas fractions (left) and share the same gFMR slope at high gas fractions (right).
    However, the \citet{zahidUniversalRelationGalactic2014} formula systematically underpredicts the metallicity in the inflow-driven regime by a factor of $(1-R)/(1-R+\eta)$, an offset that grows with $\eta$ because the constant-$\zeta$ assumption overestimates the metal loss rate when $Z_\mathrm{g} \ll Z_\mathrm{eq}$.
    The exact solution also shows that the shape of the transition from the inflow-driven regime to equilibrium depends on $\eta$, whereas the \citet{zahidUniversalRelationGalactic2014} formula adopts a fixed functional form $[1-\exp(-M_\star/M_\mathrm{g})]$ regardless of the mass-loading factor.
  }
  \label{fig:compare_zahid14}
\end{figure}

\citet{zahidUniversalRelationGalactic2014} proposed that galaxies follow a universal, redshift-independent relation between metallicity and the stellar-to-gas mass ratio,
\begin{equation}
  Z_\mathrm{g} = y/(1-R+\eta)\,[1 - \exp(-M_\star/M_\mathrm{g})],
\end{equation}
from which the MZR originates.
In Appendix~\ref{sec:zahid_derivation}, we show that this formula can be derived from our gas flow equations under four simplifying assumptions, the most consequential of which is that the net metal loss rate $\zeta_z \equiv Z_\mathrm{g}\,\eta$ is treated as a constant, which we evaluated at its equilibrium value $Z_\mathrm{eq}\,\eta = y\eta/(1-R+\eta)$.

Our $\mathcal{K}_1$--$\mathcal{K}_2$ framework recovers the same limiting behaviours---$Z_\mathrm{g} \propto M_\star/M_\mathrm{g}$ in the inflow-driven regime and $Z_\mathrm{g} \to y/(1-R+\eta)$ in equilibrium, but differs in two respects, illustrated in Fig.~\ref{fig:compare_zahid14}.
First, the constant-$\zeta$ assumption overestimates the metal loss when $Z_\mathrm{g} \ll Z_\mathrm{eq}$, causing the \citeauthor{zahidUniversalRelationGalactic2014} formula to underpredict the metallicity in the inflow-driven regime by a factor of $(1-R)/(1-R+\eta)$, an offset that grows with $\eta$.
Second, the transition shape $[1-\exp(-M_\star/M_\mathrm{g})]$ is independent of $\eta$, whereas the exact solution shows that galaxies with higher $\eta$ reach equilibrium faster at fixed gas fraction.
This $\eta$-dependence is what enables the mass-loading factor to be inferred from the combination of $Z_\mathrm{g}$ and $M_\mathrm{g}/M_\star$, a capability absent from the \citeauthor{zahidUniversalRelationGalactic2014} formula.

\subsection{Generalisation to differential mass and metal loading}
\label{sec:differential_loading}

In the standard model we assume that the outflowing gas carries the ISM metallicity, so that the same loading factor $\eta$ governs both the mass and metal outflow rates.
In general, however, the outflow may be enriched or diluted relative to the ISM \citep{peeplesConstraintsStarFormation2011, creaseyMetallicityGalacticWinds2015}.
We therefore introduce a metal-loading factor $\zeta$ defined such that the metal outflow rate is $\zeta\,Z_\mathrm{g}\,\mathrm{SFR}$, while the mass outflow rate remains $\eta\,\mathrm{SFR}$.
The case $\zeta > \eta$ corresponds to metal-enriched outflows, as found in some hydrodynamical simulations and observational analyses \citep[e.g.][]{peeplesConstraintsStarFormation2011}.
The governing equations become
\begin{align}
  \frac{\mathrm{d}M_\mg}{\mathrm{d}t}
   & = \Phi - (1-R+\eta)\,\epsilon\,M_\mg\,,
  \label{eq:dmg_general}                                           \\
  \frac{\mathrm{d}(M_\mg\,Z_\mg)}{\mathrm{d}t}
   & = y\,\epsilon\,M_\mg - (1-R+\zeta)\,Z_\mg\,\epsilon\,M_\mg\,,
  \label{eq:dmz_general}
\end{align}
which define two distinct timescales,
\begin{equation}
  \tau_\mathrm{m} \equiv \frac{1}{(1-R+\eta)\,\epsilon}\,,\qquad
  \tau_Z \equiv \frac{1}{(1-R+\zeta)\,\epsilon}\,,
\end{equation}
governing the equilibration of the gas reservoir and the metal content, respectively.
When $\zeta = \eta$, both reduce to the single timescale $\tau_\mathrm{eq}$ used throughout the main text.

Following the same procedure as in \S\,\ref{sec:analytic_approximate} (see Appendix~\ref{sec:derivation_of_the_generalisation_to_differential_mass_and_metal_loading} for the full derivation), and defining the timescale ratio $r \equiv \tau_\mathrm{m}/\tau_Z = (1-R+\zeta)/(1-R+\eta)$, the gas metallicity can be expressed as
\begin{align}
  Z_\mg & = \frac{y}{1-R}\left(\frac{M_\mg}{M_\star}\right)^{\!-1}
  \mathcal{K}^\prime_1\left(\frac{t}{\tau_{\rm m}},\,r\right)\,, \nonumber  \\[4pt]
        & \mathcal{K}^\prime_1(x,\,r) = \frac{(1-\me^{-rx})/r
    - (\me^{-x}-\me^{-rx})/(r-1)}{x - 1 + \me^{-x}}\,.
  \label{eq:K1_general}
\end{align}
The function $\mathcal{K}_2$, which relates the gas fraction to the evolutionary stage, depends only on the mass-loading and retains its original form,
\begin{equation}
  \frac{1-R+\eta}{1-R}\left(\frac{M_\mg}{M_\star}\right)^{\!-1}
  = \mathcal{K}_2(x)\,,\qquad
  \mathcal{K}_2(x) = \frac{x - 1 + \me^{-x}}{1 - \me^{-x}}\,.
  \label{eq:K2_general}
\end{equation}

The two limiting behaviours are:
\begin{itemize}
  \item \textit{Inflow-driven} ($x \to 0$): $\mathcal{K}^\prime_1 \to 1$ regardless of $r$. The metallicity reduces to $Z_\mg = y/(1-R) \times M_\star/M_\mg$, independent of both $\eta$ and $\zeta$. Metal-enriched outflows leave no imprint on galaxies that have not yet processed a significant fraction of their gas reservoir.
  \item \textit{Equilibrium} ($x \to \infty$): $\mathcal{K}^\prime_1 \to 1/(rx)$, giving $Z_\mg \to y/(1-R+\zeta)$. The equilibrium metallicity depends on the metal-loading factor $\zeta$ rather than the mass-loading factor $\eta$.
\end{itemize}
When $r = 1$ ($\zeta = \eta$), both limits reduce to those of the standard model, and equation~(\ref{eq:K1_general}) reduces to equation~(\ref{eq:master}).

In the general case, the system of equations~(\ref{eq:K1_general}) and~(\ref{eq:K2_general}) contains four quantities: $Z_\mg$, $M_\mg/M_\star$, $\eta$, and $\zeta$.
Given any three, the fourth can be determined.
In particular, if both the gas metallicity and gas fraction are measured and the mass-loading factor is independently constrained, the metal-loading factor $\zeta$ can be inferred, providing a route to measuring the metal enrichment of galactic outflows.

\subsection{The deviation from FMR at high-$z$}
\label{sub:the_deviation_from_fmr_at_high_z_} 

\begin{figure}
  \begin{center}
    \includegraphics[width=0.95\linewidth]{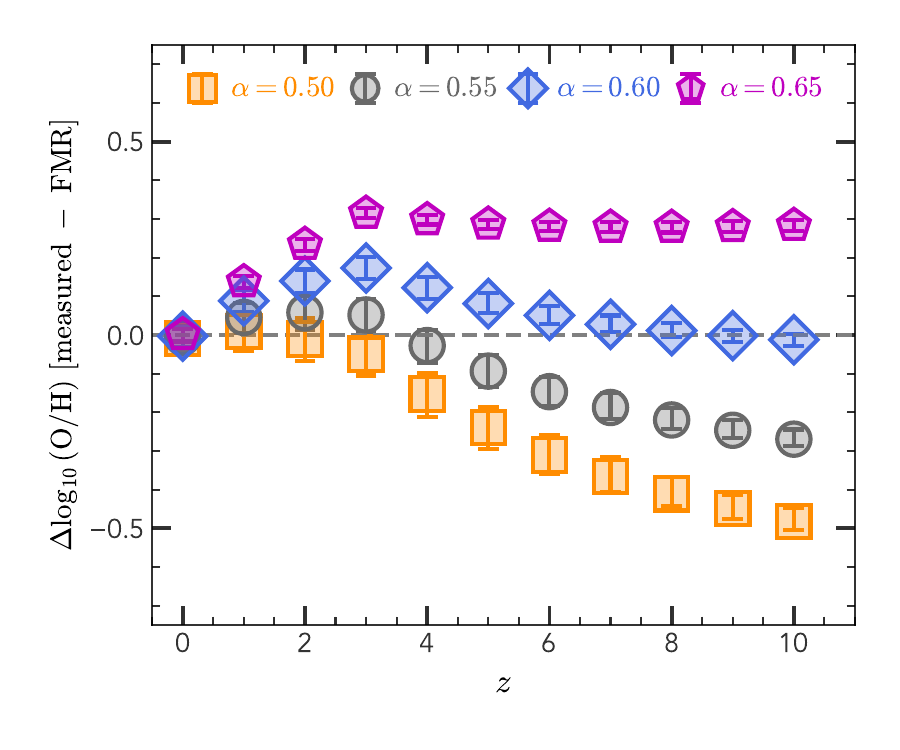}
  \end{center}
  \caption{
    Deviation of the model gas metallicity from the locally calibrated FMR around $M_\star\approx 10^9\,\rm M_\odot$, as a function of redshift, for four choices of the FMR parameter $\alpha = 0.50$, 0.55, 0.60, and 0.65.
    At each redshift, the offset $\Delta\log_{10}(\mathrm{O/H})$ is defined as the difference between the metallicity predicted by the full cosmological gas flow model and that predicted by the FMR relation calibrated at $z = 0$.
    All four parameterisations produce tight, approximately redshift-invariant FMR sequences out to $z \sim 3$ (see Appendix~\ref{sec:fmr_with_varying_alpha_}), yet diverge by up to $\sim 0.5$~dex in both directions at redshift up to 10, demonstrating the sensitivity of the extrapolated FMR prediction to the precise local calibration.
  }
  \label{fig:fmr_parameter_offset}
\end{figure}

\citet{curtiJADESInsightsLowmass2024} found that galaxies at $z > 4$ deviate from the FMR calibrated in the local Universe, with metallicities lower than predicted by $\gtrsim 0.4$~dex.
As discussed in \S\,\ref{sec:fundamental_metallicity_relation}, the redshift invariance of the FMR is not a fundamental symmetry but a contingent consequence of how the star formation efficiency $\epsilon$ depends on stellar mass and redshift.
In particular, the standard FMR is redshift-invariant only if the redshift dependence of $\epsilon$ is sufficiently regular that it can be absorbed by the FMR parameter $\alpha$.
The observation that galaxies at $z \gtrsim 4$ fall below the locally calibrated FMR therefore indicates that the star formation efficiency at these redshifts deviates from the power-law extrapolation that holds at $z \lesssim 3$, not necessarily that a new physical mechanism has emerged.

However, a more fundamental concern is that the extrapolated FMR prediction at high redshift is extremely sensitive to the local calibration itself \citep{nishigakiDREAMSIIGalaxyDemographics2025}.
To demonstrate this, Fig.~\ref{fig:fmr_parameter_offset} shows the metallicity offset between the full cosmological model and the locally calibrated FMR, for four choices of $\alpha = 0.50$, 0.55, 0.60, and 0.65, at $M_\star \approx 10^9\,\rm M_\odot$, the typical mass scale reported in \citet{curtiJADESInsightsLowmass2024}.
The corresponding FMR projections are shown in Appendix~\ref{sec:fmr_with_varying_alpha_}, where all four cases produce tight sequences with comparable scatter and approximate redshift invariance out to $z \sim 3$.
In other words, the data at $z \lesssim 3$ do not strongly discriminate among these parameterisations.

Despite this degeneracy at low redshift, the extrapolated predictions diverge dramatically at $z \gtrsim 4$: the model metallicity lies $\sim 0.5$~dex below the prediction for $\alpha = 0.50$ but $\sim 0.3$~dex above it for $\alpha = 0.65$, with $\alpha = 0.60$ remaining close to zero offset out to $z \sim 10$.
The sign and magnitude of the apparent high-redshift deviation therefore depend entirely on which locally calibrated FMR is adopted as the baseline.
This sensitivity arises because $\alpha$ controls the relative weighting of stellar mass and SFR in the FMR projection; small changes in $\alpha$ produce modest horizontal shifts at $z \lesssim 3$, where galaxies span a limited range in SFR at fixed stellar mass, but these shifts are amplified at high redshift, where the range in SFR at fixed stellar mass is orders of magnitude higher.

In summary, according to our framework, there is no reason to expect that parameterising the FMR with a single $\alpha$ remains valid at arbitrarily high redshift: whether such an $\alpha$ exists that absorbs the SFR and redshift dependence of the mass--metallicity relation is contingent on the functional form of $\epsilon(M_\star, z)$.
More importantly, before drawing conclusions about new physics from the observed deviation of high-redshift galaxies, one must first ensure that the locally calibrated FMR is determined with sufficient precision that its extrapolation to $z \gtrsim 4$ is reliable.
Current data do not yet achieve this level of precision, so interpreting the high-redshift offset as evidence for new physics may be premature.

\subsection{Stellar-to-gas metallicity difference}
\label{sub:stellar_gas_metallicity_difference}

The framework developed in this paper predicts a clear connection between the evolutionary stage of a galaxy and the difference between its gas-phase and stellar metallicities \citep[see also][]{wangOriginGalaxySizestellar2026}.
In the inflow-driven limit, the gas-phase metallicity traces the instantaneous ISM enrichment and exceeds the stellar metallicity, which records the time-integrated enrichment history, by a factor of $3/2$ (equation~\ref{eq:zgas_inflow} and \ref{eq:zstar_inflow}), corresponding to a maximum difference of $\Delta Z_{\mg,\star} \equiv \log_{10} Z_\mg - \log_{10} Z_\star = \log_{10}(3/2) \approx 0.18$~dex.
As galaxies transition toward equilibrium, a regime occupied by massive, low-sSFR systems, the gas metallicity stabilises at the equilibrium value set by the yield and mass-loading factor, the stellar metallicity converges toward it, and $\Delta Z_{\mg,\star} \rightarrow 0$.
The framework therefore predicts that $\Delta Z_{\mg,\star}$ should decrease monotonically from $\approx 0.18$~dex in low-mass, high-sSFR galaxies to $\approx 0$ in massive, low-sSFR galaxies, with the transition occurring around the characteristic equilibrium timescale $\tau_{\rm eq}$.

This prediction is in good qualitative agreement with the observational results of \citet{fraser-mckelvieSAMIGalaxySurvey2022}, who measured $\Delta Z_{\mg,\star}$ for a representative sample of star-forming galaxies in the SAMI Galaxy Survey and found that low-mass, high-sSFR galaxies show the largest $\Delta Z_{\mg,\star}$, while massive, low-sSFR galaxies approach $\Delta Z_{\mg,\star} \sim 0$ \citep[see also][]{lianMassmetallicityRelationsGas2018, boardmanCompetingEffectsRecent2025}.
Within our framework, this trend arises naturally: in the inflow-driven regime, the gas metallicity evolves rapidly in proportion to the stellar-to-gas mass ratio and exceeds the stellar metallicity, which reflects the time-averaged enrichment history.
As galaxies grow in mass and transition toward equilibrium, the gas metallicity stabilises and the gap between $Z_{\mg}$ and $Z_\star$ narrows.
We note that the predicted maximum difference of $\approx 0.18$~dex is modest compared to the observed $\Delta Z_{\mg,\star}\lesssim 0.3\,\rm dex$, which may reflect systematic offsets in the metallicity calibrations used for gas and stars; we therefore focus on the qualitative trend rather than the absolute normalisation.

\subsection{Degeneracy between star formation efficiency and mass-loading factor}
\label{sub:degeneracy}

\begin{figure}
  \begin{center}
    \includegraphics[width=0.95\linewidth]{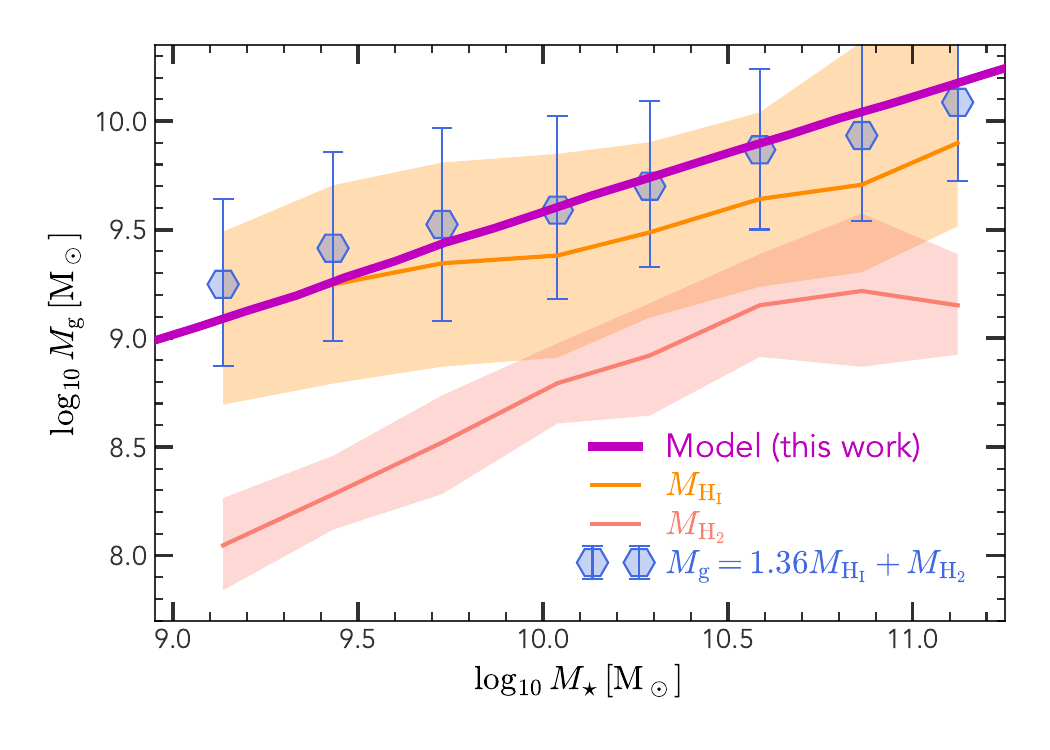}
  \end{center}
  \caption{
    Total cold gas mass, $M_{\rm g}$, as a function of stellar mass, $M_\star$, compared to the analytic model developed in this work.
    Hexagons show $M_{\rm g} \equiv 1.36\times M_{\rm HI} + M_{\rm H_2}$, the atomic plus molecular gas mass with a helium correction applied to $M_{\rm HI}$, compiled from the xGASS and xCOLD GASS surveys by \citet{saintongeColdInterstellarMedium2022a}; error bars show the $1\sigma$ scatter.
    The $M_{\rm HI}$-only and $M_{\rm H_2}$-only relations are shown separately, each with their $1\sigma$ scatter, to illustrate the relative contribution of the two gas phases across the stellar mass range.
    The model reproduces the observed $M_{\rm g}$--$M_\star$ relation without being fit directly to these data.
  }
  \label{fig:gas_mass_comparison}
\end{figure}

The two mass-continuity equations~\eqref{eq:mgas} and \eqref{eq:zgas} govern the evolution of the gas reservoir and its metal content.
Given the inflow rate $\Phi(t)$, the star formation efficiency $\epsilon(t)$, and the mass-loading factor $\eta(t)$, the system is fully determined and the gas mass history $M_\mg(t)$, gas metallicity history $Z_\mg(t)$, and stellar mass history $M_\star(t)$ can all be solved for uniquely.

Rather than modelling the inflow rate through halo accretion and cooling efficiency as we do in \S\,\ref{sub:the_cosmological_gas_flow_model}, some approaches bypass this step by assuming a parametric form for the star formation history $\mathrm{SFR}(t)$ directly \citep[e.g.][]{weinbergEquilibriumSuddenEvents2017, linConstraintsGalacticOutflows2023}.
In this case the inflow rate is determined implicitly by equation~\eqref{eq:mgas},
\begin{equation}
  \Phi(t) = \frac{\dd}{\dd t} \frac{\mathrm{SFR}(t)}{\epsilon(t)} + [1-R+\eta(t)]\,\mathrm{SFR}(t).
  \label{eq:phi_implicit}
\end{equation}
One caveat is that the inflow rate derived from equation~\eqref{eq:phi_implicit} is not guaranteed to be positive; a negative $\Phi$ would imply that pristine gas is being drained from the ISM, which has no physical counterpart in the baryon cycle.
As we argue below, this pathology arises because $\epsilon(t)$ cannot be determined from $\mathrm{SFR}(t)$ and $Z_\mg(t)$ alone, owing to the degeneracy between $\epsilon(t)$ and $\eta(t)$.

Even if the complete star formation history $\mathrm{SFR}(t)$ and gas metallicity history $Z_\mg(t)$ are both known, the system remains degenerate.
Given $\mathrm{SFR}(t)$ and $Z_\mg(t)$, the metal mass continuity equation~\eqref{eq:zgas} can be rewritten as
\begin{equation}
  \frac{\dd}{\dd t}\frac{{\rm SFR}(t) Z_\mg(t)}{\epsilon(t)} =
  y\,\mathrm{SFR}(t) - \left[1-R+\eta(t)\right]Z_\mg(t)\mathrm{SFR}(t),
  \label{eq:metal_sfr}
\end{equation}
For any assumed $\eta(t)$, equation~\eqref{eq:metal_sfr} is a first-order ODE in $M_\mg(t) = \mathrm{SFR}(t)/\epsilon(t)$ that, given an initial condition, uniquely determines $M_\mg(t)$ and hence $\epsilon(t)$.
The same star formation and metallicity histories can thus be reproduced for any choice of $\eta(t)$, with $\epsilon(t)$ adjusting accordingly.
Two physically distinct limiting cases illustrate the degeneracy: in the inflow-driven limit the metallicity is governed by the star formation efficiency, $Z_\mg(t) =\epsilon(t) y/(1-R)\times {M_\star(t)}/{\mathrm{SFR}(t)}$, while in equilibrium it is set by the mass-loading factor, $Z_\mg(t) = {y}/[1-R+\eta(t)]$.
Since both reproduce the same $\mathrm{SFR}(t)$ and $Z_\mg(t)$ by construction, neither the star formation history nor the metallicity history alone can distinguish between them.

Due to the presence of this degeneracy, approaches that start from an assumed star formation history may implicitly introduce unphysical gas stripping.
Combining equations~\eqref{eq:phi_implicit} and \eqref{eq:metal_sfr} gives
\begin{equation}
  \Phi(t) = \frac{y\,\mathrm{SFR}(t)}{Z_\mg(t)} - \frac{\mathrm{SFR}(t)}{\epsilon(t)\,Z_\mg(t)} \frac{\dd Z_\mg(t)}{\dd t}.
  \label{eq:phi_implicit2}
\end{equation}
When the gas metallicity is rising ($\dd Z_\mg/\dd t > 0$), the second term is negative and reduces the inflow rate; if $\epsilon$ is underestimated, this term is amplified and can drive $\Phi < 0$.
To see this concretely, consider a galaxy with a growing SFR and rising $Z_\mg$.
If the star formation efficiency is underestimated, the gas mass $M_\mg = \mathrm{SFR}/\epsilon$ is overestimated, so the ISM contains more gas than it should.
To match the observed $Z_\mg$ with this inflated $M_\mg$, the total metal mass $M_\mg Z_\mg$ must be larger than the star formation history can supply through nucleosynthesis alone on the required timescale.
The model compensates by suppressing the inflow of pristine gas, which would dilute $Z_\mg$, and in the extreme case requires $\Phi < 0$, draining metal-poor gas from the reservoir to artificially elevate the metallicity.

This degeneracy also raises a challenge for semi-analytic spectral fitting techniques, which attempt to model gas inflow and outflow by fitting to the spectral energy distribution of galaxies \citep[see also][]{lianMassmetallicityRelationsGas2018, zhouSemianalyticSpectralFitting2022}.
The spectral energy distribution of a galaxy is fully specified by its star formation history and metallicity history, so constraining the gas flow properties further requires first breaking the degeneracy between the star formation efficiency and the mass-loading factor.
In one such approach, \citet{zhouSemianalyticSpectralFitting2022} anchor the star formation efficiency to its $z=0$ value and infer the outflow properties from this assumption.
The resulting constraints on gas flow should therefore be interpreted with caution: it is unclear whether they originate from the spectral energy distribution itself or from the assumption used to break the degeneracy.

The degeneracy between $\epsilon$ and $\eta$ can be broken by the gas mass history $M_\mg(t)$.
Since $\epsilon(t) \equiv \mathrm{SFR}(t)/M_\mg(t)$, the ratio of the star formation rate to the gas mass directly measures the star formation efficiency at every epoch, independently of $\eta(t)$.
Once $\epsilon(t)$ is known, $\eta(t)$ follows immediately from equation~\eqref{eq:metal_sfr}.

Our cosmological gas flow model is calibrated against gas metallicity and star formation properties only, without direct constraints on the gas mass history.
We break the degeneracy by assuming that the mass-loading factor does not evolve with redshift, following the observational evidence from $z \sim 0$ to $z \sim 1.5$ \citep{heckmanSystematicPropertiesWarm2015, chisholmMassMomentumOutflow2017, schroetterMusEGAsFLOw2019, schroetterMusEGAsFLOw2024}.
This assumption is supported empirically by \citet{wanginflow3}, who use gas mass and metallicity observations at $z \sim 0$--$1$ to constrain the mass-loading factor directly and find that it evolves little over the past $\sim 8\,$Gyr.
Under this assumption, the observed redshift evolution of the mass--metallicity relation can no longer be attributed to a redshift-dependent $\eta$, and instead directly informs us that galaxies reside close to the inflow-driven limit, where the star formation efficiency governs the metallicity evolution (\S\,\ref{sub:implication_to_sam}).
As a check on the resulting model, we compare the predicted gas mass with observations: the model reproduces the observed $M_\mg$--$M_\star$ relation from xGASS and xCOLD GASS \citep{saintongeColdInterstellarMedium2022a} without being fit to these data (Fig.~\ref{fig:gas_mass_comparison}), lending confidence that the inferred star formation efficiency and mass-loading factor are physically meaningful rather than artefacts of the calibration.

\subsection{Decoupling the mass--metallicity relation from the stellar mass--halo mass relation}
\label{sub:mzr_shmr}

\begin{figure*}
  \begin{center}
    \includegraphics[width=0.95\linewidth]{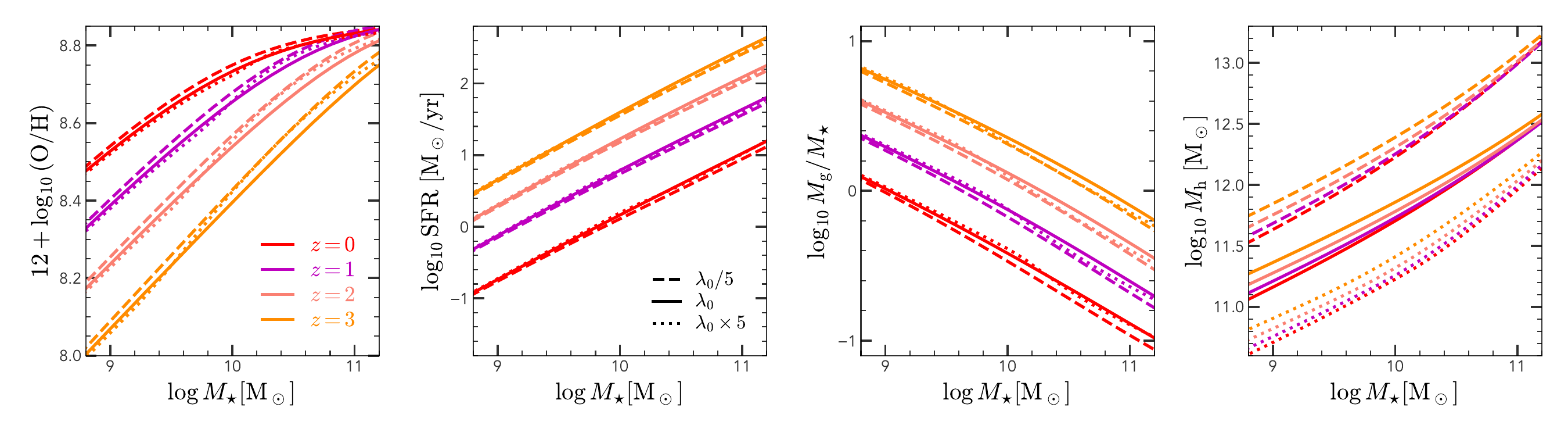}
  \end{center}
  \caption{
    Effect of varying the gas inflow rate on galaxy scaling relations.
    Solid lines show the fiducial model; dashed and dotted lines show models in which the cooling efficiency is reduced by a factor of 5 and increased by a factor of 5, respectively, at all halo masses and redshifts.
    Colours denote redshift as indicated in the leftmost panel.
    From left to right: the gas-phase mass--metallicity relation, the star-forming main sequence, the gas-to-stellar mass ratio, and the stellar mass--halo mass relation.
    The first three panels are nearly unchanged despite the order-of-magnitude variation in inflow rate, confirming that the MZR, the SFMS, and the gas fraction are governed by the star formation efficiency and the mass-loading factor rather than by the inflow rate.
    Only the stellar mass--halo mass relation (rightmost panel) responds to the change in $\Phi$, as expected: the inflow rate sets the total baryonic mass supply to the galaxy and therefore controls the normalisation of $M_\star$ at fixed $M_\mathrm{h}$, but does not enter the relationships among $M_\star$, SFR, $M_\mathrm{g}$, and $Z_\mathrm{g}$ that define the other three scaling relations.
    This demonstrates why the cooling efficiency must be calibrated independently using the stellar mass--halo mass relation (\S\,\ref{sub:mzr_shmr}).
  }
  \label{fig:inflow_rate_sensitivity}
\end{figure*}

A striking property of the gas flow model is that the gas inflow rate does not enter any scaling relation among stellar mass, star formation rate, gas mass, and metallicity.
Fig.~\ref{fig:inflow_rate_sensitivity} demonstrates this directly: varying the cooling efficiency by more than an order of magnitude at all halo masses and redshifts leaves the mass--metallicity relation, the star-forming main sequence, and the gas fraction virtually unchanged, while the stellar mass--halo mass relation shifts by the corresponding factor.

This insensitivity follows from the self-similarity of halo accretion.
In the \citet{wechslerConcentrationsDarkHalos2002} model, $\dot{M}_\mathrm{h} \propto M_\mathrm{h}$ at fixed redshift, so $\Phi \propto \lambda\,M_\mathrm{h}$ and hence $M_\star \propto \lambda\,M_\mathrm{h}$ at leading order.
Increasing $\lambda$ by a factor $f$ therefore shifts every galaxy to $f$ times higher $M_\star$, and the galaxy that now enters a given $M_\star$ bin comes from a halo of mass $M_\mathrm{h}/f$.
Since the gas supply history is unchanged at fixed $M_\star$, so are $Z_\mathrm{g}$, $M_\mathrm{g}/M_\star$, and SFR/$M_\star$.
This cancellation is exact when $\lambda$ is constant; the small residual in Fig.~\ref{fig:inflow_rate_sensitivity} arises because $\lambda(M_\mathrm{h})$ differs slightly between the original and replacement haloes.

The stellar mass--halo mass relation, by contrast, compares the baryonic outcome $M_\star$ against the dark matter halo mass $M_\mathrm{h}$, which is set by gravitational collapse and does not respond to $\lambda$.
Increasing $\Phi$ by a factor $f$ rescales $M_\star$ at fixed $M_\mathrm{h}$, directly altering $M_\star/M_\mathrm{h}$.
This is why the cooling efficiency $\lambda(M_\mathrm{h}, z)$ must be calibrated using the stellar mass--halo mass relation as an independent constraint, rather than being degenerate with $\epsilon$ and $\eta$.

This separation has an important consequence.
If the cooling efficiency $\lambda$ were a universal constant, independent of halo mass and redshift, then $M_\star/M_\mathrm{h}$ and $Z_\mathrm{g}$ would be directly proportional in both the inflow-driven limit and equilibrium, and the mass--metallicity relation and the stellar mass--halo mass relation would share the same stellar mass dependence and redshift evolution.
Observations, however, show that this proportionality does not hold.
The stellar mass--halo mass relation exhibits only weak redshift evolution out to $z \sim 3$ \citep{mosterConstraintsRelationshipStellar2010, yangEvolutionGalaxyDarkMatter2012, behrooziUNIVERSEMACHINECorrelationGalaxy2019}, whereas the mass--metallicity relation evolves strongly over the same interval \citep{maiolinoAMAZEEvolutionMassmetallicity2008, sandersMOSDEFSurveyEvolution2021, jainUniformAnalysisGasphase2025}.
Furthermore, the slope of the $M_\star/M_\mathrm{h}$--$M_\star$ relation is approximately $+0.6$ at $M_\star \sim 10^9\,\mathrm{M}_\odot$ \citep{zuMappingStellarContent2015, behrooziUNIVERSEMACHINECorrelationGalaxy2019}, which is steeper than the mass--metallicity relation ($\approx +0.3$; \citealt{curtiMassmetallicityFundamentalMetallicity2020}; \citealt{jainUniformAnalysisGasphase2025}) over the same stellar mass range.
Both discrepancies indicate that $\lambda$ cannot be a universal constant: to decouple the stellar mass and redshift dependence of metallicity from that of the stellar-to-halo mass ratio, the cooling efficiency must depend on both halo mass and redshift.

There are clear physical motivations for both dependences.
The mass dependence is naturally provided by preventative feedback.
For galaxies with $M_\star \lesssim 10^{10}\,\mathrm{M}_\odot$, neither inefficient radiative cooling \citep{whiteCoreCondensationHeavy1978} nor AGN feedback \citep{bowerDarkNemesisGalaxy2017} can effectively suppress the gas supply.
Instead, stellar feedback-driven outflows deposit energy and momentum into the halo gas, preventing it from cooling and accreting onto the galaxy \citep{luImportancePreventiveFeedback2017, wrightImpactStellarAGN2020}.
\citet{mitchellGalacticOutflowRates2020} showed in the EAGLE simulation that the mass outflow rate measured at the halo virial radius is significantly higher than that at the galaxy scale, because the overpressurised outflow entrains and heats ambient halo gas.
Importantly, the gas ejected from the ISM carries metals, whereas the entrained halo gas remains largely pristine.
This distinction means that preventative feedback suppresses the gas inflow rate without proportionally affecting the metal budget, providing precisely the mass-dependent $\lambda$ needed to steepen the $M_\star/M_\mathrm{h}$--$M_\star$ relation relative to the MZR.
The consequence of omitting this mechanism is illustrated by \citet{boseGalacticTugofwarHow2026}, who show that the \textsc{GALFORM} model \citep{coleHierarchicalGalaxyFormation2000, laceyUnifiedMultiwavelengthModel2016}, which implements only ejective supernova feedback, cannot fit the Milky Way satellite luminosity function and stellar MZR simultaneously: the feedback strength required to match the luminosity function produces a satellite stellar MZR that is too low, and vice versa, because the same mass-loading factor controls both the total gas supply and the metal budget \citep[see also][]{houConstrainingSNFeedback2016}.
Incorporating preventative feedback, as in \citet{luImportancePreventiveFeedback2017}, offers a route to resolving this tension by decoupling the two.

The redshift dependence is naturally provided by cold-mode accretion \citep{keresHowGalaxiesGet2005, dekelColdStreamsEarly2009}, in which gas penetrates directly to the galaxy centre along filaments without being shock-heated by the halo.
Cold-mode accretion is more prevalent at high redshift, when the universe is denser, effectively increasing $\lambda$ at early times.
This reconciles the weak redshift evolution of the stellar mass--halo mass relation with the strong evolution of the MZR: at high redshift, the elevated $\lambda$ means that a larger fraction of accreted baryons reach the ISM, so that the stellar-to-halo mass ratio changes only mildly even as the gas fraction (and hence metallicity) evolves substantially.

In summary, the differing mass and redshift dependences of the mass--metallicity relation and the stellar mass--halo mass relation encode information about two distinct physical processes: the mass dependence of preventative feedback and the redshift evolution of the gas cooling efficiency.
The parameterisation of $\lambda(M_\mathrm{h}, z)$ in our model (equation~\ref{eq:cooling}) captures both effects, and the need to calibrate its parameters independently of the star formation efficiency and mass-loading factor is one of the primary motivations for fitting the stellar mass--halo mass relation as a third constraint alongside the mass--metallicity relation and star-forming main sequence.

\subsection{Implications for semi-analytic models of galaxy formation}
\label{sub:implication_to_sam} 

A central result of this work is that the observed redshift evolution of the mass--metallicity relation arises naturally when galaxies reside in the inflow-driven regime, where the gas metallicity is governed by the star formation efficiency rather than by the outflow mass-loading factor (\S\,\ref{sec:inflow_driven_regime}).
Here we argue that this conclusion carries concrete implications for the feedback prescriptions adopted in semi-analytic models of galaxy formation.

Our argument rests on two empirical preconditions.
First, the relationship between the mass-loading factor $\eta$ and stellar mass does not itself evolve with redshift to a degree comparable to the evolution of the mass--metallicity relation.
Observations of galactic winds from $z \sim 0$ to $z \sim 1.5$ are broadly consistent with this assumption \citep{heckmanSystematicPropertiesWarm2015, chisholmMassMomentumOutflow2017}, although the uncertainties remain large.
Second, recycled gas---material previously ejected from the galaxy and subsequently re-accreted---does not dominate the total gas accretion rate \citep{mitchellGalacticInflowWind2020}, so that the accreting gas remains predominantly pristine.

If both conditions hold, the amplitude of the mass--metallicity relation can increase towards lower redshift only if galaxies spend a significant fraction of cosmic time in the inflow-driven regime, where $Z_\mg = y\epsilon t/2$ and the redshift evolution of the star formation efficiency is shallower than $t^{-1}$; in our cosmological gas flow model, $\epsilon \propto t^{-0.73}$ (see \S\,\ref{sub:analytic_understanding_fmr}).
Since galaxies remain inflow-driven for longer when $\tau_{\rm eq} \equiv 1/[(1-R+\eta)\epsilon]$ is large, this requires galaxies to have simultaneously low mass-loading factor and low star formation efficiency.

Alternatively, if galaxies have a high mass-loading factor, as in many semi-analytic models \citep{mitchellGalacticOutflowRates2020}, $\tau_{\rm eq}$ is short, galaxies reach equilibrium rapidly, and their metallicity is determined by $Z_{\rm eq} \approx y/(1-R+\eta)$.
Because semi-analytic models typically parameterise $\eta$ as a function of stellar mass or halo circular velocity alone \citep[e.g.][]{kauffmannFormationEvolutionGalaxies1993, coleRecipeGalaxyFormation1994, coleHierarchicalGalaxyFormation2000, finlatorOriginGalaxyMassmetallicity2008, somervilleSemianalyticModelCoevolution2008}, the predicted metallicity at fixed stellar mass inherits no explicit redshift dependence.
Consequently, these models tend to produce a mass--metallicity relation that evolves weakly or not at all with redshift \citep{luSemianalyticModelsCANDELS2014, somervilleStarFormationSemianalytic2015, guoGalaxiesEAGLEHydrodynamical2016}, in conflict with observational results \citep[e.g.][]{maiolinoAMAZEEvolutionMassmetallicity2008, jainUniformAnalysisGasphase2025}.

This tension admits at least two resolutions within the semi-analytic framework.
The more immediate fix is to allow the mass-loading factor to evolve with redshift at fixed halo mass, for instance by adopting $\eta(M_{\rm h}, z)$ rather than $\eta(M_{\rm h})$.
\citet{mitraEquilibriumModelConstraints2015} adopted precisely this approach, fitting an equilibrium model with $\eta = (M_\mathrm{h}/10^{\eta_1 + \eta_2\sqrt{z}})^{\eta_3}$ and efficient wind recycling ($t_\mathrm{rec} \sim 0.5\,$Gyr) to the MZR, stellar mass--halo mass relation, and star-forming main sequence from $z = 0$ to $2$. Their Bayesian evidence analysis confirms that the redshift dependence of $\eta$ is essential within the equilibrium framework.
Similarly, \citet{xieH2basedStarFormation2017} adopted a redshift-dependent outflow prescription in order to reproduce the observed evolution of the mass--metallicity relation.
However, there is at present no compelling observational evidence nor clear physical argument to suggest that the mass-loading factor evolves dramatically with redshift at fixed halo mass \citep{heckmanSystematicPropertiesWarm2015, chisholmMassMomentumOutflow2017, mitchellGalacticOutflowRates2020}.

This approach also faces a deeper difficulty when confronted with the FMR.
If galaxies reside in equilibrium and the MZR evolution is driven by a redshift-dependent $\eta$, then the only remaining mechanism to produce the anti-correlation between SFR and metallicity at fixed stellar mass is stochastic fluctuation in the gas accretion rate \citep{forbesOriginFundamentalMetallicity2014}.
But stochastic fluctuations and a redshift-dependent $\eta$ are physically unrelated processes, and there is no reason why their combination should produce a surface in the $(M_\star, \mathrm{SFR}, Z_{\rm g})$ space that is both tight and redshift-invariant out to $z \sim 3$.
The redshift invariance of the FMR therefore poses a fundamental challenge to any equilibrium framework in which the MZR evolution is attributed to $\eta(z)$.

A more fundamental resolution would be to lower the mass-loading factor to values closer to those measured in hydrodynamical simulations \citep{muratovGustyGaseousFlows2015, mitchellGalacticOutflowRates2020} and inferred from observations \citep{heckmanSystematicPropertiesWarm2015, leethochawalitEvolutionStellarMassMetallicity2019}.
In this regime, galaxies would naturally remain in the inflow-driven limit for a significant fraction of their evolution, and the redshift dependence of the MZR would emerge from the evolution of the gas fraction without requiring an explicit redshift-dependent $\eta$.
Within this framework, the FMR and its approximate redshift invariance follow simultaneously as consequences of the same inflow-driven physics, rather than requiring two separate and unrelated mechanisms to conspire.

Crucially, this second resolution does not require the mass-loading factor to evolve with redshift.
The critical difference is that the lower $\eta$ extends the equilibrium timescale $\tau_{\rm eq} = 1/[(1-R+\eta)\,\epsilon]$, keeping galaxies in the inflow-driven regime over a larger fraction of cosmic time.
In the inflow-driven regime, the gas metallicity evolves as $Z_\mg = y\epsilon t/2$, depending on cosmic time only through the star formation efficiency.
Since the redshift evolution of the star formation efficiency is shallower than $t^{-1}$---in our case $\epsilon \propto t^{-0.73}$ (see \S\,\ref{sub:analytic_understanding_fmr})---the product $\epsilon t$ still increases with cosmic time, so the amplitude of the mass--metallicity relation increases towards lower redshift, as observations suggest.

We now turn to the question of how these two scenarios---inflow-driven evolution with a non-evolving $\eta$, versus equilibrium-dominated evolution with an evolving $\eta$---might be distinguished observationally.
We have argued that both scenarios can reproduce the same mass--metallicity relation and stellar-to-halo mass relation, so neither of these scaling relations can break the degeneracy.
The gas mass $M_\mg$, combined with SFR, however, provides a direct discriminant.
As shown in \S\,\ref{sub:degeneracy}, $\epsilon \equiv \mathrm{SFR}/M_\mg$ is directly measurable from the ratio of the star formation rate to the gas mass, independently of $\eta$.
Once $\epsilon$ is known, $\eta$ follows from the metal continuity equation~\eqref{eq:zgas}, breaking the degeneracy completely.
Direct measurements of the gas fraction and star formation rate therefore constrain $\epsilon$ and subsequently $\eta$, providing the cleanest observational test of the two pictures.

\subsection{Other processes regulating metallicity evolution}
\label{sub:other_processes_in_regulating_metallicity_evolution}

In this work we have constructed a minimal model of galaxy chemical evolution, incorporating halo accretion, gas cooling, star formation, and stellar feedback.
Despite its simplicity, this model successfully recovers several key features of the observed gas-phase metallicity scaling relations, including the evolving mass--metallicity relation, the fundamental metallicity relation, and the gaseous fundamental metallicity relation.
Nevertheless, to isolate the core physics responsible for these relations, we have omitted a number of processes that also regulate galaxy metallicity evolution.
We discuss these in turn below, both to clarify the scope of our model and to motivate future work.

\paragraph*{Stochastic gas accretion.}
We have treated gas accretion as a smooth process described by analytic functions of halo mass and redshift.
In reality, the gas inflow history is stochastic: galaxies experience fluctuations due to variations in the large-scale accretion rate, interactions with neighbouring structures, and the clumpy nature of cold-mode accretion along cosmic web filaments \citep{keresHowGalaxiesGet2005, dekelColdStreamsEarly2009, vandevoortRatesModesGas2011}.
Several studies using cosmological simulations have shown that such fluctuations can produce an anti-correlation between SFR and gas metallicity along the history of individual galaxies: a sudden increase in the gas inflow rate simultaneously enhances star formation and dilutes the ISM \citep{yatesRelationMetallicityStellar2012, forbesOriginFundamentalMetallicity2014, torreySimilarStarFormation2018, deluciaGasAccretionRegulates2020, wangGasphaseMetallicityDiagnostic2021, maRevisitingFundamentalMetallicity2024}.
It has therefore been argued that this stochastic variability is the physical origin of the FMR \citep{forbesOriginFundamentalMetallicity2014, torreySimilarStarFormation2018, wangGasphaseMetallicityDiagnostic2021}.
However, this picture falls short in two respects.
First, stochastic fluctuations drive individual galaxies above and below their own equilibrium SFR and $Z_\mathrm{g}$, which are themselves set by the galaxy's average inflow rate and mass-loading factor; this produces an anti-correlation between SFR and $Z_\mathrm{g}$ along the history of each galaxy, but not necessarily across the population.
A population-wide anti-correlation additionally requires that galaxies at fixed stellar mass share closely aligned equilibrium values of SFR and $Z_\mathrm{g}$, a condition that has not been demonstrated.
Second, such explanations do not capture the full content of the FMR.
The FMR is not merely an anti-correlation between SFR and metallicity at fixed stellar mass; it requires that galaxies populate a well-defined surface in the $(M_\star, \mathrm{SFR}, Z_\mg)$ space, and that this surface is approximately redshift-invariant out to at least $z \sim 3$ \citep{mannucciFundamentalRelationMass2010, curtiMassmetallicityFundamentalMetallicity2020}.
Even if stochastic fluctuations can on their own produce a surface in the $(M_\star, \mathrm{SFR}, Z_{\rm g})$ space, they cannot explain why a single parameter $\alpha$ can, even approximately, absorb the scatter, nor can they explain the redshift evolution of the underlying MZR and SFMS (see also discussion in \S\,\ref{sub:implication_to_sam}).
The stochastic picture, as currently developed, does not yet constitute a complete explanation of the FMR.

\paragraph*{Hierarchical assembly and mergers.}
We model each galaxy as evolving along a single main branch, accreting gas smoothly from the intergalactic medium.
Real galaxies, particularly massive ones, are assembled hierarchically through mergers \citep{laceyMergerRatesHierarchical1993, rodriguez-gomezStellarMassAssembly2016}.
A merger brings together gas and stars that formed and enriched in two lower-mass progenitors, producing a chemical history that differs from that of a galaxy reaching the same final stellar mass through purely in-situ star formation.
Minor mergers, which dominate the merger rate by number, tend to deposit metal-poor stellar populations into the outskirts of massive galaxies, potentially steepening radial metallicity gradients \citep{hirschmannStellarAccretionOrigin2015}.
Major mergers can drive gas inflows toward the galactic centre, triggering starbursts that alter both the gas and stellar metallicity on short timescales \citep{rupkeGasphaseOxygenGradients2010, torreyMetallicityEvolutionInteracting2012}.
These effects may contribute to the scatter in the MZR and the FMR at the massive end, where the merger history becomes increasingly important.

\paragraph*{Gas recycling and metal-enriched inflows.}
We assume that outflowing gas and metals are permanently removed from the system once ejected.
In practice, ejected material may remain bound to the halo and return to the ISM on relatively short timescales \citep{oppenheimerFeedbackSupermassiveBlack2020, mitchellGalacticInflowWind2020}.
Such recycling means that subsequent gas accretion is not pristine but carries a non-negligible metallicity, altering the effective yield of the system.
This recycling channel plays a central role in semi-analytic models \citep[e.g.][]{coleHierarchicalGalaxyFormation2000, bowerBreakingHierarchyGalaxy2006, laceyUnifiedMultiwavelengthModel2016}, where it is used to compensate for the high mass-loading factors required to reproduce the observed luminosity function.

\paragraph*{Variation in star formation efficiency and mass-loading factor.}
We assume that the star formation efficiency $\epsilon$ and mass-loading factor $\eta$ depend only on the mass scale of the galaxy or halo at a given redshift.
In reality, both quantities vary across galaxy populations at fixed stellar mass.
For instance, compact galaxies exhibit higher star formation efficiencies than diffuse galaxies at the same stellar mass, owing to their elevated stellar and gas surface densities and the star formation law \citep{kennicuttStarFormationMilky2012}.
\citet{wangOriginGalaxySizestellar2026} argued that this variation in star formation efficiency is the primary driver of the anti-correlation between galaxy size and stellar metallicity, in contrast to the traditional interpretation that compact galaxies are more metal-rich because their deeper potential wells suppress mass-loading \citep{ellisonCluesOriginMassMetallicity2008, sanchez-menguianoMoreFundamentalFundamental2024, maRevisitingFundamentalMetallicity2024}.
As we showed in Fig.~\ref{fig:fmr_physics}, the dependence of star formation efficiency on stellar mass and redshift directly shapes the parameterisation of the FMR, determining the optimal value of $\alpha$ and any residual mass or redshift dependence.
Capturing the full diversity of star formation efficiency across galaxy populations would require modelling galaxy structure, which is beyond the scope of this work.

\paragraph*{AGN feedback.}
Outflows driven by active galactic nuclei (AGN) represent another important process absent from our minimal model.
AGN feedback becomes significant once the central supermassive black hole has grown sufficiently massive, which occurs at a characteristic halo mass of $\sim 10^{12}\,\mathrm{M}_\odot$, corresponding to a stellar mass of $\sim 10^{10}\,\mathrm{M}_\odot$ \citep{bowerDarkNemesisGalaxy2017}.
Hydrodynamical simulations have shown that efficient AGN feedback can suppress the gas-phase metallicity in galaxies above this mass scale by preferentially ejecting metal-enriched gas from the galaxy centre \citep{derossiGalaxyMetallicityScaling2017, wangEnvironmentalDependenceMassMetallicity2023}.
Our model deliberately does not attempt to match the gas-phase metallicity at $M_\star \gtrsim 10^{10}\,\mathrm{M}_\odot$, where the calibrated model systematically overestimates the observed values.
This residual is qualitatively consistent with the expected effect of ejective AGN feedback, which would suppress the metallicity of massive galaxies and improve agreement with the data without altering the results at lower masses.

\paragraph*{Environmental effects on satellite galaxies.}
Finally, satellite galaxies are subject to environmental processes that we have not modelled.
Starvation, the cessation of fresh gas supply after a galaxy is accreted into a larger halo, eliminates the dilution effect of pristine inflow and allows continued star formation to enrich the ISM to higher metallicities \citep{larsonEvolutionDiskGalaxies1980, pengMassEnvironmentDrivers2010, pengMASSENVIRONMENTDRIVERS2012, pengStrangulationPrimaryMechanism2015}.
Ram-pressure stripping preferentially removes gas from the galaxy outskirts, where metallicity gradients are typically negative, effectively elevating the integrated gas metallicity \citep{gunnInfallMatterClusters1972, baheOriginEnhancedMetallicity2017}.
Both observations and hydrodynamical simulations confirm that satellite galaxies at fixed stellar mass tend to be more metal-rich than centrals \citep{pasqualiGasphaseMetallicityCentral2012, baheOriginEnhancedMetallicity2017,  wangEnvironmentalDependenceMassMetallicity2023}, consistent with both of these mechanisms operating in concert.
These environmental effects would need to be included in any model aiming to reproduce the full metallicity distribution across all galaxy environments.

\paragraph*{}
Including all of these effects is necessary for a comprehensive understanding of galaxy chemical evolution.
However, the complexity of these processes and their interplay precludes analytic treatment of all of them simultaneously.
Cosmological galaxy formation models, whether semi-analytic models or hydrodynamical simulations, are required to incorporate these effects self-consistently and to assess their relative importance across cosmic time.

\section{Summary}
\label{sec:summary}

The fundamental metallicity relation implies that star-forming galaxies populate a redshift-invariant surface in the three-dimensional space of stellar mass, star formation rate, and gas metallicity.
Despite its observational robustness, the physical origin of this surface and the reason for its approximate redshift invariance have remained unclear for many years.
Previous theoretical frameworks either assume equilibrium, which erases the dependence of metallicity on the accretion history and, therefore, cannot produce the FMR, or reproduce it numerically without identifying the underlying physics.
Here, we have departed from the equilibrium assumption and shown that the transition from the inflow-driven regime to equilibrium provides a unified framework for understanding the mass--metallicity relation, the fundamental metallicity relation, and the gaseous fundamental metallicity relation simultaneously.

Our main results are as follows:

\begin{enumerate}

  \item
        We constructed a minimal cosmological gas flow model governed by two mass-continuity equations, with gas inflow set by halo accretion and cooling efficiency ($\lambda$) that depends on halo mass and redshift, star formation efficiency ($\epsilon$) that depends on stellar mass and redshift, and mass-loading factor ($\eta$) that depends on stellar mass.
        Calibrated to three observational constraints, the mass--metallicity relation (which evolves with redshift), the star-forming main sequence, and the stellar mass--halo mass relation (\S\,\ref{ssub:calibration}; Fig.~\ref{fig:mzr_redshift_evolution}), the model predicts both the standard FMR (\S\,\ref{sub:fmr_in_the_cosmological_gas_flow_model}; Fig.~\ref{fig:fmr_parameter_fitting}) and the gaseous FMR (\S\,\ref{sub:gaseous_fmr_in_the_cosmological_gas_flow_model}; Fig.~\ref{fig:gaseous_fmr}) as direct  consequences that emerge without further tuning.

  \item
        Through controlled experiments that progressively simplify the assumptions about $\epsilon$ and $\eta$, we showed that in a universe where both quantities are universal constants, the FMR reduces to a single, universal sequence between $Z_\mg$ and $M_\star/\mathrm{SFR}$, whose shape directly traces the transition from the inflow-driven regime (steep, low $M_\star/\mathrm{SFR}$) to equilibrium (flat, high $M_\star/\mathrm{SFR}$).
        This is the physical origin of the FMR (\S\,\ref{sub:the_physics_of_fmr}; Fig.~\ref{fig:fmr_physics}).

  \item
        Reintroducing the mass and redshift dependence of $\epsilon$ and $\eta$ produces offsets among different stellar mass bins and redshifts.
        The FMR parameter $\alpha < 1$ acts as a horizontal shift that realigns these offset sequences.
        Its specific value encodes the combined mass and redshift dependence of $\epsilon$ and $\eta$, and the redshift invariance of the FMR is not a fundamental symmetry but a contingent consequence of how these quantities happen to depend on stellar mass and redshift (\S\,\ref{sub:the_physics_of_fmr}).

  \item
        The gaseous FMR, defined in the $(M_\star, M_\mg, Z_\mg)$ space, is more fundamental than the standard FMR.
        In the inflow-driven limit, $Z_\mg$ depends on $M_\star/M_\mg$ independently of the star formation efficiency, and the approach to equilibrium is governed by the gas fraction and mass-loading factor without further dependence on $\epsilon$.
        Varying $\epsilon$ with stellar mass or redshift therefore degrades the standard FMR but leaves the gFMR largely intact (\S\,\ref{sub:the_physics_of_fmr}; Fig.~\ref{fig:gfmr_physics}).

  \item
        We derived an analytic solution for the ideal gas flow model with constant $\Phi$ (mass inflow rate), $\epsilon$, and $\eta$.
        This solution provides closed-form expressions (equations~\ref{eq:master} and \ref{eq:t2tau}) relating the gas metallicity, the gas fraction, and the mass-loading factor through two functions $\mathcal K_1$ and $\mathcal K_2$ of the evolutionary stage, $t/\tau_\mathrm{eq}$ (Fig.\,\ref{fig:K123}).
        Given any two of $Z_\mg$, $M_\mg/M_\star$, and $\eta$, the third can be determined.
        Despite its simplifying assumptions, the analytic solution reproduces the full cosmological model (\S\,\ref{sec:analytic_approximate}; Fig.~\ref{fig:ideal_model_explain}).

  \item
        The origin of the FMR parameterisation can be understood analytically.
        When the mass-loading factor is dynamically unimportant, the ideal model approximation reduces to a universal monotonic relation between $Z_\mg$ and $M_\star/M_\mg$ (Fig.~\ref{fig:no_eta}), which forms the backbone of the FMR.
        If the star formation efficiency can be approximated as a power law in stellar mass and SFR, $\epsilon \propto M_\star^a\,\mathrm{SFR}^b$, then a single parameter $\alpha = (1-b)/(1+a)$ maps $M_\star/M_\mg$ onto the FMR projection $\xi = \log_{10}M_\star - \alpha\log_{10}\mathrm{SFR}$, collapsing all galaxies onto this universal relation at any given epoch.
        Redshift invariance then follows from the approximation $\mathrm{sSFR} \propto 1/t$, which provides a population-level mapping between cosmic time and observable galaxy properties: the redshift dependence of $\epsilon$ can be re-expressed as a dependence on $\mathrm{sSFR}$, preserving the same power-law structure across epochs so that a single value of $\alpha$ approximately absorbs offsets in both stellar mass and redshift (\S\,\ref{sub:analytic_understanding_fmr}).

\end{enumerate}

In the discussion, we further showed that:

\begin{enumerate}
  \setcounter{enumi}{6}

  \item
        The equilibrium approximation for the metal yield, $y \gg \epsilon^{-1}\,\dd Z_\mg/\dd t$, which underlies the gas regulator model, is quantitatively justified only for massive galaxies at low redshift.
        The majority of star-forming galaxies at $z \gtrsim 1$ reside in or near the inflow-driven regime, where the time-derivative term remains significant (\S\,\ref{sub:revisiting_the_equilibrium_model}; Fig.~\ref{fig:dZdt}).

  \item
        The framework of \citet{peeplesConstraintsStarFormation2011}, widely used to infer the mass-loading factor from observed scaling-relation slopes, conflates total (Lagrangian) derivatives with partial (Eulerian) derivatives.
        This identification is valid in the inflow-driven regime but breaks down near equilibrium, precisely where outflow constraints are most sought (\S\,\ref{sub:revisiting_citep_peeplesconstraintsstarformation_}; Figs~\ref{fig:demo_derivative} and~\ref{fig:derivative_alpha_ratio}).

  \item
        The ``universal metallicity relation'' of \citet{zahidUniversalRelationGalactic2014}, which relates $Z_\mg$ to $M_\star/M_\mg$ via a fixed functional form $[1-\exp(-M_\star/M_\mg)]$, can be derived from our framework under the assumption that the metal loss rate is constant.
        This assumption underpredicts the metallicity in the inflow-driven regime by a factor of $(1-R)/(1-R+\eta)$ and erases the dependence of the transition shape on the mass-loading factor that the $\mathcal{K}_1$--$\mathcal{K}_2$ system retains (\S\,\ref{sub:revisiting_citet_zahiduniversalrelationgalactic_}; Fig.~\ref{fig:compare_zahid14}).

  \item
        We also derive a generalisation of the analytic framework to the case where the mass- and metal-loading factors differ ($\eta \neq \zeta$).
        In the inflow-driven limit, the metallicity remains $Z_\mg \propto M_\star/M_\mg$ independently of both $\eta$ and $\zeta$; in equilibrium, it converges to $y/(1-R+\zeta)$ rather than $y/(1-R+\eta)$.
        Given any three of $Z_\mg$, $M_\mg/M_\star$, $\eta$, and $\zeta$, the fourth can be determined, providing a route to inferring the metal enrichment of galactic outflows from observations (see \S\,\ref{sec:differential_loading}).

  \item
        The deviation of high-redshift galaxies from the locally calibrated FMR is extremely sensitive to the precise value of the FMR parameter $\alpha$: parameterisations that are nearly indistinguishable at $z \lesssim 3$ diverge by up to $\sim 0.5$~dex in both directions at $z \gtrsim 4$.
        Before interpreting high-redshift offsets as evidence for new physics, the local FMR must be calibrated to sufficient precision that its extrapolation is reliable (\S\,\ref{sub:the_deviation_from_fmr_at_high_z_}; Fig.~\ref{fig:fmr_parameter_offset}).

  \item
        The framework predicts that the gas-to-stellar metallicity difference, $\Delta Z_{\mg,\star} \equiv \log_{10} Z_{\mg} - \log_{10} Z_\star$, should decrease monotonically from a maximum of $\log_{10}(3/2) \approx 0.18$~dex in the inflow-driven regime to zero at equilibrium, naturally explaining the observed trend whereby low-mass, high-sSFR galaxies show the largest $\Delta Z_{\mg,\star}$ and massive, low-sSFR galaxies approach $\Delta Z_{\mg,\star} \sim 0$ in \citet{fraser-mckelvieSAMIGalaxySurvey2022} (\S\,\ref{sub:stellar_gas_metallicity_difference}).

  \item
      For a given star formation history $\mathrm{SFR}(t)$ and gas metallicity history $Z_\mg(t)$, the star formation efficiency $\epsilon(t)$ and the mass-loading factor $\eta(t)$ remain degenerate: any choice of $\eta(t)$ uniquely determines a corresponding $\epsilon(t)$ that reproduces the same observables, and the two cannot be disentangled from $\mathrm{SFR}(t)$ and $Z_\mg(t)$ alone.
      This degeneracy can be broken by direct measurements of the gas mass history $M_\mg(t)$, since $\epsilon(t) \equiv \mathrm{SFR}(t)/M_\mg(t)$ is then directly measurable independently of $\eta(t)$.
      Although our model is calibrated against gas metallicity and star formation properties alone, without direct constraints on the gas mass history, it reproduces the observed stellar-to-gas mass relation at $z\sim0$ (Fig.~\ref{fig:gas_mass_comparison}), lending confidence that the inferred $\epsilon(t)$ and $\eta(t)$ are physically meaningful rather than artefacts of the calibration.
      In approaches that assume a parametric star formation history, this degeneracy introduces a practical hazard: a low-biased $\epsilon(t)$ overestimates the gas mass, requiring the model to suppress the inflow of pristine gas to match the observed $Z_\mg$, and in the extreme case drives the implied inflow rate negative --- an unphysical draining of metal-poor gas from the ISM with no counterpart in the baryon cycle (\S\,\ref{sub:degeneracy}).

  \item
        The gas inflow rate does not enter any scaling relation among $M_\star$, SFR, $M_\mg$, and $Z_\mg$: varying it by an order of magnitude leaves the MZR, the SFMS, and the gas fraction virtually unchanged, while shifting the stellar mass--halo mass relation by the corresponding factor (\S\,\ref{sub:mzr_shmr}, Fig.~\ref{fig:inflow_rate_sensitivity}).
        The cooling efficiency $\lambda(M_\mh, z)$ must therefore be calibrated independently using the stellar mass--halo mass relation.

  \item
        Semi-analytic models that adopt high mass-loading factors drive galaxies into equilibrium at all redshifts, predicting a non-evolving mass--metallicity relation in tension with observations.
        This tension can be resolved by simultaneously reducing the mass-loading factor and the star formation efficiency, extending the equilibrium timescale $\tau_{\rm eq} = 1/[(1-R+\eta)\epsilon]$ and allowing galaxies to reside in the inflow-driven regime for a larger fraction of cosmic time (\S\,\ref{sub:implication_to_sam}).

  \item
        The minimal model omits several processes that regulate metallicity evolution: stochastic gas accretion, hierarchical assembly and mergers, gas recycling and metal-enriched inflows, variations in $\epsilon$ and $\eta$, ejective AGN feedback, and environmental effects on satellite galaxies.
        Incorporating these effects self-consistently requires cosmological galaxy formation models, whether semi-analytic or hydrodynamical (\S\,\ref{sub:other_processes_in_regulating_metallicity_evolution}).
        In particular, we identify two unresolved problems with explaining the FMR through stochastic gas accretion alone: first, a population-wide anti-correlation between SFR and $Z_\mathrm{g}$ requires that galaxies at fixed stellar mass share closely aligned equilibrium values of SFR and $Z_\mathrm{g}$, a condition that has not been demonstrated; second, stochastic fluctuations do not explain why a single parameter $\alpha$ can project the $(M_\star, \mathrm{SFR}, Z_\mathrm{g})$ surface onto a redshift-invariant sequence, nor do they explain the redshift evolution of the underlying MZR.

\end{enumerate}

The mass--metallicity relation and the fundamental metallicity relation are among the tightest scaling relations in extragalactic astronomy, yet previous theoretical frameworks have relied on assumptions that obscure the connection between these relations and the underlying physics, limiting their use as quantitative constraints on galaxy evolution.
By establishing a unified analytic framework that derives both relations from first principles and connects their shape and parameterisation directly to the star formation efficiency and mass-loading factor, this work provides the foundation to turn metallicity scaling relations into precision probes of the baryon cycle across cosmic time.

\section*{Acknowledgements}

KW thanks Andrew Pontzen, Joop Schaye, Rob Crain, Kyle Oman, Evgenii Chaikin, Cheng Li, Shude Mao, Yong Shi, Romeel Dav\'e, Mark Swinbank, Yunjing Wu, Mingyu Li, Ivan Baldry, Andreea Font, Dirk Scholte, Vivienne Wild, Nicole Marcelina Gountanis, Robert Yates, Guinevere Kauffmann, Azadeh Fattahi, Alexander H. Riley, Hiranya Peiris, Xihan Ji, Xiaohu Yang, Cheqiu Lyu, Francesco Shankar, and Yangyao Chen for inspiring discussions at different stages of this work.
KW acknowledges the use of Claude (Anthropic) as a writing and research aid in the preparation of this manuscript, including literature searches, drafting assistance, and language editing; all scientific content, analysis, and conclusions are the authors' own.

This work is supported by the Science and Technology Facilities Council (STFC) through grant ST/X001075/1.
SB is supported by the UK Research and Innovation (UKRI) Future Leaders Fellowship [grant number MR/V023381/1 and UKRI2044].
This work is co-funded by the European Union (Widening Participation, ExGal-Twin, GA 101158446). Views and opinions expressed are however those of the author(s) only and do not necessarily reflect those of the European Union. Neither the European Union nor the granting authority can be held responsible for them. 
NFB acknowledges support from Science and Technologies Facilities Council (STFC) grant ST/Y00275X/1.
Y.P. acknowledges support from the National Natural Science Foundation of China (NSFC) under grant Nos. 12125301 and 12192222, and from the New Cornerstone Science Foundation through the XPLORER PRIZE.

This work used the DiRAC@Durham facility managed by the Institute for Computational Cosmology on behalf of the STFC DiRAC HPC Facility (www.dirac.ac.uk).
The equipment was funded by BEIS capital funding via STFC capital grants ST/K00042X/1, ST/P002293/1, ST/R002371/1 and ST/S002502/1, Durham University and STFC operations grant ST/R000832/1.
DiRAC is part of the National e-Infrastructure.

This research made use of NASA's Astrophysics Data System for bibliographic information.

\section*{Data Availability}
The data underlying this article will be shared on reasonable request to the corresponding author.



\bibliography{bibtex.bib} 
\bibliographystyle{mnras}



\appendix

\section{Analytic solutions of the ideal gas flow model}
\label{sec:ideal_model_derivation}

Here we derive the closed-form solutions for the ideal gas flow model (equations~\ref{eq:mgas}--\ref{eq:zstar}), in which the inflow rate $\Phi$, star formation efficiency $\epsilon$, and mass-loading factor $\eta$ are all constant, and the initial conditions are $M_\mg(0) = Z_\mg(0) = 0$. We define the equilibrium timescale
\begin{equation}
  \tau_\mathrm{eq} \equiv \frac{1}{(1-R+\eta)\,\epsilon}\,.
  \label{eq:tau_eq_def}
\end{equation}

\subsection{Gas mass}

The gas-mass continuity equation (equation~\ref{eq:mgas}) reads
\begin{equation}
  \frac{\mathrm{d} M_\mg}{\mathrm{d} t} + \frac{M_\mathrm{g}}{\tau_\mathrm{eq}} = \Phi\,.
  \label{eq:Mg_ode}
\end{equation}
This is a first-order linear ODE with constant coefficients.
Multiplying both sides by the integrating factor $\mathrm{e}^{t/\tau_\mathrm{eq}}$ gives
\begin{equation}
  \frac{\mathrm{d}}{\mathrm{d} t}
  \!\left[M_\mathrm{g}\,\mathrm{e}^{t/\tau_\mathrm{eq}}\right]
  = \Phi\,\mathrm{e}^{t/\tau_\mathrm{eq}}\,.
\end{equation}
which yields
\begin{equation}
  M_\mathrm{g}(t) = \Phi\,\tau_\mathrm{eq}\left(1 - \mathrm{e}^{-t/\tau_\mathrm{eq}}\right).
\end{equation}

\subsection{Gas metallicity}

Defining the gas-phase metal mass $M_Z \equiv M_\mathrm{g}\,Z_\mathrm{g}$, the metal-mass continuity equation (equation~\ref{eq:zgas}) becomes
\begin{equation}
  \frac{\mathrm{d} M_Z}{\mathrm{d} t} + \frac{M_Z}{\tau_\mathrm{eq}} = y\,\epsilon\,M_\mathrm{g}(t)\,.
  \label{eq:MZ_ode}
\end{equation}
This is again a first-order linear ODE, now with a known source term. Substituting the solution for $M_\mathrm{g}(t)$ (equation \ref{eq:mgas_ana}) and applying the integrating factor $\mathrm{e}^{t/\tau_\mathrm{eq}}$,
\begin{equation}
  \frac{\mathrm{d}}{\mathrm{d} t}
  \!\left[M_Z\,\mathrm{e}^{t/\tau_\mathrm{eq}}\right]
  = y\,\epsilon\,\Phi\,\tau_\mathrm{eq}
  \left(\mathrm{e}^{t/\tau_\mathrm{eq}} - 1\right).
\end{equation}
so that
\begin{equation}
  M_Z(t)
  = y\,\epsilon\,\Phi\,\tau_\mathrm{eq}
  \left[\tau_\mathrm{eq}
    - \left(t + \tau_\mathrm{eq}\right)\mathrm{e}^{-t/\tau_\mathrm{eq}}\right].
  \label{eq:MZ_solution}
\end{equation}
The gas metallicity follows as $Z_\mathrm{g} = M_Z/M_\mathrm{g}$,
\begin{equation}
  Z_\mathrm{g}(t) = y\,\epsilon\left(\tau_\mathrm{eq} - \frac{t}{\mathrm{e}^{t/\tau_\mathrm{eq}} - 1}\right).
\end{equation}

\subsection{Stellar mass}

The stellar mass is the time integral of the net star formation rate,
\begin{align}
  M_\star(t)
   & = \int_0^t (1-R)\,\epsilon\,M_\mathrm{g}(t')\,\mathrm{d}t' \nonumber \\
   & = (1-R)\,\epsilon\,\Phi\,\tau_\mathrm{eq}
  \int_0^t \left(1 - \mathrm{e}^{-t'/\tau_\mathrm{eq}}\right)\mathrm{d}t'\,.
\end{align}
which yields
\begin{equation}
  M_\star(t) = (1-R)\,\epsilon\,\tau_\mathrm{eq}\,\Phi \left[t - \tau_\mathrm{eq}\!\left(1 - \mathrm{e}^{-t/\tau_\mathrm{eq}}\right)\right].
\end{equation}

\subsection{Stellar metallicity}

The mass-weighted stellar metallicity is
\begin{equation}
  Z_\star(t)
  = \frac{(1-R)}{M_\star(t)} \int_0^t \epsilon\,M_\mathrm{g}(t')\,Z_\mathrm{g}(t')\,\mathrm{d}t' \\
\end{equation}
Substituting and writing $\tau \equiv \tau_\mathrm{eq}$ for brevity,
\begin{align}
  \int_0^t \epsilon\,M_\mathrm{g}(t')\,Z_\mathrm{g}(t')\,\mathrm{d}t'
   & = y\,\epsilon^2\,\Phi\,\tau
  \int_0^t \left[\tau - (t'+\tau)\,\mathrm{e}^{-t'/\tau}\right]\mathrm{d}t'                  \\
   & = y\,\epsilon^2\,\Phi\,\tau^2 \left[t - 2\tau + (t+2\tau)\,\mathrm{e}^{-t/\tau}\right].
\end{align}
Dividing by $M_\star/(1-R) = \epsilon\,\tau\,\Phi\left[t - \tau(1-\mathrm{e}^{-t/\tau})\right]$ gives
\begin{equation}
  Z_\star(t)
  = y\,\epsilon\,\tau_\mathrm{eq}\;\frac{t - 2\tau_\mathrm{eq}
    + (t+2\tau_\mathrm{eq})\,\mathrm{e}^{-t/\tau_\mathrm{eq}}}
  {t - \tau_\mathrm{eq} + \tau_\mathrm{eq}\,\mathrm{e}^{-t/\tau_\mathrm{eq}}}\,.
\end{equation}

\section{Deriving $Z_\mg$}
\label{sec:the_derivation_of_equation_eqref_eq_zgas_lilly_}

Here we derive the exact expression for the gas metallicity (equation~\ref{eq:zgas_lilly}) from the two mass-continuity equations governing the gas reservoir (equation~\ref{eq:mgas}) and its metal content (equation~\ref{eq:zgas}).
Note that our yield $y$ is defined as the metal mass produced per unit total star formation, which differs from the convention in \citet{lillyGASREGULATIONGALAXIES2013} by a factor of $(1-R)$.

We begin by expanding the time derivative of the gas metallicity $Z_\mathrm{g} \equiv M_{\rm Z}/M_\mathrm{g}$,
\begin{equation}
  \frac{\mathrm{d} Z_\mathrm{g}}{\mathrm{d} t}
  = \frac{1}{M_\mathrm{g}}\left[\frac{\mathrm{d} M_{\rm Z}}{\mathrm{d} t}
    - Z_\mathrm{g}\,\frac{\mathrm{d} M_\mathrm{g}}{\mathrm{d} t}\right].
  \label{eq:dZdt_expand}
\end{equation}
The first term is given directly by the metal-mass continuity equation,
\begin{equation}
  \frac{\mathrm{d} M_{\rm Z}}{\mathrm{d} t}
  = y\,\epsilon\,M_\mathrm{g} - Z_\mathrm{g}\,(1-R+\eta)\,\epsilon\,M_\mathrm{g}.
  \label{eq:dMZdt}
\end{equation}
To express the second term, we relate $\mathrm{d}M_\mathrm{g}/\mathrm{d}t$ to the gas fraction $\mu \equiv M_\mathrm{g}/M_\star$.
Taking the logarithmic derivative of $\mu$ gives
\begin{equation}
  \frac{\mathrm{d}\ln\mu}{\mathrm{d} t}
  = \frac{1}{M_\mathrm{g}}\frac{\mathrm{d} M_\mathrm{g}}{\mathrm{d} t}
  - \frac{1}{M_\star}\frac{\mathrm{d} M_\star}{\mathrm{d} t}.
  \label{eq:dlnmu}
\end{equation}
Since $\mathrm{d}M_\star/\mathrm{d}t = (1-R)\,\epsilon\,M_\mathrm{g}$, this can be rearranged to obtain
\begin{equation}
  \frac{\mathrm{d} M_\mathrm{g}}{\mathrm{d} t}
  = \left[\mu(1-R) + \epsilon^{-1}\frac{\mathrm{d}\ln\mu}{\mathrm{d} t}\right]
  \epsilon\,M_\mathrm{g}.
  \label{eq:dMgdt_mu}
\end{equation}
Substituting equations~\eqref{eq:dMZdt} and~\eqref{eq:dMgdt_mu} into equation~\eqref{eq:dZdt_expand} yields
\begin{align}
  \frac{\mathrm{d} Z_\mathrm{g}}{\mathrm{d} t}
   & = y\,\epsilon
  - Z_\mathrm{g}\,(1-R+\eta)\,\epsilon
  - Z_\mathrm{g}\left[\mu(1-R)\,\epsilon
    + \frac{\mathrm{d}\ln\mu}{\mathrm{d} t}\right].
\end{align}
Dividing both sides by $\epsilon$ and collecting the $Z_\mathrm{g}$ terms,
\begin{equation}
  \epsilon^{-1}\frac{\mathrm{d} Z_\mathrm{g}}{\mathrm{d} t}
  = y - Z_\mathrm{g}\left[(1-R+\eta) + (1-R)\,\mu
    + \epsilon^{-1}\frac{\mathrm{d}\ln\mu}{\mathrm{d} t}\right].
\end{equation}
Solving for $Z_\mathrm{g}$ gives the exact expression,
\begin{equation}
  Z_\mathrm{g}
  = \frac{y - \epsilon^{-1}\,\mathrm{d}Z_\mathrm{g}/\mathrm{d}t}
  {(1-R+\eta) + (1-R)\,\mu + \epsilon^{-1}\,\mathrm{d}\ln\mu/\mathrm{d}t}.
  \label{eq:zgas_lilly2}
\end{equation}
This expression is exact and holds at every instant along a galaxy's evolution, regardless of whether the system is in equilibrium.
The equilibrium approximation adopted by \citet{lillyGASREGULATIONGALAXIES2013} and \citet{feldmannEquilibriumViewDust2015} consists of dropping the $\epsilon^{-1}\,\mathrm{d}Z_\mathrm{g}/\mathrm{d}t$ term in the numerator, which is justified only when $y \gg \epsilon^{-1}\,\mathrm{d}Z_\mathrm{g}/\mathrm{d}t$ (see Fig.~\ref{fig:dZdt}).

Equation~(\ref{eq:zgas_lilly2}) can be recast into the form used by \citet{peeplesConstraintsStarFormation2011} by converting the time derivatives into derivatives with respect to stellar mass.
Since $\mathrm{d}M_\star/\mathrm{d}t = (1-R)\,\epsilon\,M_\mathrm{g}$, any time derivative can be written as
\begin{equation}
  \epsilon^{-1}\frac{\mathrm{d}(\cdot)}{\mathrm{d}t}
  = (1-R)\,M_\mathrm{g}\,\frac{\mathrm{d}(\cdot)}{\mathrm{d}M_\star}
  = (1-R)\,\mu\,\frac{\mathrm{d}(\cdot)}{\mathrm{d}\ln M_\star}\,,
  \label{eq:chain_rule}
\end{equation}
where $\mu \equiv M_\mathrm{g}/M_\star$.
Applying this to the two time-derivative terms in equation~(\ref{eq:zgas_lilly}):
\begin{align}
  \epsilon^{-1}\frac{\mathrm{d}Z_\mathrm{g}}{\mathrm{d}t}
   & = (1-R)\,\mu\,Z_\mathrm{g}\,\frac{\mathrm{d}\ln Z_\mathrm{g}}{\mathrm{d}\ln M_\star}\,,
  \label{eq:dZdt_to_dZdM}                                                                    \\[4pt]
  \epsilon^{-1}\frac{\mathrm{d}\ln\mu}{\mathrm{d}t}
   & = (1-R)\,\mu\left(\frac{\mathrm{d}\ln M_\mathrm{g}}{\mathrm{d}\ln M_\star} - 1\right).
  \label{eq:dlnmudt_to_dM}
\end{align}
Substituting into equation~(\ref{eq:zgas_lilly}), and it reads
\begin{equation}
  Z_\mathrm{g}
  = \frac{y - (1-R)\,\mu\,Z_\mathrm{g}\,\dfrac{\mathrm{d}\ln Z_\mathrm{g}}{\mathrm{d}\ln M_\star}}
  {(1-R+\eta) + (1-R)\,\mu\,\dfrac{\mathrm{d}\ln M_\mathrm{g}}{\mathrm{d}\ln M_\star}}\,.
\end{equation}
Solving for $Z_\mathrm{g}$ yields
\begin{align}
  Z_\mathrm{g}       & = \frac{y}{(1-R+\eta) + \alpha_\mathrm{PS}\,\mu}\,,                        \\[4pt]
  \alpha_\mathrm{PS} & \equiv (1-R)\left[\frac{\mathrm{d}\ln M_\mathrm{g}}{\mathrm{d}\ln M_\star}
    + \frac{\mathrm{d}\ln Z_\mathrm{g}}{\mathrm{d}\ln M_\star}\right],
\end{align}
which is equations~(\ref{eq:zgas_peeples}) and~(\ref{eq:alpha_peeples}).
We emphasise that the derivatives appearing in $\alpha_\mathrm{PS}$ are total (Lagrangian) derivatives along an individual galaxy's evolutionary track, not partial (Eulerian) derivatives across the population at fixed cosmic time.
The distinction between the two is discussed in \S\,\ref{sub:revisiting_citep_peeplesconstraintsstarformation_}.

\section{Halo accretion rate}
\label{sec:halo_accretion_rate}

We adopt the mass accretion history parameterisation of \citet{wechslerConcentrationsDarkHalos2002}, in which the halo mass at scale factor $a$ is given by
\begin{equation}
  M_{\rm h}(a) = M_{\rm h,0}\exp\left[-\frac{8a_0}{c}\left(\frac{a_0}{a} - 1\right)\right],
\end{equation}
where $M_{\rm h,0}$ is the halo mass at the present-day scale factor $a_0 = 1$ and $c$ is the concentration parameter of the final halo.
The corresponding mass accretion rate follows from the chain rule,
\begin{align}
  \frac{\mathrm{d} M_{\rm h}}{\mathrm{d} t}
   & = \frac{\mathrm{d} M_{\rm h}}{\mathrm{d} a}\frac{\mathrm{d} a}{\mathrm{d} t}
  = M_{\rm h}(a)\,\frac{8a_0}{c}\,\frac{a_0}{a^2}\,a\,H(a) \nonumber              \\
   & = \frac{8a_0^2\,H(a)}{c\,a}\,M_{\rm h}(a),
\end{align}
where $H(a) = H_0\sqrt{\Omega_{\rm m}\,a^{-3} + \Omega_\Lambda}$.
Expressing this in terms of redshift gives
\begin{align}
  \frac{\mathrm{d} M_{\rm h}}{\mathrm{d} t}
   & = \frac{8H_0}{c}\,M_{\rm h}\,(1+z)\sqrt{\Omega_{\rm m}(1+z)^3 + \Omega_\Lambda} \nonumber \\
   & = \frac{572.72\;\mathrm{M}_\odot\,\mathrm{yr}^{-1}}{c}\,
  \frac{M_{\rm h}}{10^{12}\,\mathrm{M}_\odot}\,(1+z)\sqrt{\Omega_{\rm m}(1+z)^3 + \Omega_\Lambda},
\end{align}
where the numerical coefficient assumes $H_0 = 70\;\mathrm{km\,s^{-1}\,Mpc^{-1}}$.
We adopt $c = 12$, which closely reproduces the mean halo growth histories reported by \citet{fakhouriMergerRatesMass2010}.

\section{The impact of Eddington bias on the slope of FMR}
\label{sec:the_impact_of_eddington_bias_on_the_slope_of_fmr}

\begin{figure}
  \begin{center}
    \includegraphics[width=0.95\linewidth]{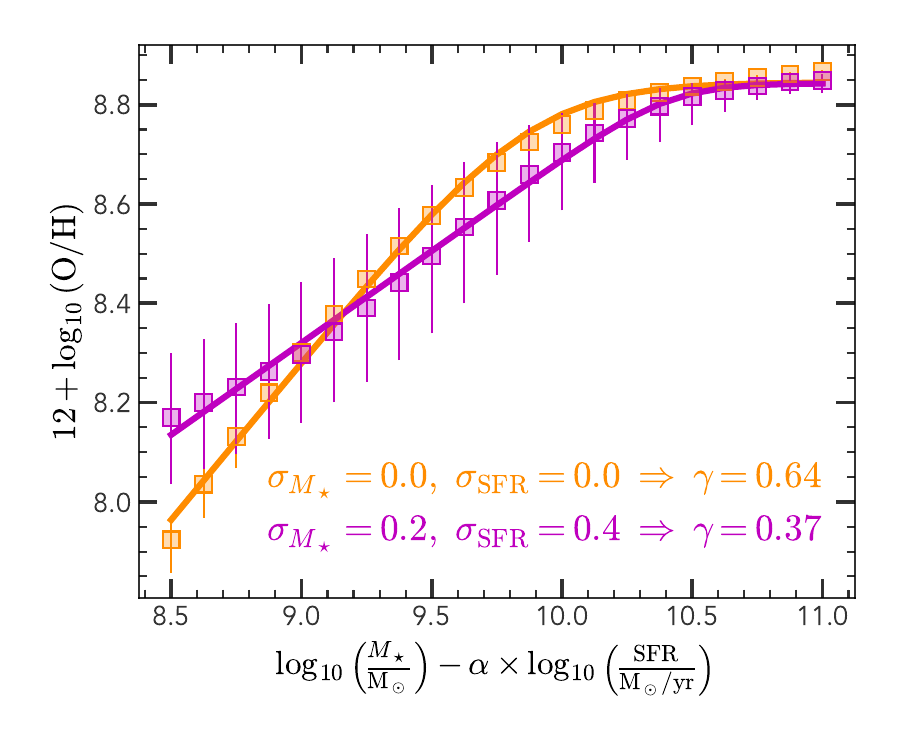}
  \end{center}
  \caption{
      Mock FMR recovered from the model predictions with and without observational scatter added to $M_\star$ and SFR before fitting.
      Orange squares show the noise-free relation, for which the best-fitting slope recovers the intrinsic value, $\gamma = 0.64$.
      Magenta squares show the same underlying relation after adding Gaussian scatter of $\sigma_{M_\star} = 0.2$ dex and $\sigma_{\rm SFR} = 0.4$ dex to $M_\star$ and SFR, respectively, before re-fitting; the best-fitting slope flattens to $\gamma = 0.37$.
      Error bars show the $16^{\rm th}-84^{\rm th}$ range in each bin.
      Adding realistic observational uncertainty to $M_\star$ and SFR is therefore sufficient to flatten the intrinsic slope towards the shallower value reported observationally.
  }
  \label{fig:fmr_edd_bias}
\end{figure}

To test whether Eddington bias can account for the discrepancy between our best-fitting slope, $\gamma = 0.64$, and the value of $\gamma = 0.30$ reported by \citet{curtiMassmetallicityFundamentalMetallicity2020}, we construct a mock sample from the model predictions and refit equation~\eqref{eq:fmr_fitting} after adding observational scatter to $M_\star$ and SFR.
We first evaluate the model on a grid of $M_\star$ and redshift, generating a noise-free mock catalogue of $(M_\star, \mathrm{SFR}, Z_\mg)$ triples that follow the intrinsic FMR by construction.
We then perturb $\log_{10}M_\star$ and $\log_{10}\mathrm{SFR}$ independently with Gaussian scatter of $\sigma_{M_\star} = 0.2$~dex and $\sigma_{\rm SFR} = 0.4$~dex, values representative of typical observational uncertainties in stellar mass and SFR indicators, and refit the perturbed catalogue using the same functional form and fitting procedure applied to the real data.

Fig.~\ref{fig:fmr_edd_bias} shows the result.
Without added scatter, the fit recovers the intrinsic slope, $\gamma = 0.64$.
Once observational scatter is added, the best-fitting slope flattens to $\gamma = 0.37$, close to the observed value of $0.30$.
This flattening arises from an asymmetry in how scatter moves galaxies between bins.
Scatter in $M_\star$ and SFR is isotropic, but because low-mass galaxies vastly outnumber high-mass galaxies, more low-$\xi$ galaxies scatter upward into the high-$\xi$ tail than high-$\xi$ galaxies scatter downward.
This asymmetry causes the fitted slope to be shallower than the intrinsic value.
The direction and approximate magnitude of this effect match the discrepancy between our predicted slope and the observed slope of \citet{curtiMassmetallicityFundamentalMetallicity2020}.

\section{FMR with varying $\alpha$}
\label{sec:fmr_with_varying_alpha_}

\begin{figure*}
  \begin{center}
    \includegraphics[width=0.95\linewidth]{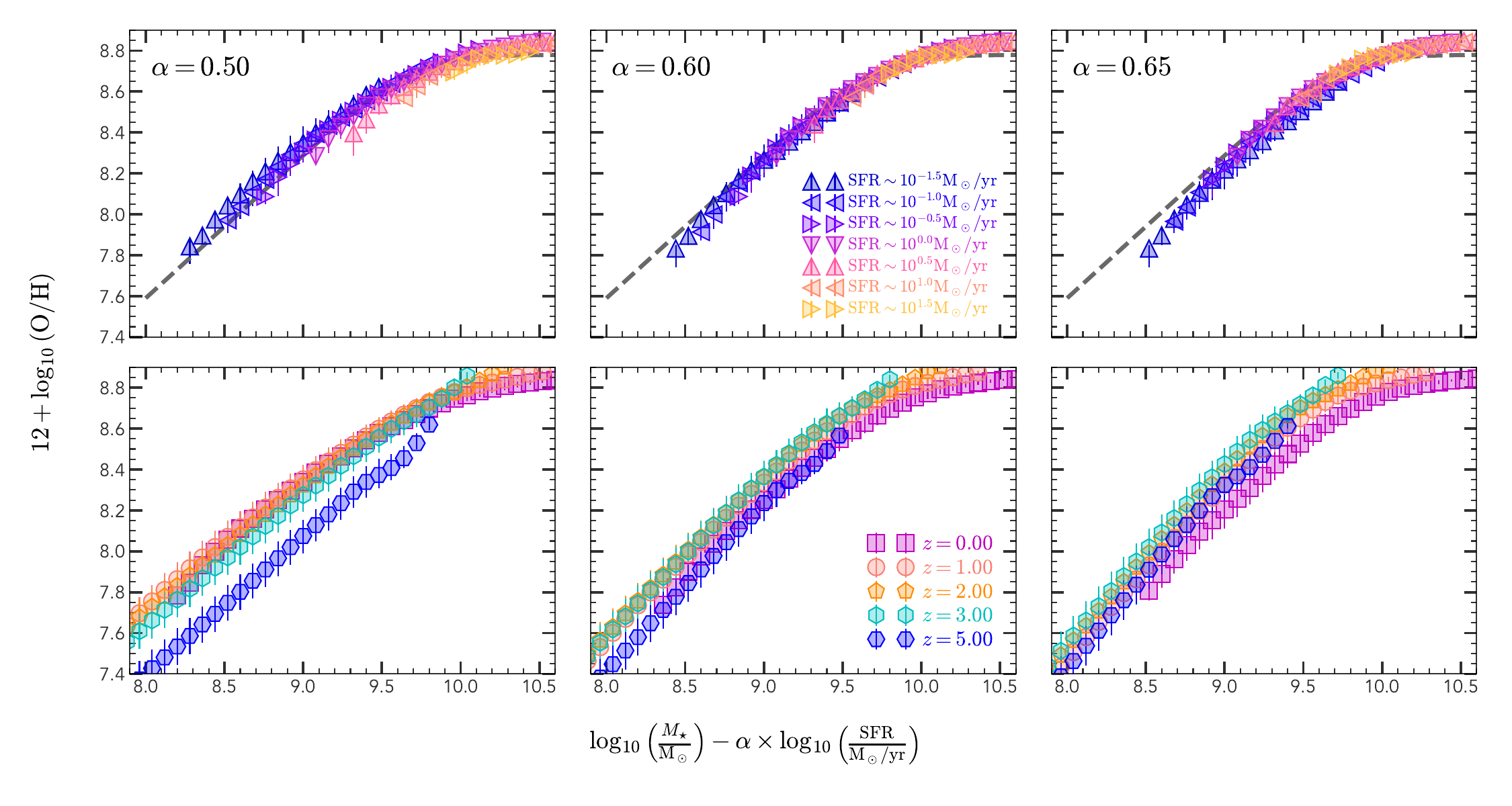}
  \end{center}
  \caption{
    Gas metallicity as a function of the FMR parameter $\log_{10}(M_\star/\mathrm{M}_\odot) - \alpha \log_{10}(\mathrm{SFR}/\mathrm{M}_\odot\,\mathrm{yr}^{-1})$ predicted by the cosmological gas flow model, for $\alpha = 0.50$ (left), $\alpha=0.60$ (middle), and $\alpha = 0.65$ (right).
    \textbf{Upper row}: results at $z = 0$, colour-coded by SFR bin.
    \textbf{Lower row}: the same projection applied across redshifts $z = 0$--5.
    The dashed line in each panel shows the best-fitting relation using the parametrisation of \citet{curtiMassmetallicityFundamentalMetallicity2020} with $\alpha=0.55$ for reference.
    All three values of $\alpha$ produce tight sequences at $z \lesssim 3$, but their behaviour differs at $z = 5$: for $\alpha = 0.50$, the $z = 5$ points fall systematically below the locally calibrated relation, while for $\alpha = 0.60-0.65$, they remain close to it.
    This demonstrates that the apparent magnitude of any high-redshift deviation from the FMR depends sensitively on the choice of $\alpha$, even though the low-redshift data do not strongly discriminate between the two parameterisations.
  }
  \label{fig:fmr_parameter_fitting_varying_alpha}
\end{figure*}

Fig.~\ref{fig:fmr_parameter_fitting_varying_alpha} presents the FMR projection for three values of $\alpha$ that span the fiducial value $\alpha = 0.55$ adopted in the main text (Fig.~\ref{fig:fmr_parameter_fitting}).
For $\alpha = 0.50$ (left column), the SFR receives less weight in the projection.
At $z = 0$ (upper left), the different SFR bins show a mild residual ordering at the low-$\xi$ end, with low-SFR galaxies lying slightly above the mean relation and high-SFR galaxies slightly below.
The redshift sequences in the lower left panel collapse well out to $z \sim 3$, but the $z = 5$ points fall visibly below the dashed line by $\sim 0.3$--0.5~dex, producing a systematic negative offset at high redshift.

Increasing $\alpha$ to 0.60-0.65 (middle and right columns) shifts high-SFR galaxies further to the left relative to low-SFR galaxies, tightening the collapse of SFR bins at $z = 0$ (upper right).
In the lower right panel, the different redshift sequences separate more visibly than at $\alpha = 0.50$: the $z = 5$ points now sit above the $z = 0$ locus at the low-$\xi$ end rather than below it, and the locally calibrated dashed line passes through the high-redshift data rather than above them.
The sign of the apparent high-redshift offset has therefore reversed between $\alpha = 0.50$ and $\alpha = 0.65$.

These trends follow directly from the physics discussed in \S\,\ref{sec:fundamental_metallicity_relation}.
A lower $\alpha$ undercompensates for the redshift evolution of the star formation efficiency, causing high-redshift galaxies to appear metal-poor relative to the local baseline.
A higher $\alpha$ overcompensates, producing an epoch-dependent spread in the opposite sense.
The fiducial value $\alpha = 0.55$ is intermediate and yields the tightest overall collapse, but the residual differences among all three parameterisations are small at $z \lesssim 3$ relative to observational uncertainties.
As shown in Fig.~\ref{fig:fmr_parameter_offset}, this low-redshift degeneracy amplifies into systematic offsets of up to $\sim 0.5$~dex when extrapolated to $z \gtrsim 4$, underscoring the need for a precisely calibrated local FMR before high-redshift deviations can be physically interpreted.

\section{Derivation of the \texorpdfstring{$\mathcal{K}_1$, $\mathcal{K}_2$, and $\mathcal{K}_3$}{K1, K2, and K3} functions}
\label{sec:k1k2k3}

Here we derive the closed-form expressions for the gas metallicity (equation~\ref{eq:master}), the evolutionary stage (equation~\ref{eq:t2tau}), and the stellar metallicity (equation~\ref{eq:zstar_master}), starting from the ideal gas flow model solutions (equations~\ref{eq:mgas_ana}--\ref{eq:mstar_ana}).
Throughout, we define the dimensionless evolutionary stage $x \equiv t/\tau_{\rm eq}$ and use the equilibrium timescale $\tau_{\rm eq} \equiv 1/[(1-R+\eta)\epsilon]$.

From equations~\eqref{eq:mgas_ana} and \eqref{eq:mstar_ana}, the ratio $M_{\rm g}/M_\star$ is
\begin{align}
  \frac{M_{\rm g}}{M_\star}
   & = \frac{\Phi\tau_{\rm eq}(1 - \me^{-x})} {(1-R)\epsilon\tau_{\rm eq}\Phi [x - (1 - \me^{-x})]} \\
   & = \frac{1-\me^{-x}}{(1-R)\epsilon\tau_{\rm eq} [x-(1-\me^{-x})]}.
  \label{eq:app_mu}
\end{align}
Since $\tau_{\rm eq} = 1/[(1-R+\eta)\epsilon]$, we have $(1-R)\epsilon\tau_{\rm eq} = (1-R)/(1-R+\eta)$, so equation~\eqref{eq:app_mu} rearranges to
\begin{equation}
  \frac{1-R+\eta}{1-R} \left(\frac{M_{\rm g}}{M_\star}\right)^{-1} = \frac{x - (1 - \me^{-x})}{1 - \me^{-x}} \equiv \mathcal{K}_2(x),
  \label{eq:app_k2}
\end{equation}
which is equation~\eqref{eq:t2tau}.

From equation~\eqref{eq:zgas_ana}, the gas metallicity reads
\begin{equation}
  Z_{\rm g} = y\epsilon \tau_{\rm eq}\left(1 - \frac{x}{\me^x - 1}\right) =  y\epsilon\tau_{\rm eq}\,
  \frac{1 - \me^{-x} - xe^{-x}}{1 - \me^{-x}}.
  \label{eq:app_zg_intermediate}
\end{equation}
Combining equation~\eqref{eq:app_mu} to eliminate $\epsilon\tau_{\rm eq}$, we can have
\begin{equation}
  Z_{\rm g} =
  \frac{y}{1-R} \left(\frac{M_{\rm g}}{M_\star}\right)^{-1} \frac{1 - \me^{-x} - xe^{-x}} {x - (1 - \me^{-x})}
  \equiv
  \frac{y}{1-R} \left(\frac{M_{\rm g}}{M_\star}\right)^{-1} \mathcal{K}_1(x),
  \label{eq:app_k1}
\end{equation}
where
\begin{equation}
  \mathcal{K}_1(x) \equiv
  \frac{1 - \me^{-x} - xe^{-x}} {x - (1 - \me^{-x})},
  \label{eq:app_k1_def}
\end{equation}
which is equation~\eqref{eq:master}.

From equation~\ref{eq:zstar_ana}, the mass-weighted stellar metallicity is
\begin{equation}
  Z_\star = \frac{y\epsilon\tau_{\rm eq}\, [x - 2 + (x+2)\me^{-x}]} {x - (1 - \me^{-x})}.
  \label{eq:app_zstar_intermediate}
\end{equation}
Combining with equation~\eqref{eq:app_mu} to eliminate $\epsilon\tau_{\rm eq}$, we can have
\begin{equation}
  Z_\star =
  \frac{y}{1-R}
  \left(\frac{M_{\rm g}}{M_\star}\right)^{-1}
  \mathcal{K}_3(x),
  \label{eq:app_zstar}
\end{equation}
where
\begin{equation}
  \mathcal{K}_3(x) \equiv \frac{1 - \mathcal K_1(x)} {\mathcal K_2(x)}.
  \label{eq:app_k3_def}
\end{equation}

\section{Derivation of the generalisation to differential mass and metal loading}
\label{sec:derivation_of_the_generalisation_to_differential_mass_and_metal_loading}

The gas mass and stellar mass depend only on $\tau_\mathrm{m}$ and are given by equations~(\ref{eq:mgas_ana}) and~(\ref{eq:mstar_ana}) with the replacement $\tau_\mathrm{eq} \to \tau_\mathrm{m}$. The metal-mass equation~(\ref{eq:dmz_general}) is a first-order linear ODE in $M_Z \equiv M_\mathrm{g}\,Z_\mathrm{g}$ with a different decay timescale $\tau_Z$,
\begin{equation}
  \frac{\mathrm{d}M_Z}{\mathrm{d}t} + \frac{M_Z}{\tau_Z}
  = y\,\epsilon\,\Phi\,\tau_\mathrm{m}\!\left(1 - \mathrm{e}^{-t/\tau_\mathrm{m}}\right).
\end{equation}
Solving with the integrating factor $\mathrm{e}^{t/\tau_Z}$ and the initial condition $M_Z(0) = 0$ gives
\begin{equation}
  M_Z(t) = y\,\epsilon\,\Phi\,\tau_\mathrm{m}
  \left[\frac{\tau_\mathrm{m}}{r}\!\left(1-\mathrm{e}^{-r x}\right)
    - \frac{\tau_\mathrm{m}}{r-1}\!\left(\mathrm{e}^{-x} - \mathrm{e}^{-r x}\right)\right],
  \label{eq:MZ_general}
\end{equation}
where $x \equiv t/\tau_\mathrm{m}$ and $r \equiv \tau_\mathrm{m}/\tau_Z = (1-R+\zeta)/(1-R+\eta)$. The gas metallicity follows as $Z_\mathrm{g} = M_Z/M_\mathrm{g}$,
\begin{equation}
  Z_\mathrm{g} = y\,\epsilon\,\tau_\mathrm{m}\;\frac{(1 - \mathrm{e}^{-rx})/r
    - (\mathrm{e}^{-x} - \mathrm{e}^{-rx})/(r-1)}{1 - \mathrm{e}^{-x}}\,.
  \label{eq:Zg_general}
\end{equation}

\section{Deriving the Zahid et al. (2014) universal metallicity relation}
\label{sec:zahid_derivation}

Here we show how the universal metallicity relation of \citet{zahidUniversalRelationGalactic2014} emerges from the gas flow model (equations~\ref{eq:mgas} and~\ref{eq:zgas}) under a series of simplifying assumptions.
Converting from time to stellar mass using $\mathrm{d}M_\star = \epsilon\,M_\mathrm{g}\,\mathrm{d}t$ \footnote{Here $\dd M_\star$ is the newly formed stellar mass without subtracting the returned mass following the notation in \citet{zahidUniversalRelationGalactic2014}.}, and dividing equation~\eqref{eq:zgas} by $\epsilon\,M_\mathrm{g}^2$, the exact metallicity derivative is
\begin{equation}
  \frac{\mathrm{d}Z_\mathrm{g}}{\mathrm{d}M_\star} = \frac{y - Z_\mathrm{g}\,(1-R+\eta)}{M_\mathrm{g}} - \frac{Z_\mathrm{g}}{M_\mathrm{g}}\,\frac{\mathrm{d}M_\mathrm{g}} {\mathrm{d}M_\star}\,.
  \label{eq:zahid_exact}
\end{equation}
The first assumption is to neglect the second term, i.e.\ to assume that metal production dominates over the effect of a changing gas reservoir:
\begin{equation}
  \frac{\mathrm{d}Z_\mathrm{g}}{\mathrm{d}M_\star}
  \approx \frac{y - Z_\mathrm{g}\,(1-R) - Z_\mathrm{g}\,\eta}
  {M_\mathrm{g}}\,.
  \label{eq:zahid_approx1}
\end{equation}
The second assumption is that the net metal loss rate $\zeta \equiv Z_\mathrm{g}\,\eta$ is constant, motivated empirically by the observed proportionality between the total oxygen expelled and stellar mass \citep{zahidUniversalRelationGalactic2014}.
This allows the metal loss term to be absorbed into a constant net yield $y_N \equiv y - \zeta$, so that equation~(\ref{eq:zahid_approx1}) becomes
\begin{equation}
  \frac{\mathrm{d}Z_\mathrm{g}}{\mathrm{d}M_\star} \approx \frac{y_N - Z_\mathrm{g}\,(1-R)}{M_\mathrm{g}}\,.
  \label{eq:zahid_ode}
\end{equation}
The third assumption is that the gas mass follows a power law in stellar mass, $M_\mathrm{g} = G\,M_\star^g$.
Defining $W \equiv y_N - Z_\mathrm{g}\,(1-R)$, so that $\mathrm{d}W = -(1-R)\,\mathrm{d}Z_\mathrm{g}$, equation~(\ref{eq:zahid_ode}) becomes separable,
\begin{equation}
  \frac{\mathrm{d}W}{W} = -\frac{(1-R)\,\mathrm{d}M_\star}
  {G\,M_\star^g}\,.
\end{equation}
Integrating from $M_\star = 0$ (where $Z_\mathrm{g} = 0$ and hence $W = y_N$) to $M_\star$,
\begin{equation}
  W = y_N \exp\!\left(-\frac{(1-R)\,M_\star^{1-g}}
    {G\,(1-g)}\right).
\end{equation}
Since $M_\mathrm{g} = G\,M_\star^g$, we have $M_\star^{1-g}/G = M_\star/M_\mathrm{g}$, and solving for $Z_\mathrm{g}$ gives
\begin{equation}
  Z_\mathrm{g} = \frac{y_N}{1-R}\left[1 - \exp\!\left( -\frac{1-R}{1-g}\,\frac{M_\star}{M_\mathrm{g}}\right)\right].
  \label{eq:zahid_prefactor}
\end{equation}
The fourth and final assumption is that $(1-R)/(1-g) \approx 1$, since $R \approx g \approx 0.5$, and that the net yield can be written as $y_N/(1-R) = y/(1-R+\eta)$.
Equation~(\ref{eq:zahid_prefactor}) then reduces to
\begin{equation}
  Z_\mathrm{g} = \frac{y}{1-R+\eta}\left[1 - \exp\!\left(
    -\frac{M_\star}{M_\mathrm{g}}\right)\right],
  \label{eq:zahid_universal}
\end{equation}
which is the universal metallicity relation of \citet{zahidUniversalRelationGalactic2014} written in our notation, where we have identified their free parameter $y_N/(1-R)$ with $y/(1-R+\eta)$ by evaluating the constant metal loss rate $\zeta$ at its equilibrium value.
In the gas-rich limit ($M_\star/M_\mathrm{g} \ll 1$), equation~(\ref{eq:zahid_universal}) reduces to $Z_\mathrm{g} \approx y/(1-R+\eta) \times M_\star/M_\mathrm{g}$, which is lower than the exact gFMR (equation~\ref{eq:gas_fmr_ideal}) by a factor of $(1-R)/(1-R+\eta)$.
This offset arises because treating $\zeta$ as a constant at its equilibrium value overestimates the metal loss rate when $Z_\mathrm{g} \ll Z_\mathrm{eq}$: in the inflow-driven regime outflows are negligible and the true metal loss $Z_\mathrm{g}\,\eta \ll Z_\mathrm{eq}\,\eta$.
In the gas-depleted limit ($M_\star/M_\mathrm{g} \gg 1$), $Z_\mathrm{g} \to y/(1-R+\eta)$, recovering the equilibrium metallicity (equation~\ref{eq:zgas_eq}).


\bsp  
\label{lastpage}
\end{document}